\documentclass[bibyear]{aa} 

\usepackage{lscape}
\usepackage{float}
\usepackage{graphicx}
\usepackage{txfonts}
\begin{document} 

\title{(Sub)millimetre interferometric imaging of a sample of COSMOS/AzTEC submillimetre galaxies IV. Physical properties derived from spectral energy distributions}

   \author{O.~Miettinen\inst{1}, I.~Delvecchio\inst{1}, V.~Smol\v{c}i\'{c}\inst{1}, M.~Novak\inst{1}, M.~Aravena\inst{2}, A.~Karim\inst{3}, E.~J.~Murphy\inst{4,5}, E.~Schinnerer\inst{6}, P.~Capak\inst{7}, O.~Ilbert\inst{8}, H.~T.~Intema\inst{9,10}, C.~Laigle\inst{11}, and H.~J.~McCracken\inst{11}}

   \institute{Department of Physics, University of Zagreb, Bijeni\v{c}ka cesta 32, HR-10000 Zagreb, Croatia \\ \email{oskari@phy.hr} \and N\'ucleo de Astronom\'{\i}a, Facultad de Ingenier\'{\i}a, Universidad Diego Portales, Av. Ej\'ercito 441, Santiago, Chile \and Argelander-Institut f\"{u}r Astronomie, Universit\"{a}t Bonn, Auf dem H\"{u}gel 71, D-53121 Bonn, Germany \and Infrared Processing and Analysis Center, California Institute of Technology, MC 314-6, Pasadena, CA 91125, USA \and National Radio Astronomy Observatory, 520 Edgemont Road, Charlottesville, VA 22903, USA \and Max-Planck-Institut f\"{u}r Astronomie, K\"{o}nigstuhl 17, 69117 Heidelberg, Germany \and Spitzer Science Center, 314-6 California Institute of Technology, Pasadena, CA 91125, USA \and Aix Marseille Universit\'e, CNRS, LAM (Laboratoire d'Astrophysique de Marseille), UMR 7326, 13388, Marseille, France \and Leiden Observatory, Leiden University, P.O. Box 9531, NL-2300 RA Leiden, the Netherlands \and National Radio Astronomy Observatory, 1003 Lopezville Road, Socorro, NM 87801-0387, USA \and Sorbonne Universit\'es, UPMC Universit\'e Paris 6 et CNRS, UMR 7095, Institut d'Astrophysique de Paris, 98 bis Boulevard Arago, 75014 Paris, France}

   \date{Received ; accepted}

\authorrunning{Miettinen et al.}
\titlerunning{Physical properties of a sample of COSMOS SMGs}

\abstract {Submillimetre galaxies (SMGs) in the early universe are the potential antecedents of the most massive galaxies we see in the present-day universe. An important step towards quantifying this galactic evolutionary connection is to investigate the fundamental physical properties of SMGs, like their stellar mass content ($M_{\star}$) and star formation rate (SFR).} {We attempt to characterise the physical nature of a 1.1~mm-selected, flux-limited, and interferometrically followed up sample of SMGs in the COSMOS field.} {We used the latest release of the {\tt MAGPHYS} code to fit the multiwavelength (UV to radio) spectral energy distributions (SEDs) of 16 of the target SMGs, which lie at redshifts $z \simeq 1.6-5.3$. We also constructed the pure radio SEDs of our SMGs using three different radio bands (325~MHz, 1.4~GHz, and 3~GHz). Moreover, since two SMGs in our sample, AzTEC1 and AzTEC3, benefit from previous $^{12}$C$^{16}$O line observations, we studied their properties in more detail.} {The median and 16th--84th percentile ranges of $M_{\star}$, infrared ($8-1\,000~\mu$m) luminosity ($L_{\rm IR}$), SFR, dust temperature ($T_{\rm dust}$), and dust mass ($M_{\rm dust}$) were derived to be $\log(M_{\star}/{\rm M}_{\sun})=10.96^{+0.34}_{-0.19}$, $\log(L_{\rm IR}/{\rm L}_{\sun})=12.93^{+0.09}_{-0.19}$, ${\rm SFR}=856^{+191}_{-310}$~${\rm M}_{\sun}~{\rm yr}^{-1}$, $T_{\rm dust}=40.6^{+7.5}_{-8.1}$~K, and $\log(M_{\rm dust}/{\rm M}_{\sun})=9.17^{+0.03}_{-0.33}$, respectively. We found that $63\%$ of our target SMGs lie above the galaxy main-sequence by more than a factor of 3, and hence are starbursts. The 3~GHz radio sizes we have previously measured for the target SMGs were compared with the present $M_{\star}$ estimates, and we found that the $z>3$ SMGs are fairly consistent with the mass--size relationship of $z\sim2$ compact, quiescent galaxies (cQGs). The median radio spectral index is found to be $\alpha=-0.77^{+0.28}_{-0.42}$. The median IR-radio correlation parameter is found to be $q=2.27^{+0.27}_{-0.13}$, which is lower than measured locally (median $q=2.64$). The gas-to-dust mass ratio for AzTEC1 is derived to be $\delta_{\rm gdr}=90^{+23}_{-19}$, while that for AzTEC3 is $33^{+28}_{-18}$. AzTEC1 is found to have a sub-Eddington SFR surface density (by a factor of $2.6^{+0.2}_{-0.1}$), while AzTEC3 appears to be an Eddington-limited starburster. The gas reservoir in these two high-$z$ SMGs would be exhausted in only $\sim86$ and $19$~Myr at the current SFR, respectively.} {A comparison of the {\tt MAGPHYS}-based properties of our SMGs with those of equally bright 870~$\mu$m-selected, ALMA followed-up SMGs in the ECDFS field (the ALESS SMGs), suggests that the two populations share fairly similar physical characteristics, including the $q$ parameter. The somewhat higher $L_{\rm dust}$ for our sources (factor of $1.9^{+9.3}_{-1.6}$ on average) can originate in the longer selection wavelength of 1.1~mm. Although the derived median $\alpha$ is consistent with a canonical synchrotron spectral index, some of our SMGs exhibit spectral flattening or steepening, which can be attributed to different cosmic-ray energy gain and loss mechanisms. A hint of negative correlation is found between the 3~GHz size and the level of starburstiness, and hence cosmic-ray electrons in more compact starbursts might be more susceptible to free-free absorption. Some of the derived low and high $q$ values (compared to the local median) could be the result of a specific merger/post-starburst phase of galaxy evolution. Overall, our results, like the $M_{\star}$--3~GHz radio size analysis and comparison with the stellar masses of $z\sim2$ cQGs, in concert with the star formation properties of AzTEC1 and 3, support the scenario where $z>3$ SMGs evolve into today's giant, gas-poor ellipticals.}


   \maketitle
%

\section{Introduction}

Submillimetre galaxies or SMGs (e.g. \cite{smail1997}; \cite{hughes1998}; 
\cite{barger1998}; \cite{eales1999}) are a population of some of the most extreme, dusty star-forming galaxies in the universe, and have become one of the prime targets for studying massive galaxy evolution across cosmic time (for a recent review, see \cite{casey2014}). Abundant evidence has emerged that high-redshift ($z\gtrsim3$) SMGs are the potential antecedents of the $z \sim 2$ compact, quiescent galaxies (cQGs), which ultimately evolve into the present-day massive ($M_{\star}\geq10^{11}$~M$_{\odot}$), gas-poor elliptical galaxies (e.g. \cite{swinbank2006}; \cite{fu2013}; \cite{toft2014}; \cite{simpson2014}). A better, quantitative understanding of the interconnected physical processes that drive the aforementioned massive galaxy evolution requires us to determine the key physical properties of SMGs, like the stellar mass ($M_{\star}$) and star formation rate (SFR). Fitting the observed multiwavelength spectral energy distributions (SEDs) of SMGs provides an important tool for this purpose. The physical characteristics derived through SED fitting for a well-defined sample of SMGs -- as done in the present study -- can provide new, valuable insights into the evolutionary path from the $z\gtrsim3$ SMG phase to local massive ellipticals. However, these studies are exacerbated by the fact that high-redshift SMGs are also the most dust-obscured objects in the early universe (e.g. \cite{dye2008}; \cite{simpson2014}).

Further insight into the nature of SMGs can be gained by studying the infrared (IR)-radio correlation (e.g. \cite{helou1985}; \cite{yun2001}) of this galaxy population. On the basis of the relative strength of the continuum emission in the IR and radio wavebands, the IR-radio correlation can provide clues to the evolutionary (merger) stage of a starbursting SMG (\cite{bressan2002}), or it can help identify radio-excess active galactic nuclei (AGN) in SMGs (e.g. \cite{delmoro2013}). Moreover, because submillimetre-selected galaxies have been identified over a wide redshift range, from $z\sim0.1$ (e.g. \cite{chapman2005}) to $z=6.34$ (\cite{riechers2013}), it is possible to examine whether the IR-radio correlation of SMGs has evolved across cosmic time. A potentially important bias in the IR-radio correlation studies is the assumption of a single radio spectral index (usually the synchrotron spectral index ranging from $\alpha=-0.8$ to $-0.7$, where $\alpha$ is defined at the end of this section) for all individual sources in the sample. Hence, the sources that have steep ($\alpha \lesssim -1$), flat ($\alpha \approx 0$), or inverted ($\alpha >0$) radio spectra will be mistreated under the simplified assumption of a canonical synchrotron spectral index (see e.g. \cite{thomson2014}). As done in the present work, this can be circumvented by constructing the radio SEDs of the sources when there are enough radio data points available, and derive the radio spectral index values for each individual source.

Ultimately, a better understanding of the physics of star formation in SMGs (and galaxies in general) requires us to investigate the properties of their 
molecular gas content -- the raw material for star formation. Two of our target SMGs benefit from previous $^{12}$C$^{16}$O spectral line observations (\cite{riechers2010}; \cite{yun2015}), which, when 
combined with their SED-based properties derived here, enable us to investigate a multitude of their interstellar medium (ISM) and star formation properties.

In this paper, we study the key physical properties of a sample of SMGs in the Cosmic Evolution Survey 
(COSMOS; \cite{scoville2007}) deep field through fitting their panchromatic SEDs. The layout of this paper is as follows. 
In Sect.~2, we describe our SMG sample, previous studies of their properties, and the employed observational data. The SED analysis and its 
results are presented in Sect.~3. A comparison with previous literature and discussion of the results are presented in Appendix~C and Sect.~4, respectively (Appendices~A and B contain photometry tables and 
details of our target sources). The two high-redshift SMGs in our sample that benefit from CO observations are described in more detail in Appendix~D. In Sect.~5, we summarise the results and present our conclusions. The cosmology adopted in the present work corresponds to a spatially flat $\Lambda$CDM (Lambda cold dark matter) universe with the present-day dark energy density parameter $\Omega_{\Lambda}=0.70$, total (dark+luminous baryonic) matter density parameter $\Omega_{\rm m}=0.30$, and a Hubble constant of $H_0=70$~km~s$^{-1}$~Mpc$^{-1}$. A Chabrier (2003) Galactic-disk initial mass function (IMF) is adopted in the analysis. Throughout this paper we define the radio spectral index, $\alpha$, as $S_{\nu}\propto \nu^{\alpha}$, where $S_{\nu}$ is the flux density at frequency $\nu$. 


\section{Data}

\subsection{Source sample: the JCMT/AzTEC 1.1~mm-selected SMGs}

The target SMGs of the present study were first uncovered by the $\lambda_{\rm obs}=1.1$~mm survey of a COSMOS subfield (0.15~deg$^2$ or 7.5\% of the full 
2~deg$^2$ COSMOS field) carried out with the AzTEC bolometer array on the James Clerk Maxwell Telescope (JCMT) by Scott et al. (2008). 
The angular resolution of these observations was $18\arcsec$ (full-width at half maximum or FWHM). The 30 brightest SMGs that comprise our parent flux-limited sample were found to have de-boosted flux densities of $S_{\rm 1.1\, mm}\geq3.3$~mJy, which correspond to signal-to-noise ratios of S/N$_{\rm 1.1\, mm}\geq4.0$ (see Table~1 in \cite{scott2008}). The 15 brightest SMGs, called AzTEC1--15 (S/N$_{\rm 1.1\, mm}=4.6-8.3$), were followed up with the Submillimetre Array (SMA) at 890~$\mu$m ($2\arcsec$ 
resolution) by Younger et al. (2007, 2009; see also \cite{younger2008}, 2010; \cite{smolcic2011}; \cite{riechers2014}; M.~Aravena et al., in prep.); all the SMGs were interferometrically confirmed. Miettinen et al. (2015a; hereafter Paper~I) 
presented the follow-up imaging results of AzTEC16--30 (S/N$_{\rm 1.1\, mm}=4.0-4.5$) obtained with the Plateau de Bure Interferometer 
(PdBI) at $\lambda_{\rm obs}=1.3$~mm ($\sim 1\farcs8$ resolution). In Paper~I, we combined our results with the Younger et al. (2007, 2009) SMA survey results, and 
concluded that $\sim25\%$ of the 30 single-dish detected sources AzTEC1--30 are resolved into multiple (two to three) components at an angular resolution of about $\sim2\arcsec$, 
making the total number of interferometrically identified SMGs to be 39 among the 30 target sources (but see Appendix~B.3 herein for a revised fraction). Moreover, the median redshift of the full sample of these interferometrically identified SMGs was determined to be $z=3.17\pm0.27$, 
where the quoted error refers to the standard error of the median computed as $\sigma_{\rm median}=1.253\times \sigma/\sqrt{N}$, 
where $\sigma$ is the sample standard deviation, and $N$ is the size of the sample (e.g. \cite{lupton1993}). This high median redshift of 
our target SMGs can be understood to be caused by the long observed-wavelength of $\lambda_{\rm obs}=1.1$~mm at which the sources were identified 
(\cite{bethermin2015}; see also \cite{strandet2016}). The corresponding median rest-frame wavelength probed by 1.1~mm observations, 
$\lambda_{\rm rest}\simeq264$~$\mu$m, is very close to that of the classic 850~$\mu$m-selected SMGs lying at a median redshift of 
$z\simeq2.2$ ($\lambda_{\rm rest}\simeq266$~$\mu$m; \cite{chapman2005}).

Miettinen et al. (2015b; hereafter Paper~II) found that $\sim46\%$ of the present target SMGs are associated with 
$\nu_{\rm obs}=3$~GHz radio emission on the basis of the observations taken by the Karl G.~Jansky Very Large Array (VLA)-COSMOS 3~GHz 
Large Project, which is a sensitive ($1\sigma$ noise of 2.3~$\mu$Jy~beam$^{-1}$), high angular resolution ($0\farcs75$) survey 
(PI: V.~Smol\v{c}i\'{c}; \cite{smolcic2016b}). In Paper~II, we focused on the spatial extent of the radio-emitting regions 
of these SMGs, and derived a median deconvolved angular FWHM major axis size 
of $0\farcs54\pm0\farcs11$. For a subsample of 15 SMGs with available spectroscopic or photometric redshifts we 
derived a median linear major axis FWHM of $4.2\pm0.9$~kpc. In a companion paper by Smol{\v c}i{\'c} et al. (2016a; hereafter Paper~III), 
we present the results of the analysis of the galaxy overdensities hosting our 1.1~mm-selected AzTEC SMGs. In the present follow-up study, 
we derive the fundamental physical properties of our SMGs, including $M_{\star}$, total infrared (IR) luminosity 
($\lambda_{\rm rest}=8-1\,000$~$\mu$m), SFR, and dust mass. In addition, we study the centimetre-wavelength radio SEDs of the sources, and address the 
relationship between the IR and radio luminosities, i.e. the IR-radio correlation among the target SMGs. These provide an important addition 
to the previously determined redshift and 3~GHz size distributions (Papers~I and II), and allow us to characterise further the nature of these SMGs. 

The target SMGs, their coordinates, and redshifts are tabulated in Table~\ref{table:sample}. The ultraviolet (UV)--radio SEDs in the present work are analysed for a subsample of 16 (out of 39) SMGs whose redshift could have been determined through spectroscopic or photometric methods (i.e. not only a lower $z$ limit), and that have a counterpart in the employed photometric catalogues described in Sect.~2.2 below. We note that additional nine sources (AzTEC2, 11-N, 14-W, 17b, 18, 19b, 23, 26a, and 29b) have a $z_{\rm spec}$ or $z_{\rm phot}$ value available, but they either do not have sufficiently wide multiwavelength coverage to derive a reliable UV--radio SED, or no meaningful SED fit could otherwise be obtained (see Appendix~B.2 for details). The remaining 14 sources have only lower redshift limits available (due to the lack of counterparts at other wavelengths). As we have already pointed out in Papers~I and II (see references therein), AzTEC1--30 have not been detected in X-rays, and hence do not appear to harbour any strong AGN. In Paper~II, we found that the $\nu_{\rm obs}=3$~GHz radio emission from our SMGs is powered by processes related to star formation rather than by AGN activity (the brightness temperatures were found to be $T_{\rm B} \ll 10^4$~K). This is further supported by the fact that none of these SMGs were detected with the Very Long Baseline Array (VLBA) observations at a high, milliarcsec resolution at $\nu_{\rm obs}=1.4$~GHz (N.~Herrera Ruiz et al., in prep.). Furthermore, in the present paper we find no evidence of radio-excess emission that would imply the presence of AGN activity (Sect.~4.4). We also note that Riechers et al. (2014) did not detect the highly-excited $J=16-15$ CO line (the upper-state energy $E_{\rm up}/k_{\rm B}=751.72$~K) towards AzTEC3 in their Atacama Large Millimetre/submillimetre Array (ALMA) observations, which is consistent with no AGN contributing 
to the heating of the gas. 

\begin{table}[H]
\renewcommand{\footnoterule}{}
\caption{Source list. The 16 sources for which a UV--radio SED could be properly fit are highlighted in boldface (see Sect.~3.1).}
{\scriptsize
\begin{minipage}{1\columnwidth}
\centering
\label{table:sample}
\begin{tabular}{c c c c c}
\hline\hline 
Source ID & $\alpha_{2000.0}$ & $\delta_{2000.0}$ & Redshift\tablefootmark{a} & $z$ reference\tablefootmark{a}\\
       & [h:m:s] & [$\degr$:$\arcmin$:$\arcsec$] & & \\ 
\hline
\textbf{AzTEC1} &  09 59 42.86 & +02 29 38.2 & $z_{\rm spec}=4.3415$ & 1 \\[1ex]
AzTEC2 & 10 00 08.05 & +02 26 12.2 & $z_{\rm spec}=1.125$ & 2 \\[1ex]
\textbf{AzTEC3} &  10 00 20.70 & +02 35 20.5 & $z_{\rm spec}=5.298$ & 3\\[1ex]
\textbf{AzTEC4} &  09 59 31.72 & +02 30 44.0 & $z_{\rm phot}=1.80_{-0.61}^{+5.18}$ & 4 \\[1ex]
\textbf{AzTEC5} &  10 00 19.75 & +02 32 04.4 & $z_{\rm phot}=3.70_{-0.53}^{+0.73}$ & 4 \\[1ex]
AzTEC6 & 10 00 06.50 & +02 38 37.7 & $z_{\rm radio/submm}>3.52$ & 5 \\[1ex]
\textbf{AzTEC7} &  10 00 18.06 & +02 48 30.5 & $z_{\rm phot}=2.30\pm0.10$ & 6 \\[1ex]
\textbf{AzTEC8} &  09 59 59.34 & +02 34 41.0 & $z_{\rm spec}=3.179$ & 7 \\[1ex]
\textbf{AzTEC9} &  09 59 57.25 & +02 27 30.6 & $z_{\rm phot}=4.60_{-0.58}^{+0.43}$ & 4 \\[1ex]
\textbf{AzTEC10} &  09 59 30.76 & +02 40 33.9 & $z_{\rm phot}=2.79_{-1.29}^{+1.86}$ & 6 \\[1ex]
AzTEC11-N\tablefootmark{b} & 10 00 08.91 & +02 40 09.6 & $z_{\rm spec}=1.599$ & 8  \\[1ex]
\textbf{AzTEC11-S}\tablefootmark{b} & 10 00 08.94 & +02 40 12.3 & $z_{\rm spec}=1.599$ & 8 \\[1ex]
\textbf{AzTEC12} &  10 00 35.29 & +02 43 53.4 & $z_{\rm phot}=2.90_{-0.18}^{+0.31}$ & 4 \\[1ex]
AzTEC13 & 09 59 37.05 & +02 33 20.0 & $z_{\rm radio/submm}>4.07$ & 5 \\[1ex]
AzTEC14-E\tablefootmark{c} & 10 00 10.03 & +02 30 14.7 & $z_{\rm radio/submm}>2.95$ & 5\\[1ex]
AzTEC14-W\tablefootmark{c} & 10 00 09.63 & +02 30 18.0 & $z_{\rm phot}=1.30_{-0.36}^{+0.12}$ & 6 \\[1ex]
\textbf{AzTEC15} & 10 00 12.89 & +02 34 35.7 & $z_{\rm phot}=2.80_{-1.27}^{+2.45}$ & 4  \\[1ex]
AzTEC16 & 09 59 50.069 & +02 44 24.50 & $z_{\rm radio/submm}>2.42$ & 5 \\[1ex]
\textbf{AzTEC17a} & 09 59 39.194 & +02 34 03.83 & $z_{\rm phot}=2.96_{-0.06}^{+0.06}$ & 9 \\[1ex]
AzTEC17b & 09 59 38.904 & +02 34 04.69 & $z_{\rm phot}=4.14_{-1.73}^{+0.87}$ & 5 \\[1ex]
AzTEC18 & 09 59 42.607 & +02 35 36.96 & $z_{\rm phot}=3.00_{-0.17}^{+0.19}$ & 5 \\[1ex]
\textbf{AzTEC19a} & 10 00 28.735 & +02 32 03.84 & $z_{\rm phot}=3.20_{-0.45}^{+0.18}$ & 5 \\[1ex]
AzTEC19b & 10 00 29.256 & +02 32 09.82 & $z_{\rm phot}=1.11\pm0.10$ & 5 \\ [1ex]
AzTEC20 & 10 00 20.251 & +02 41 21.66 & $z_{\rm radio/submm}>2.35$ & 5 \\[1ex]
\textbf{AzTEC21a} & 10 00 02.558 & +02 46 41.74 & $z_{\rm phot}=2.60_{-0.17}^{+0.18}$ & 5 \\[1ex]
\textbf{AzTEC21b} & 10 00 02.710 & +02 46 44.51 & $z_{\rm phot}=2.80_{-0.16}^{+0.14}$ & 5 \\[1ex]
AzTEC21c & 10 00 02.856 & +02 46 40.80 & $z_{\rm radio/submm}>1.93$ & 5  \\[1ex]
AzTEC22 & 09 59 50.681 & +02 28 19.06 & $z_{\rm radio/submm}>3.00$ & 5  \\ [1ex]
AzTEC23 & 09 59 31.399 & +02 36 04.61 & $z_{\rm phot}=1.60_{-0.50}^{+0.28}$ & 5 \\[1ex]
AzTEC24a\tablefootmark{d} & 10 00 38.969 & +02 38 33.90 & $z_{\rm radio/submm}>2.35$ & 5  \\[1ex]
\textbf{AzTEC24b}\tablefootmark{e} & 10 00 39.410 & +02 38 46.97 & $z_{\rm phot}=1.90_{-0.25}^{+0.12}$ & 4 \\[1ex]
AzTEC24c\tablefootmark{d}  & 10 00 39.194 & +02 38 54.46 & $z_{\rm radio/submm}>3.17$ & 5  \\[1ex]
AzTEC25\tablefootmark{f} & \ldots & \ldots & \ldots & \ldots\\[1ex]
AzTEC26a & 09 59 59.386 & +02 38 15.36 & $z_{\rm phot}=2.50_{-0.14}^{+0.24}$ & 5 \\[1ex]
AzTEC26b & 09 59 59.657 & +02 38 21.08 & $z_{\rm radio/submm}>1.79$ & 5 \\[1ex]
AzTEC27 & 10 00 39.211 & +02 40 52.18 & $z_{\rm radio/submm}>4.17$ & 5 \\[1ex]
AzTEC28 & 10 00 04.680 & +02 30 37.30 & $z_{\rm radio/submm}>3.11$ & 5 \\[1ex]
AzTEC29a & 10 00 26.351 & +02 37 44.15 & $z_{\rm radio/submm}>2.96$ & 5 \\[1ex]
AzTEC29b & 10 00 26.561 & +02 38 05.14 & $z_{\rm phot}=1.45_{-0.38}^{+0.79}$ & 5 \\[1ex]
AzTEC30 & 10 00 03.552 & +02 33 00.94 & $z_{\rm radio/submm}>2.51$ & 5 \\[1ex]
\hline 
\end{tabular} 
\tablefoot{The coordinates given in columns~(2) and (3) for AzTEC1--15 refer to the SMA 890~$\mu$m peak position (\cite{younger2007}, 2009), while those for AzTEC16--30 are the PdBI 1.3~mm peak positions (Paper~I). 
\tablefoottext{a}{The $z_{\rm spec}$, $z_{\rm phot}$, and $z_{\rm radio/submm}$ values are the spectroscopic redshift, optical-near-IR photometric redshift, and the redshift derived using the Carilli-Yun redshift indicator (\cite{carilli1999}, 2000). The $z$ references in the last column are as follows: $1=$\cite{yun2015}; $2=$M.~Balokovi\'c et al., in prep.; $3=$\cite{riechers2010} and \cite{capak2011}; $4=$A forthcoming paper on the redshift distribution of the ALMA-detected ASTE/AzTEC SMGs (D.~Brisbin et al., in prep.); $5=$Paper~I; $6=$\cite{smolcic2012}; $7=$D.~A.~Riechers et al., in prep.; $8=$M.~Salvato et al., in prep.; $9=$COSMOS2015 catalogue (\cite{laigle2016}; see our Appendix~B.1).}\tablefoottext{b}{AzTEC11 was resolved into two 890~$\mu$m sources (N and S) by Younger et al. (2009). The two components are probably physically related, i.e. are at the same redshift (see discussion in Paper~II).}\tablefoottext{c}{AzTEC14 was resolved into two 890~$\mu$m sources (E and W) by Younger et al. (2009). The eastern component appears to lie at a higher redshift than the western one (\cite{smolcic2012}).}\tablefoottext{d}{The PdBI 1.3~mm source candidates AzTEC24a and 24c were not detected in the ALMA 1.3~mm imaging of AzTEC24 (M.~Aravena et al., in prep.), and hence are very likely to be spurious.}\tablefoottext{e}{The position of AzTEC24b was revised through ALMA 1.3~mm observations to be $\alpha_{2000.0}=10^{\rm h} 00^{\rm m} 39\fs294$, $\delta_{2000.0}=+02\degr 38\arcmin 45\farcs10$, i.e. $2\farcs55$ away from the PdBI 1.3~mm feature (M.~Aravena et al., in prep.).}\tablefoottext{f}{AzTEC25 was not detected in the 1.3~mm PdBI observations (Paper~I).}
}
\end{minipage} 
}
\end{table}

\subsection{Multiwavelength photometric data}

Our SMGs lie within the COSMOS field, and hence benefit from rich panchromatic datasets across the electromagnetic spectrum (from X-rays to radio). To construct the SEDs of our sources, we employed the most up-to-date photometric catalogue COSMOS2015, which consists of extensive ground and space-based photometric data in the optical to mid-IR wavelength range (\cite{laigle2016}; see also \cite{capak2007}; \cite{ilbert2009}). 

The wide-field imager, MegaCam (\cite{boulade2003}), mounted on the 3.6~m Canada-France-Hawaii Telescope (CFHT), was used to perform deep $u^*$-band 
(effective wavelength $\lambda_{\rm eff}=3\,911$~\AA) observations. Most of the wavelength bands were observed using the Subaru Prime Focus Camera (Suprime-Cam) mounted 
on the 8.2~m Subaru telescope (\cite{miyazaki2002}; \cite{taniguchi2007}, 2015). These include the six broad-band filters $B$, $g^+$, $V$, $r$, $i^+$, and $z^{++}$, 
the 12 intermediate-band filters IA427, IA464, IA484, IA505, IA527, IA574, IA624, IA679, IA709, IA738, IA767, and IA827, and the two narrow bands NB711 and NB816.
The Subaru/Hyper Suprime-Cam (\cite{miyazaki2012}) was used to perform observations in its HSC-$Y$ band (central wavelength $\lambda_{\rm cen}=0.98$~$\mu$m). 
Near-infrared imaging of the COSMOS field in the $Y$ (1.02~$\mu$m), $J$ (1.25~$\mu$m), $H$ (1.65~$\mu$m), and $K_{\rm s}$ (2.15~$\mu$m) bands is being collected by 
the Ultra\-VISTA survey (\cite{mccracken2012}; \cite{ilbert2013})\footnote{The data products are produced by TERAPIX; see {\tt http://terapix.iap.fr}}. 
The Ultra\-VISTA data used in the present work correspond to the data release version 2 (DR2). The Wide-field InfraRed Camera (WIRCam; \cite{puget2004}) on the CFHT was also used for $H$- and $K_{\rm s}$-band imaging. Mid-infrared observations were obtained with the Infrared Array Camera (IRAC; 3.6--8.0 $\mu$m; \cite{fazio2004}) and 
the Multiband Imaging Photometer for \textit{Spitzer} (MIPS; 24--160~$\mu$m; \cite{rieke2004}) on board the \textit{Spitzer} Space Telescope as part of the COSMOS \textit{Spitzer} survey 
(S-COSMOS; \cite{sanders2007}). The IRAC 3.6~$\mu$m and 4.5~$\mu$m observations used here were taken by the \textit{Spitzer} Large Area Survey with 
Hyper Suprime-Cam (SPLASH) during the warm phase of the mission (PI: P.~Capak; see \cite{steinhardt2014}). Far-infrared (100, 160, and 250~$\mu$m) to submm (350 and 500~$\mu$m) 
\textit{Herschel}\footnote{\textit{Herschel} is an ESA space observatory with science instruments provided by European-led Principal Investigator consortia and with important 
participation from NASA.} continuum observations were performed as part of the Photodetector Array Camera and Spectrometer (PACS) Evolutionary Probe (PEP; 
\cite{lutz2011}) and the \textit{Herschel} Multi-tiered Extragalactic Survey (HerMES\footnote{{\tt http://hermes.sussex.ac.uk}}; 
\cite{oliver2012}) programmes. 

From the ground-based single-dish telescope data, we used the deboosted JCMT/AzTEC 1.1~mm flux densities reported by Scott et al. (2008; their Table~1). Moreover, for 
three of our SMGs (AzTEC5, 9, and 19a) we could use the deboosted JCMT/Submillimetre Common User Bolometer Array (SCUBA-2) 450~$\mu$m and 850~$\mu$m flux densities 
from Casey et al. (2013) (for AzTEC9 only a deboosted 850~$\mu$m flux density was available). More importantly, our SMGs benefit from interferometrically observed (sub)mm flux 
densities. Among AzTEC1--15, we used the 890~$\mu$m flux densities measured with the SMA by Younger et al. (2007, 2009), while for sources among AzTEC16--30 we used the PdBI 1.3~mm flux densities from Paper~I. AzTEC1 was observed at 870~$\mu$m with ALMA during the second early science campaign (Cycle~1 ALMA project 2012.1.00978.S; PI: A.~Karim), and its 870~$\mu$m flux density -- as measured through a two-dimensional elliptical Gaussian fit -- is $S_{\rm 870\, \mu m}=14.12\pm0.25$~mJy. We also used the PdBI 3~mm flux density for AzTEC1 from Smol\v{c}i\'{c} et al. (2011; $S_{\rm 3\, mm}=0.30\pm0.04$~mJy). Riechers et al. (2014) used ALMA to measure the 1~mm flux density of AzTEC3 ($S_{\rm 1\, mm}=6.20\pm0.25$~mJy). Finally, we employed the 1.3~mm flux densities from the ALMA follow-up survey (Cycle~2 ALMA project 2013.1.00118.S; PI: M.~Aravena) by M.~Aravena et al. (in prep.) of 129 SMGs uncovered in the Atacama Submillimetre Telescope Experiment (ASTE)/AzTEC 1.1~mm survey (\cite{aretxaga2011}). Among the ALMA 1.3~mm-detected SMGs there are nine sources in common with the current SED target sources (AzTEC1, 4, 5, 8, 9, 11-S, 12, 15, and 24b; moreover, AzTEC2, 6, and 11-N were detected with ALMA at $\lambda_{\rm obs}=1.3$~mm).

To construct the radio SEDs for our SMGs, we employed the 325~MHz observations taken by the Giant Meterwave Radio Telescope (GMRT)-COSMOS survey (A.~Karim et al., in prep.). 
We also used the radio-continuum imaging data at 1.4~GHz taken by the VLA (\cite{schinnerer2007}, 2010), and at 3~GHz taken by the VLA-COSMOS 3~GHz Large 
Project (PI: V.~Smol\v{c}i\'{c}; \cite{smolcic2016b}; see also Paper~II). Hence, we could build the radio SEDs of our SMGs using data points 
at three different frequencies. 

A selected compilation of mid-IR to mm flux densities of our SMGs are listed in Table~\ref{table:fluxes}, while the GMRT and VLA radio flux densities are tabulated in 
Table~\ref{table:radio}. Because of the large beam size (FWHM) of \textit{Herschel}/PACS ($6\farcs7$ and $11\arcsec$ at 100 and 160~$\mu$m, respectively) and SPIRE 
($18\arcsec$, $25\arcsec$, and $36\arcsec$ at 250, 350, and 500~$\mu$m, respectively) observations, the \textit{Herschel} flux densities were derived using a point-spread-function-fitting method, guided by the known position of \textit{Spitzer}/MIPS 24~$\mu$m sources, i.e. we used the 24~$\mu$m-prior based photometry  (\cite{magnelli2012}) given as part of the COSMOS2015 catalogue (\cite{laigle2016}) whenever possible. Because AzTEC1, 3, 4, 8, 9, 10, and 17a are reported as non-detections at 24~$\mu$m in the COSMOS2015 catalogue, we adopted their \textit{Herschel} flux densities from the PACS and SPIRE blind catalogues\footnote{{\tt http://irsa.ipac.caltech.edu/Missions/cosmos.html}}.

\section{Analysis and results}

\subsection{Spectral energy distributions from UV to radio wavelengths}

\subsubsection{Method}

To characterise the physical properties of our SMGs, we constructed their UV to radio SEDs using the multiwavelength data described in Sect.~2.2. 
The observational data were modelled using the Multiwavelength Analysis of Galaxy Physical Properties code {\tt MAGPHYS} 
(\cite{dacunha2008})\footnote{{\tt MAGPHYS} is publicly available, and can be retrieved at {\tt http://www.iap.fr/magphys/magphys/MAGPHYS.html}}. 
The commonly used {\tt MAGPHYS} code has been described in a number of papers (e.g. \cite{dacunha2008}, 2010; \cite{smith2012}; \cite{berta2013}; \cite{rowlands2014a}; 
\cite{hayward2015}; \cite{smolcic2015}; \cite{dacunha2015}), to which we refer the reader for a detailed explanation. Very briefly, {\tt MAGPHYS} is based on a simple 
energy balance argument: the UV-optical photons emitted by young stars are absorbed by dust grains in star-forming regions and the diffuse ISM, and the absorbed energy, 
which heats the grains, is then thermally re-emitted in the IR. 

Here we have made use of a new calibration of {\tt MAGPHYS}, which is optimised to fit simultaneously 
the UV--radio SEDs of $z > 1$ star-forming galaxies, and hence better suited to derive the physical properties of SMGs than the previous 
versions of the code (see \cite{dacunha2015}). The modifications in the updated version include extended prior distributions of star formation history and dust optical thickness, 
and the addition of intergalactic medium absorption of UV photons. A simple radio emission component is also taken into account by assuming 
a far-IR($42.5-122.5$~$\mu$m)-radio correlation with a $q_{\rm FIR}$ distribution centred at $q_{\rm FIR}=2.34$ (the mean value derived by 
Yun et al. (2001) for $z\leq0.15$ galaxies detected with the \textit{Infrared Astronomical Satellite}), and a scatter of $\sigma(q_{\rm FIR})=0.25$ to 
take possible variations into account (see Sect.~3.5 herein). The thermal free-free emission spectral index in {\tt MAGPHYS} is fixed at $\alpha_{\rm ff}=-0.1$, while that of the non-thermal synchrotron emission is fixed at $\alpha_{\rm synch}=-0.8$. The thermal fraction at rest-frame 1.4~GHz is assumed to be $10\%$. We note that these assumptions might be invalid for individual SMGs (see our Sects.~3.2, 4.3, and 4.4).
 The SED models used here assume that the interstellar dust is predominantly heated by the radiation powered by star formation activity, 
while the possible, though presumably weak, AGN contribution is not taken into account; as mentioned in Sect.~2.1, our SMGs do not exhibit any clear 
signatures of AGN in the X-ray or radio emission. An AGN contamination is expected to mainly affect the stellar mass determination by yielding an overestimated value 
(see \cite{hayward2015}; \cite{dacunha2015}). We note that Hayward \& Smith (2015) found that {\tt MAGPHYS} recovers most physical parameters of their simulated galaxies well, 
hence favouring the usage of this SED modelling code. On the other hand, Micha{\l}owski et al. (2014) found that {\tt MAGPHYS}, when employing the Bruzual \& Charlot (2003) stellar population models and a Chabrier (2003) IMF, yields stellar masses that are, on average, 0.1~dex (factor of 1.26) higher than the true values of their simulated SMGs.\footnote{The stellar emission library we have used is built on the unpublished 2007 update of the Bruzual \& Charlot (2003) models (referred to as CB07), where the treatment of thermally pulsating asymptotic giant branch stars has been improved (see \cite{bruzual2007}).}  

\subsubsection{SED results}

\begin{figure*}
\begin{center}
\includegraphics[width=0.2465\textwidth]{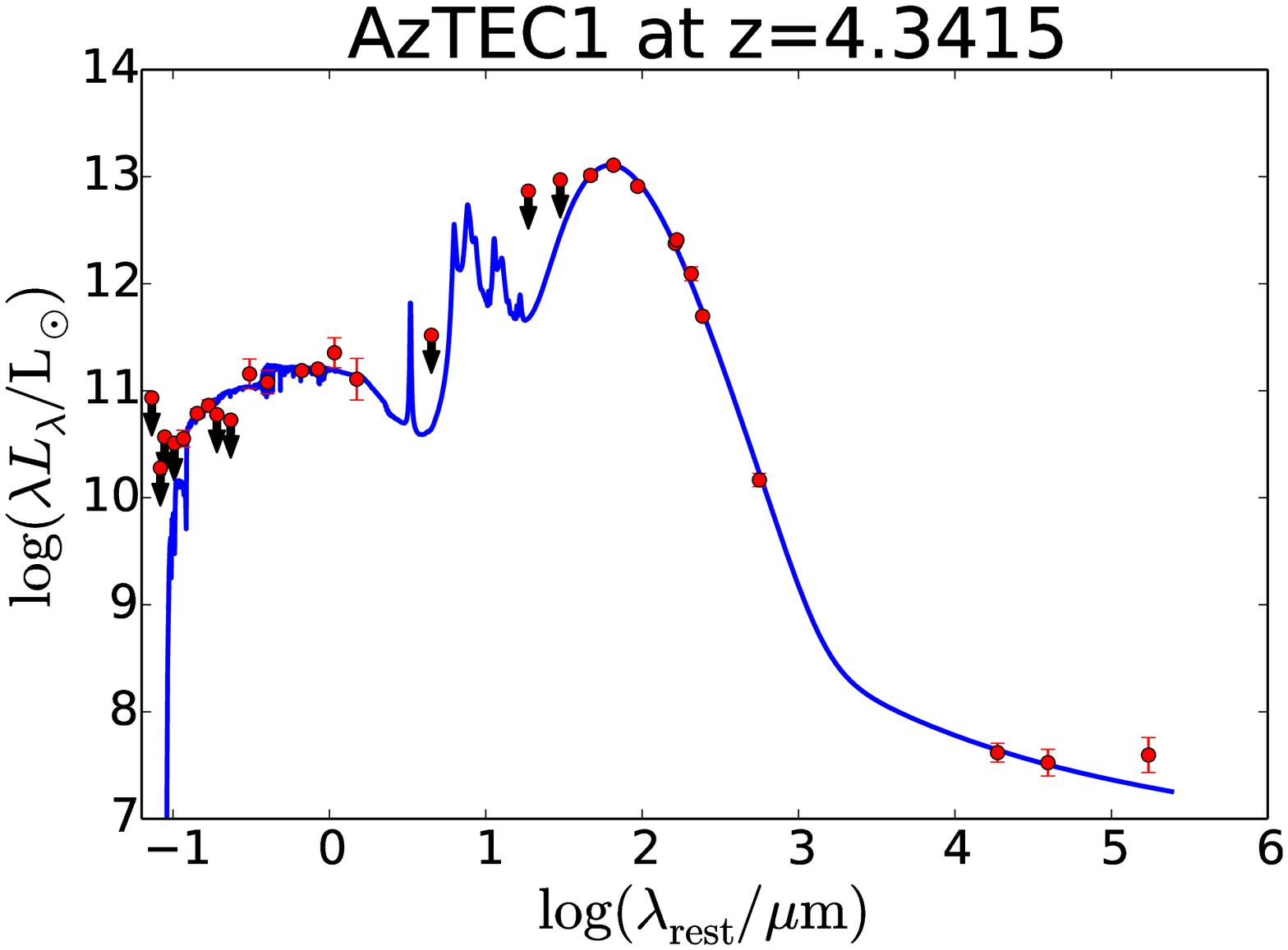}
\includegraphics[width=0.2465\textwidth]{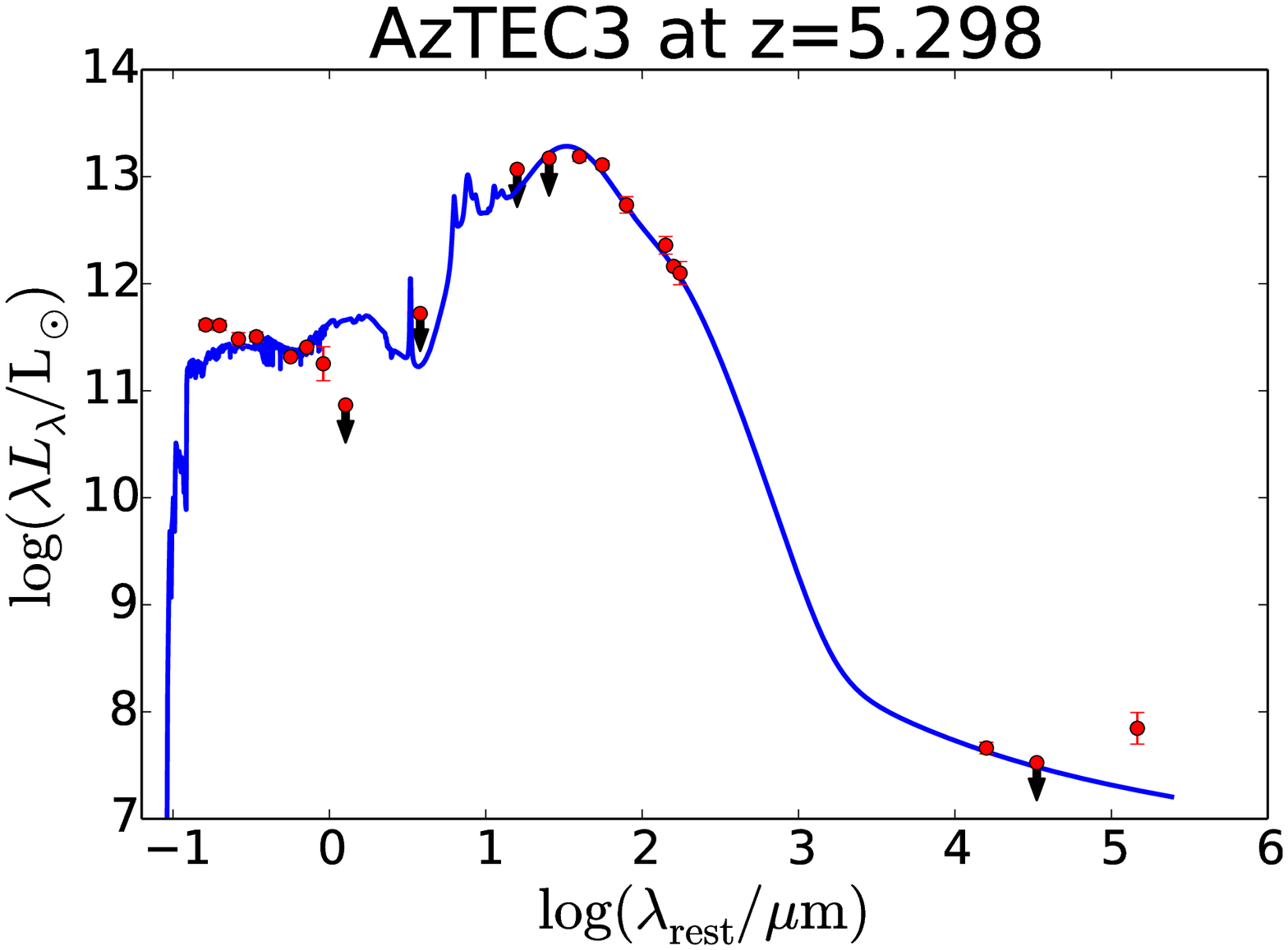}
\includegraphics[width=0.2465\textwidth]{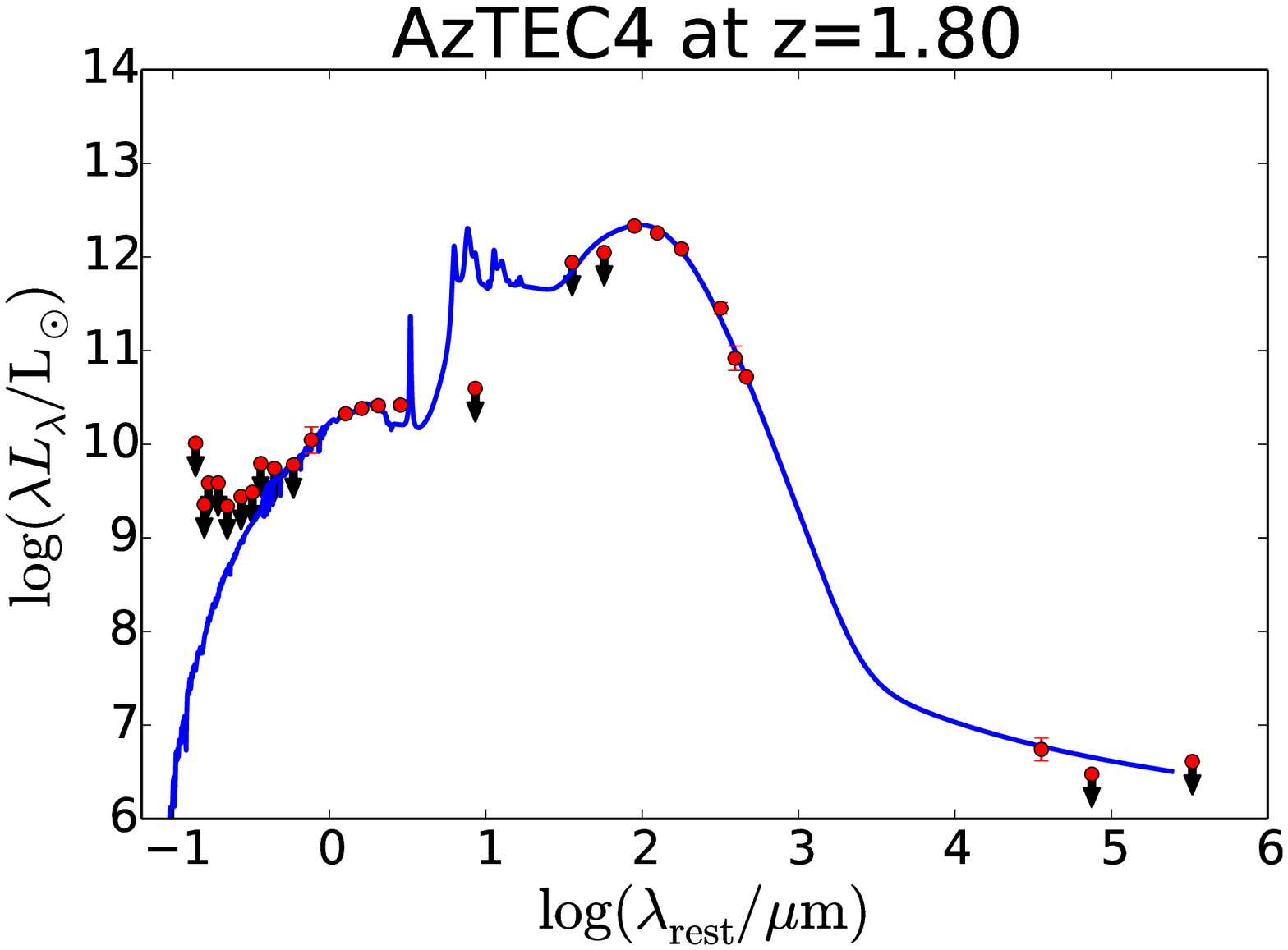}
\includegraphics[width=0.2465\textwidth]{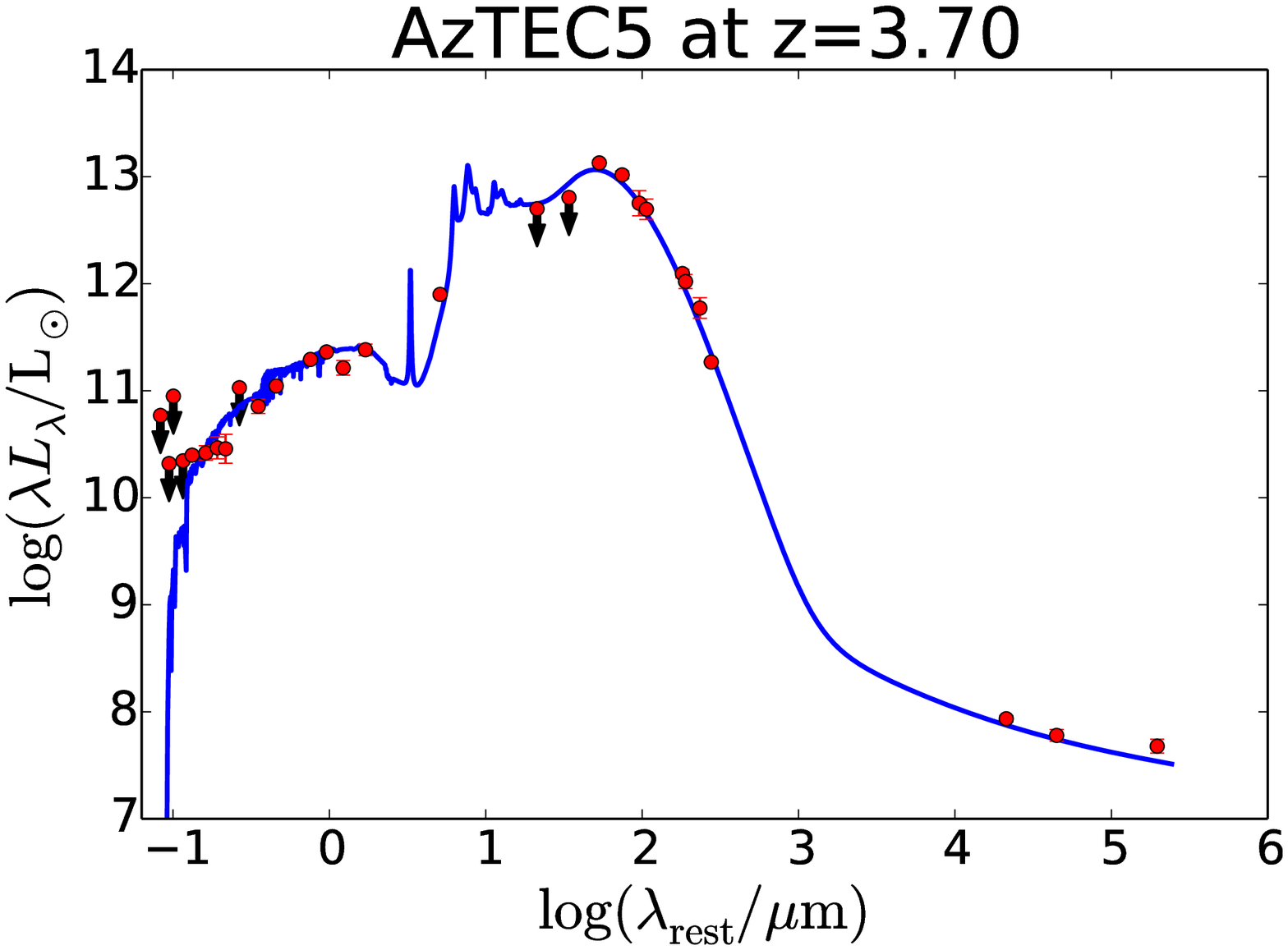}
\includegraphics[width=0.2465\textwidth]{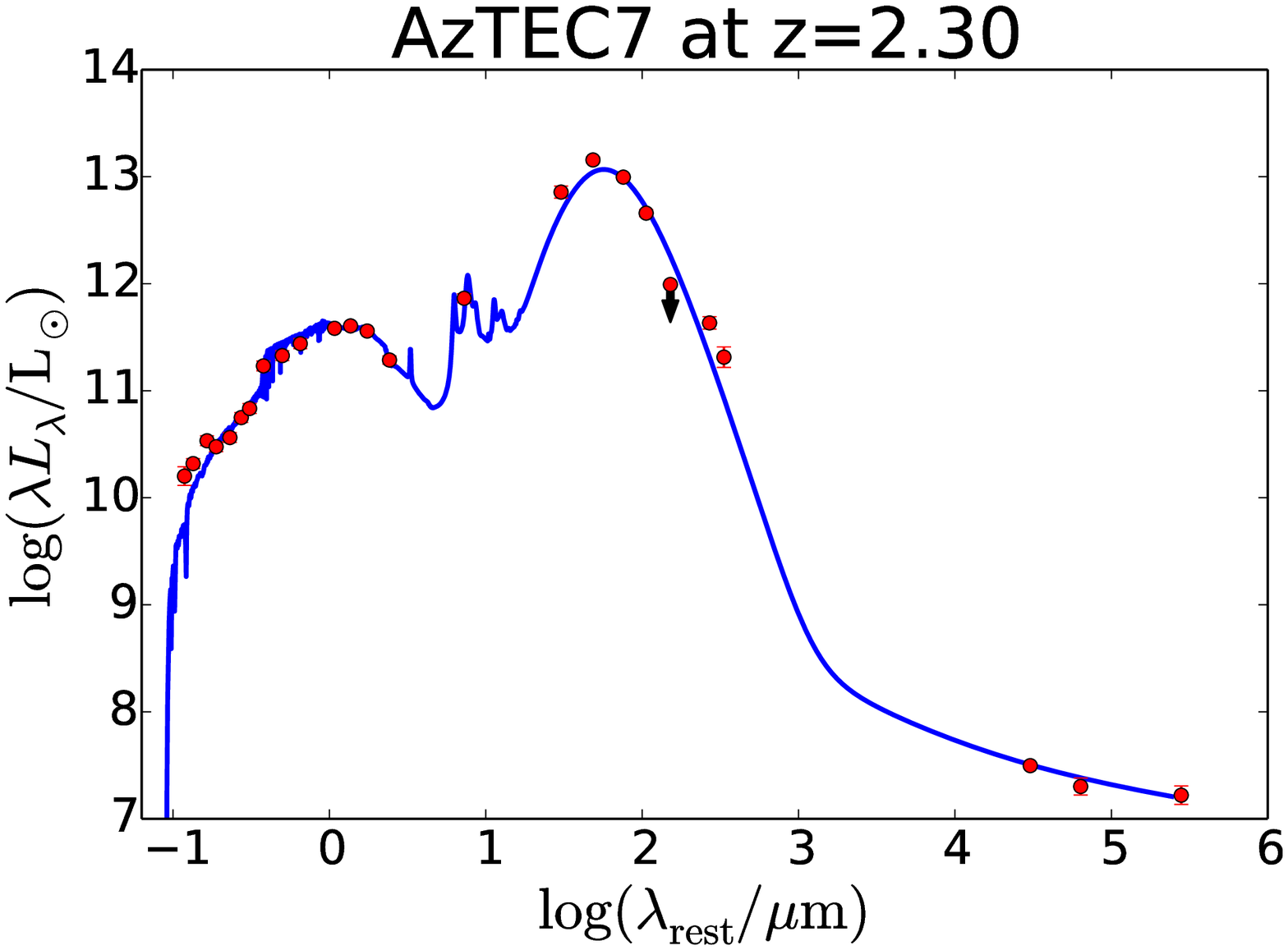}
\includegraphics[width=0.2465\textwidth]{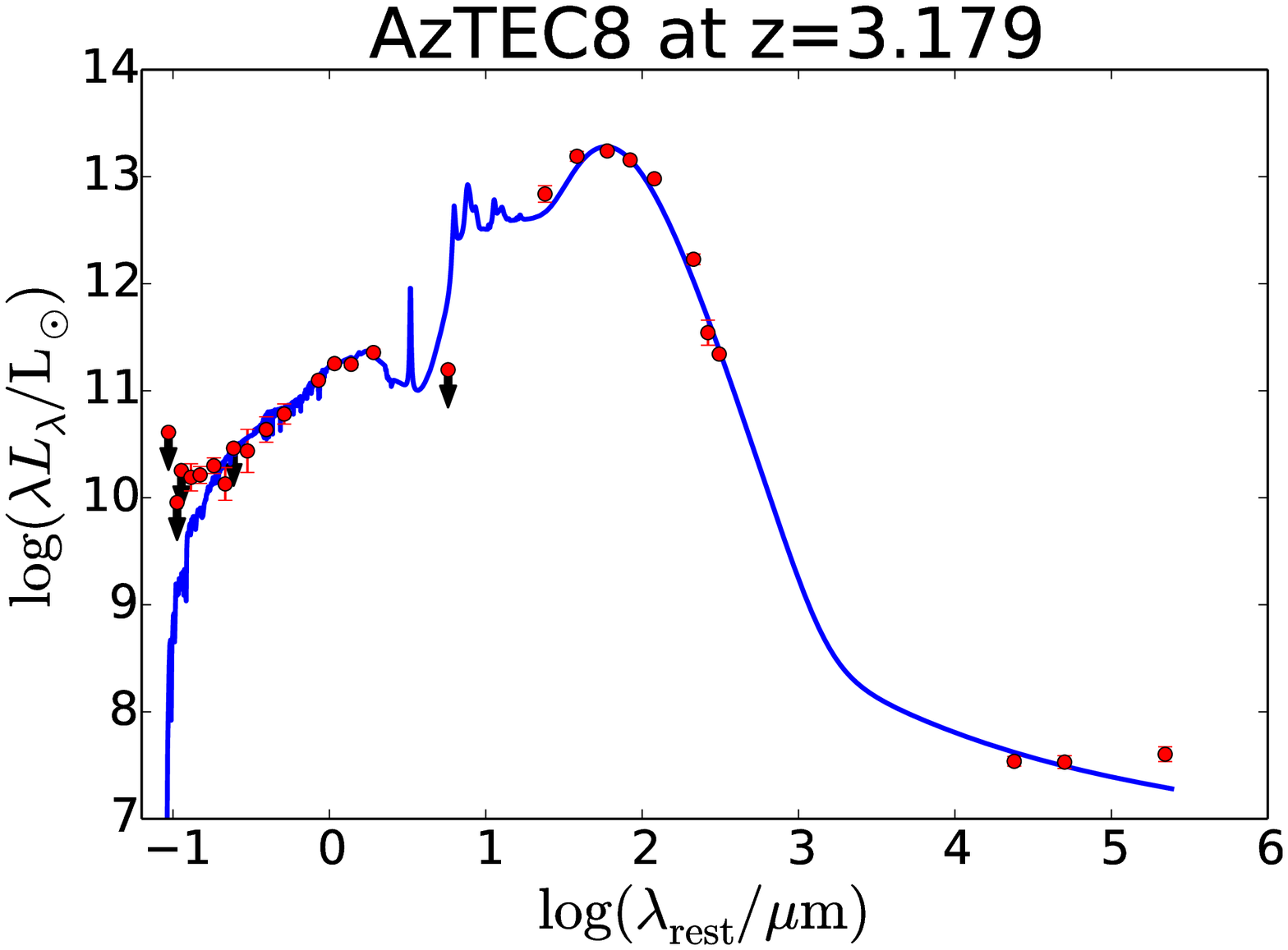}
\includegraphics[width=0.2465\textwidth]{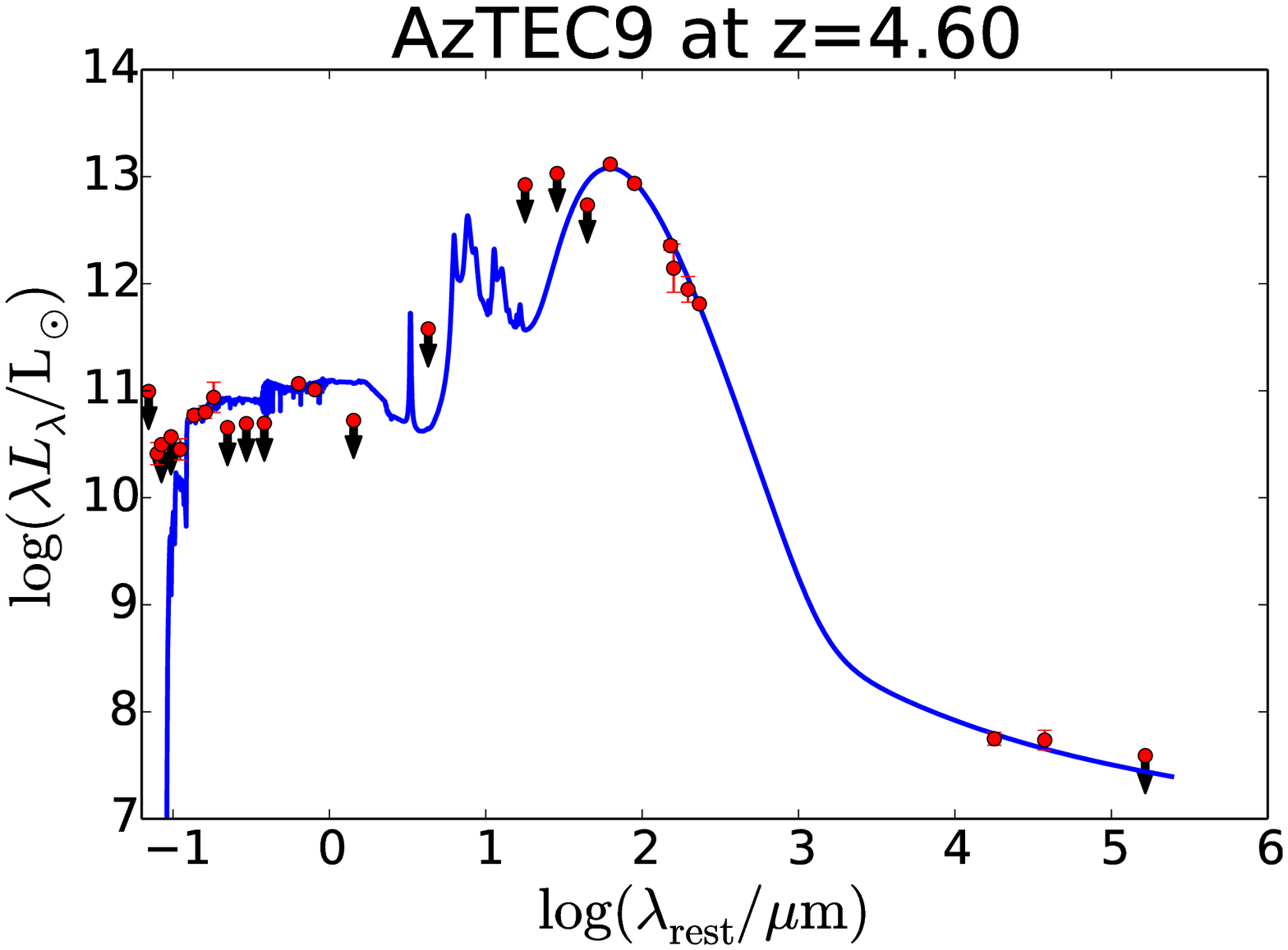}
\includegraphics[width=0.2465\textwidth]{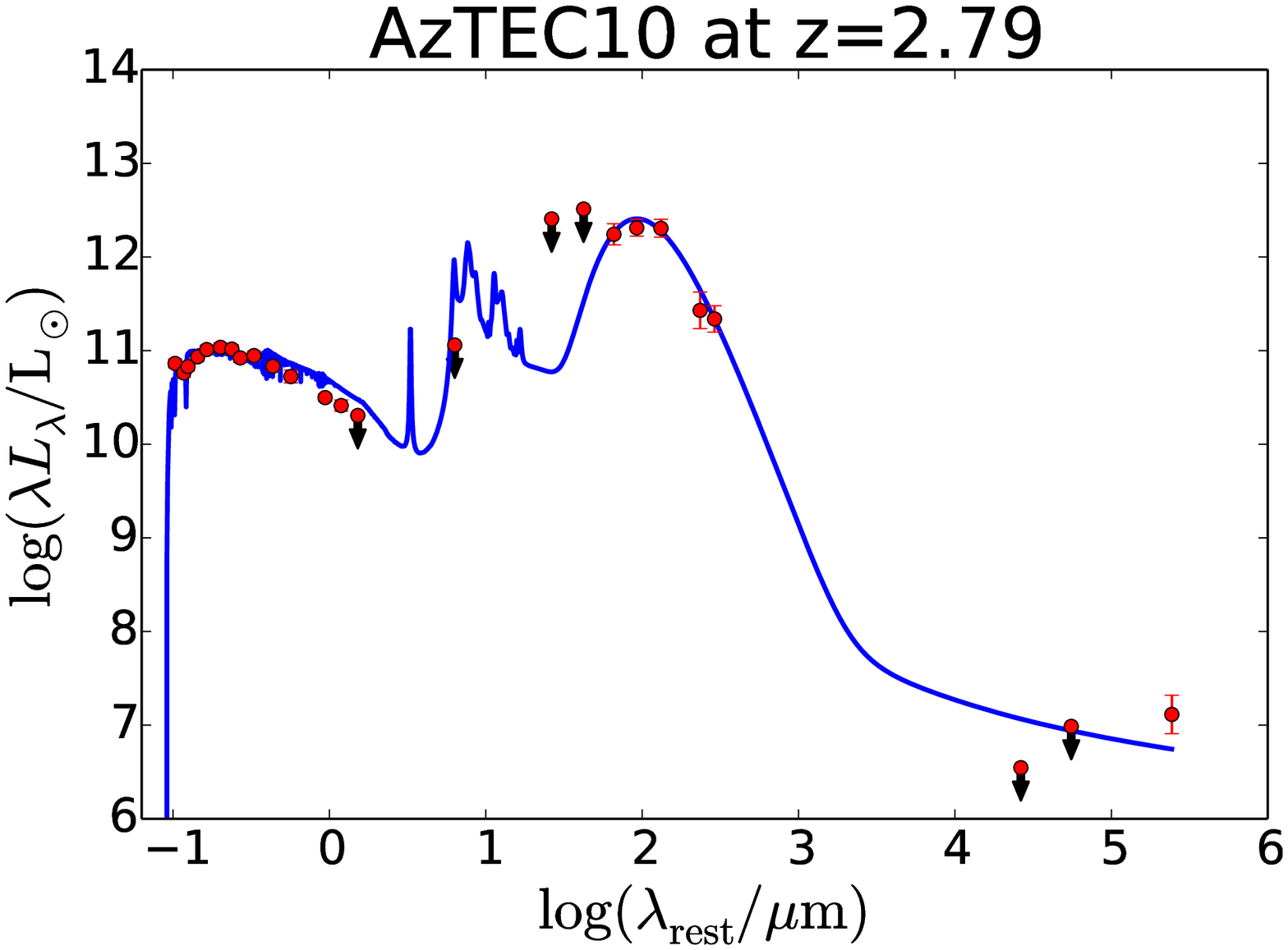}
\includegraphics[width=0.2465\textwidth]{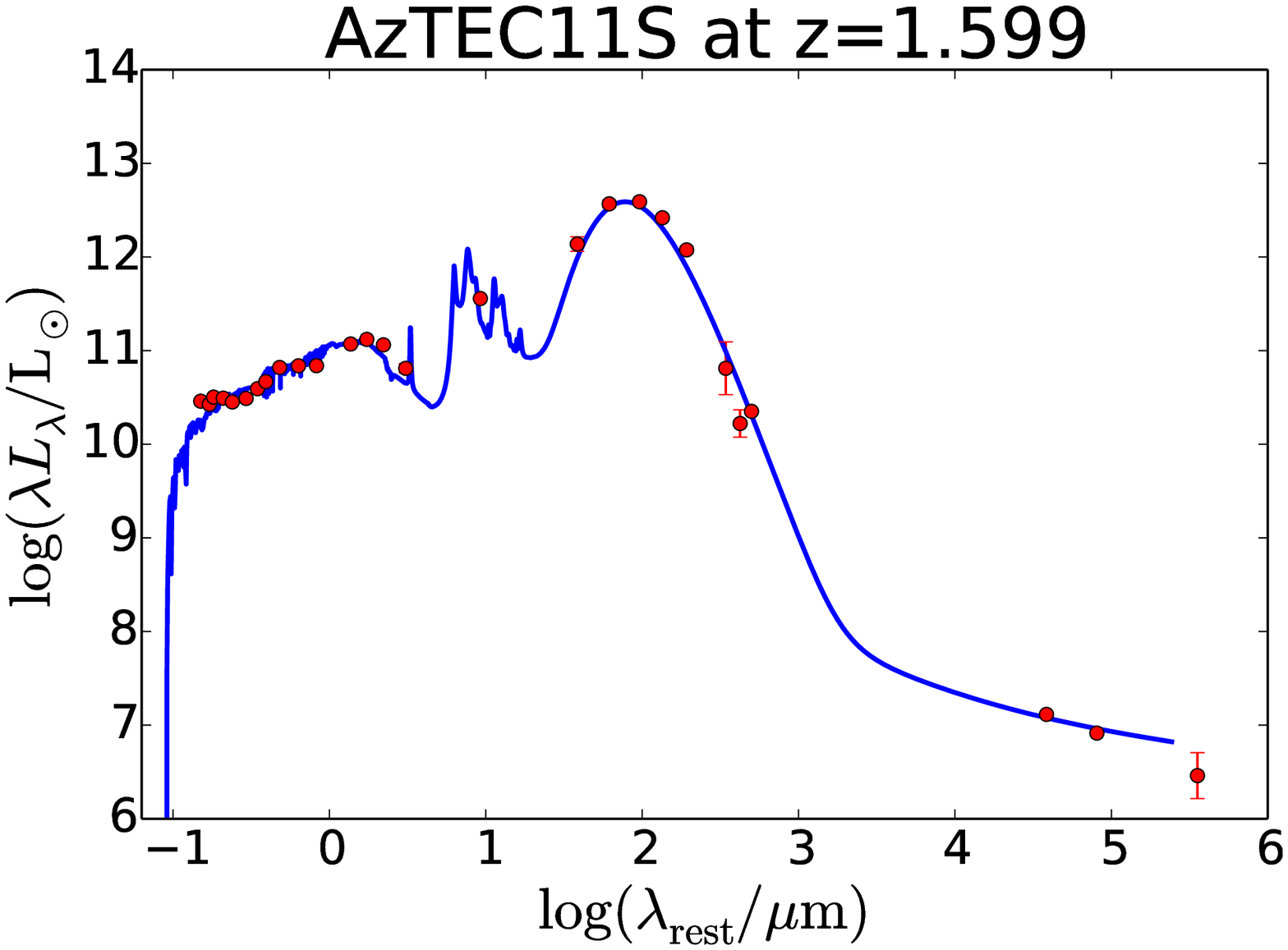}
\includegraphics[width=0.2465\textwidth]{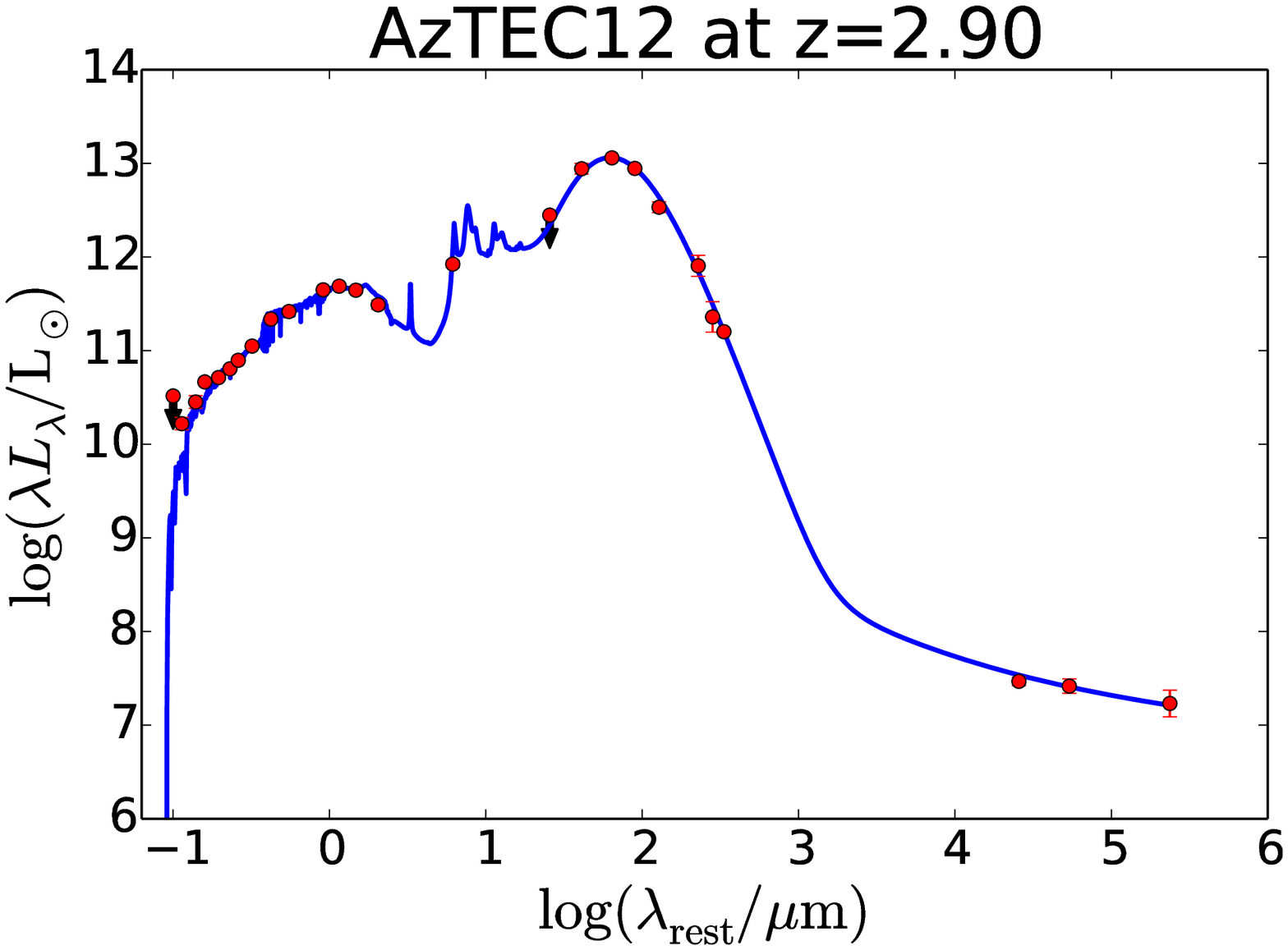}
\includegraphics[width=0.2465\textwidth]{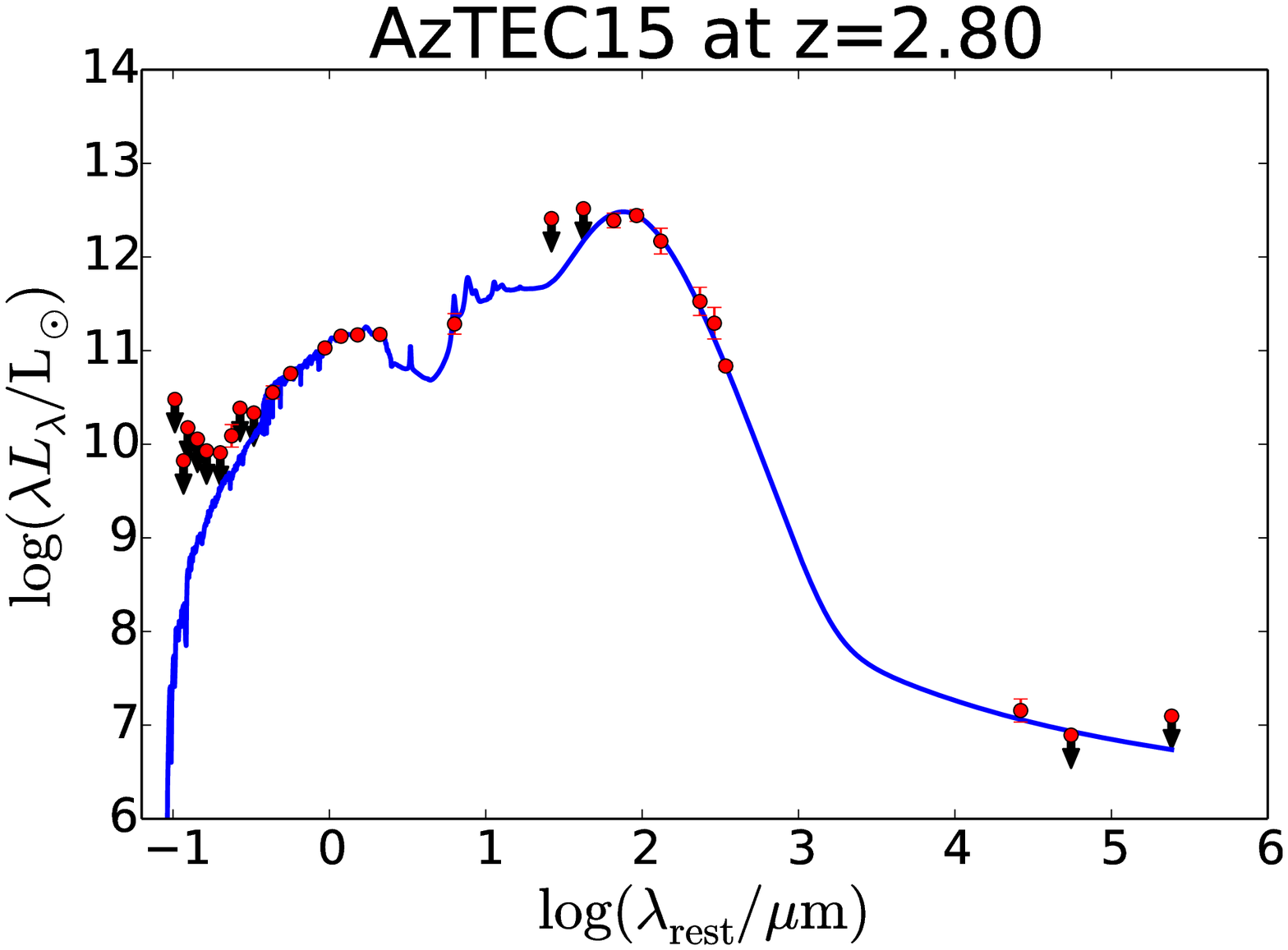}
\includegraphics[width=0.2465\textwidth]{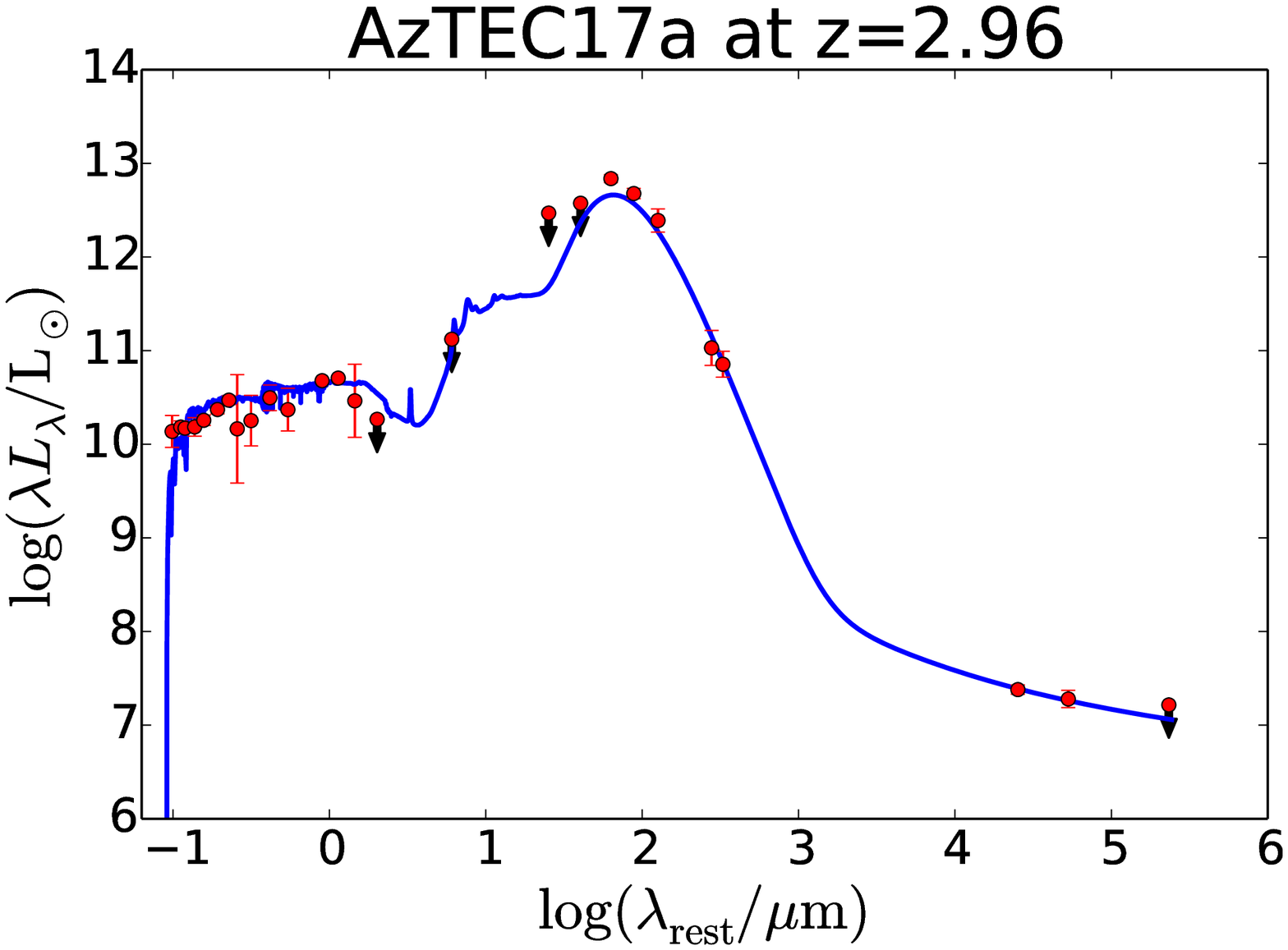}
\includegraphics[width=0.2465\textwidth]{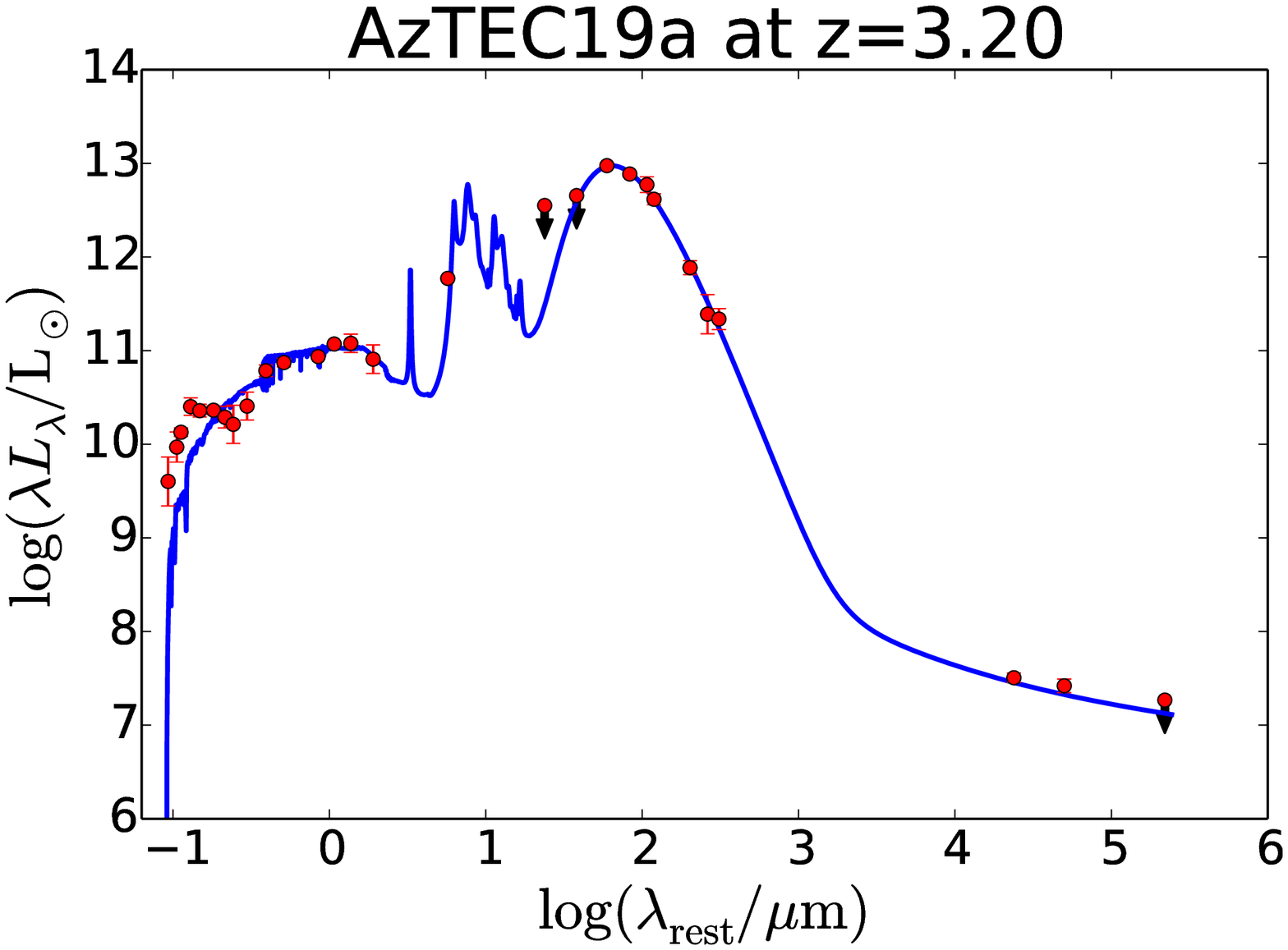}
\includegraphics[width=0.2465\textwidth]{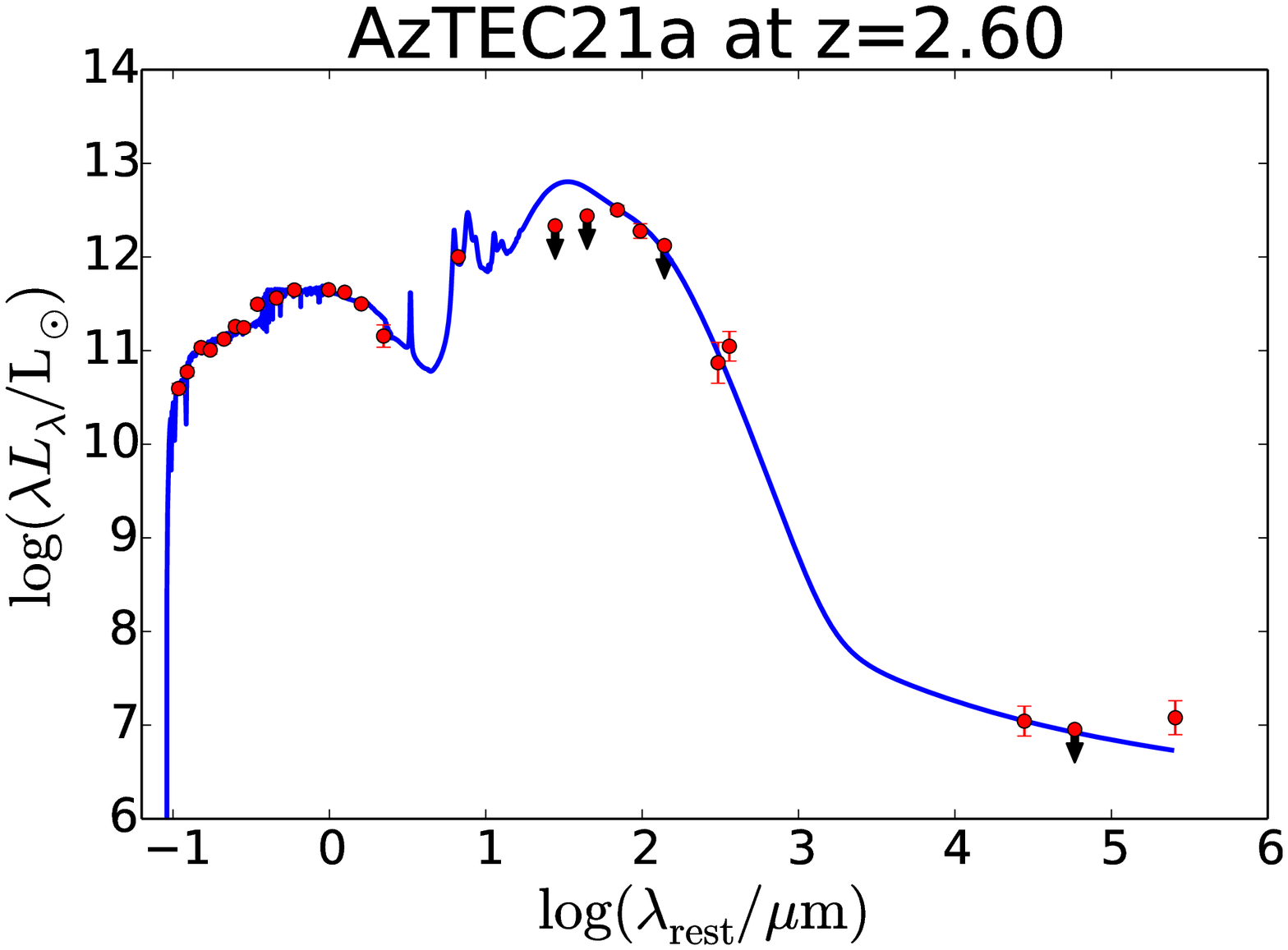}
\includegraphics[width=0.2465\textwidth]{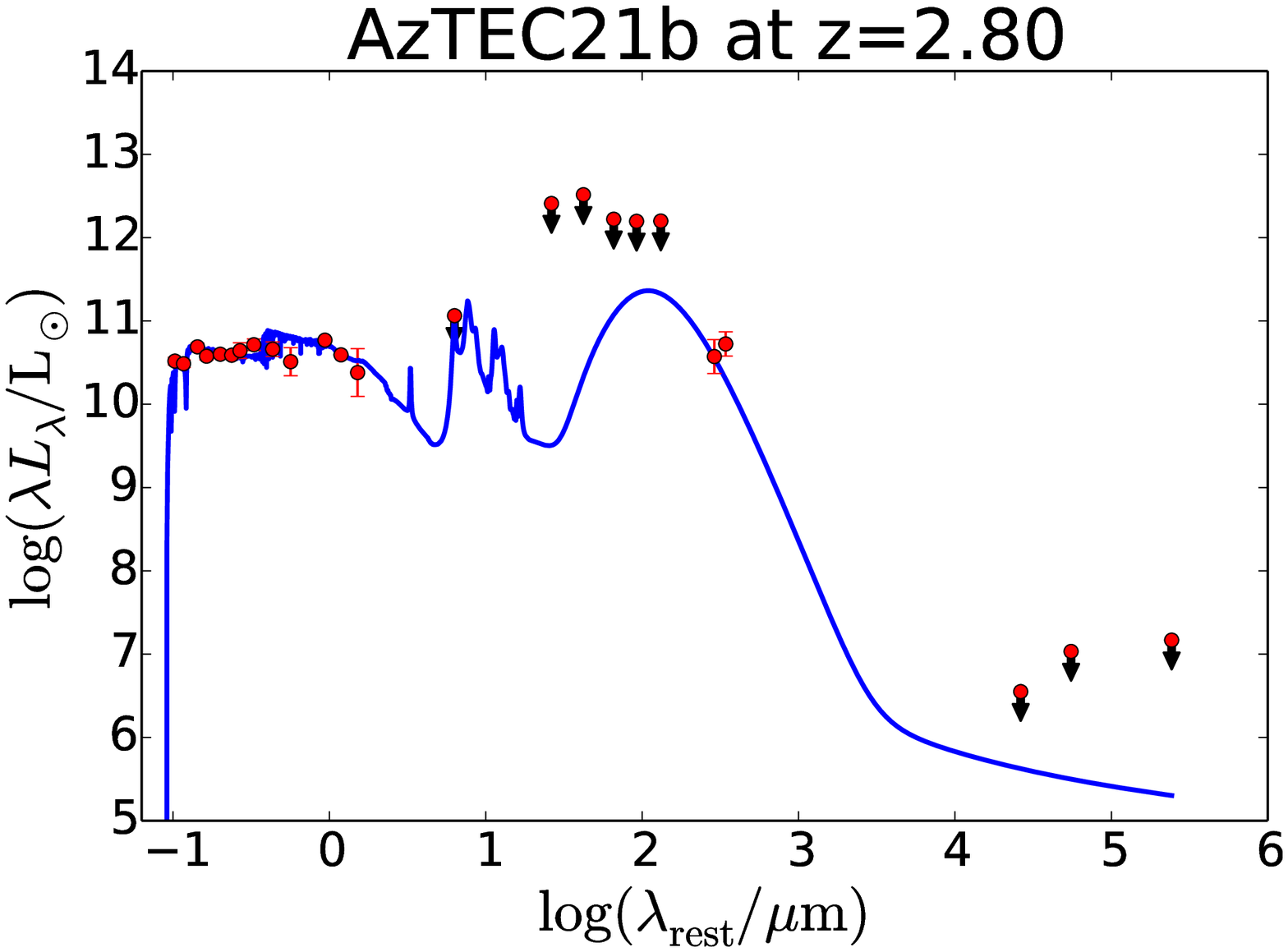}
\includegraphics[width=0.2465\textwidth]{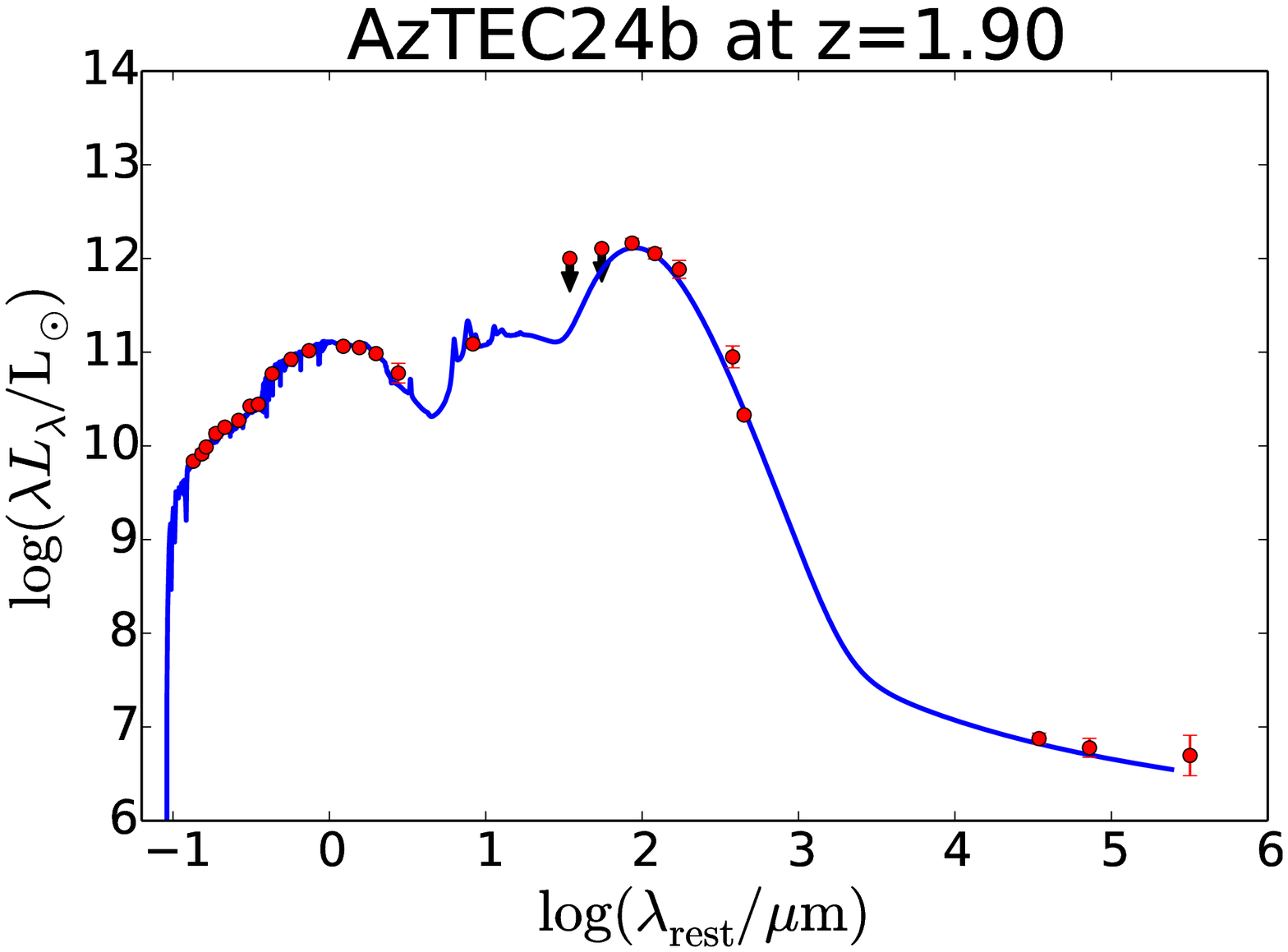}
\caption{Best-fit panchromatic (UV--radio) rest-frame SEDs of 16 of our target SMGs. The source ID and redshift are shown on top of each panel. The red points with vertical error bars represent the observed photometric data, and those with downward pointing arrows mark the $3\sigma$ upper flux density limits (taken into account in the fits). The blue line is the best-fit {\tt MAGPHYS} model SED from the high-$z$ library (\cite{dacunha2015}). We note that all the SMGs except AzTEC21b are detected in at least one radio frequency (see Fig.~\ref{figure:radio} for the pure radio SEDs).}
\label{figure:seds}
\end{center}
\end{figure*}

The resulting SEDs are shown in Fig.~\ref{figure:seds}, and the corresponding SED parameters are given in Table~\ref{table:sed}. We note that the
intermediate and narrow-band Subaru photometry were not used because their effective wavelengths are comparable to those of the broad-band filters, they pass  
only a small portion of the spectrum, and they can be sensitive to optical spectral line features not modelled by {\tt MAGPHYS}. Following da Cunha et al. (2015), 
the flux density upper limits were taken into account by setting the value to zero, and using the upper limit value (here $3\sigma$) as the flux density error. 
As can be seen in Fig.~\ref{figure:seds}, in a few cases the best-fit model disagrees with some of the observed photometric data points or $3\sigma$ upper limits. For example, the \textit{Herschel}/PACS flux density upper limits (set to $3\sigma$) for AzTEC4, 5, and 21a lie slightly 
below the best SED-fit line. As discussed in detail in Paper~I, some of our SMGs have uncertain redshifts. Indeed, our initial SED analysis 
showed that the redshifts we previously adopted for AzTEC9 and 17a might be underestimated, and the revised redshifts of these SMGs are 
described in Appendix~B.1. Moreover, the SED for AzTEC3 was fit using only photometry at and longward of $Y$ band ($\lambda_{\rm rest}=1\,620$~\AA) as the shorter wavelength photometry is likely contaminated/dominated by an unrelated foreground 
($z_{\rm phot}\simeq1$) galaxy as detailed in Appendix~D.2. Finally, as described in Appendix~B.2, we could not obtain a meaningful SED fit for the following five SMGs: AzTEC2, 6, 11-N, 19b, and 26a. 

To calculate the total SFR (0.1--100~M$_{\sun}$) averaged over the past 100~Myr (column~(4) in Table~\ref{table:sed}), 
we used the standard Kennicutt (1998) relationship scaled to a Chabrier (2003) IMF. 
The resulting $L_{\rm IR}-{\rm SFR}$ relationship is given by ${\rm SFR}=10^{-10}\times L_{\rm IR}[{\rm L}_{\sun}]\, {\rm M}_{\sun}~{\rm yr}^{-1}$. 
We note that the Kennicutt (1998) calibration assumes an optically thick starburst, and it does not account for contributions 
from old stellar populations (see \cite{bell2003}). The {\tt MAGPHYS} code also gives the SFR as an output, and 
the model allows for the heating of the dust by old stellar populations. We found a fairly good agreement with the 100~Myr-averaged SFRs calculated from $L_{\rm IR}$ and those directly resulting from the SED fit: the ratio ${\rm SFR}_{\rm IR}/{\rm SFR}_{\rm MAGPHYS}$ was found to range from 0.94 to 3.43 with a median of $1.42^{+0.53}_{-0.22}$, where the $\pm$errors represent the 16th--84th percentile range. 
We note that when this comparison was done using the ${\rm SFR}_{\rm MAGPHYS}$ values averaged over the past 10~Myr (rather than 100~Myr), the ${\rm SFR}_{\rm IR}/{\rm SFR}_{\rm MAGPHYS}$ ratio was found to lie between 0.70 and 1.36, with a median of $0.90^{+0.20}_{-0.13}$ (consistent with \cite{dacunha2015}). In column~(5) in Table~\ref{table:sed}, we give the specific SFR, defined by ${\rm sSFR}\equiv{\rm SFR}/M_{\star}$. The quantity sSFR is unaffected by the adopted stellar IMF (in the case where the newly-forming stars have a same IMF as the pre-existing stellar population). The SFR with respect to that of a main-sequence galaxy of the same stellar mass is given in column~(6) in Table~\ref{table:sed}, and will be described in Sect.~3.3. Finally, we note that the dust temperature given in column~(7) is a new {\tt MAGPHYS} output parameter in the latest version, and refers to an average, dust luminosity-weighted temperature (see Eq.~(8) in \cite{dacunha2015} for the formal definition).

Most of the SMGs analysed here have only a photometric redshift estimate available (the following 12 sources: AzTEC4, 5, 7, 9, 10, 12, 15, 17a, 19a, 21a, 21b, and 24b). As shown in Table~\ref{table:sample}, some of the photometric redshift uncertainties (here reported as the 99\% confidence interval; see e.g. Paper~I) are large, and we took those uncertainties into account by fitting the source SED over the quoted range of redshifts using a fine redshift grid of $\Delta z = 0.01$. We computed the 16th--84th percentile range of the resulting distribution for each {\tt MAGPHYS} output parameter listed in Table~\ref{table:sed}, and propagated those values as the uncertainty estimates on the physical parameters. The uncertainties derived using this approach should be interpreted as lower limits to the true uncertainties. 
We note that da Cunha et al. (2015) left the redshift as a free parameter in their {\tt MAGPHYS} analysis, in which case the derived photometric redshift uncertainties could be directly and self-consistently included in the uncertainties of all other output parameters. However, this option is not yet possible in the publicly available version of the {\tt MAGPHYS} high-$z$ extension (E.~da Cunha, priv.~comm.).

\begin{table*}[!htb]
\caption{Results of {\tt MAGPHYS} SED modelling of the target SMGs.}
{\small
\begin{minipage}{2\columnwidth}
\centering
\renewcommand{\footnoterule}{}
\label{table:sed}
\begin{tabular}{c c c c c c c c}
\hline\hline 
Source ID & $\log(M_{\star}/{\rm M}_{\sun})$ & $\log(L_{\rm IR}/{\rm L}_{\sun})$ & SFR [${\rm M}_{\sun}~{\rm yr}^{-1}$] & sSFR [${\rm Gyr}^{-1}$] & ${\rm SFR}/{\rm SFR}_{\rm MS}$ & $T_{\rm dust}$ [K] & $\log(M_{\rm dust}/{\rm M}_{\sun})$ \\[1ex]
\hline
AzTEC1 & $10.88^{+0.01}_{-0.01}$ & $13.21^{+0.01}_{-0.01}$ & $1\,622^{+38}_{-37}$ & $21.4^{+1.0}_{-1.0}$ & $6.6^{+0.2}_{-0.1}$ & $41.4^{+0.9}_{-1.2}$ & $9.19^{+0.04}_{-0.04}$ \\ [1ex]
AzTEC3 & $10.95^{+0.01}_{-0.02}$ & $13.44^{+0.01}_{-0.01}$ & $2\,754^{+64}_{-62}$ & $30.9^{+2.2}_{-1.4}$ & $8.7^{+0.2}_{-0.2}$ & $66.0^{+5.1}_{-8.4}$ & $9.21^{+0.34}_{-0.27}$ \\ [1ex]
AzTEC4 & $10.96^{+0.54}_{-0.27}$ & $12.58^{+1.13}_{-0.06}$ & $380^{+4749}_{-49}$ & $4.2^{+100.6}_{-3.1}$ & $3.2^{+40.1}_{-0.4}$ & $31.5^{+36.5}_{-3.3}$ & $9.64^{+0.03}_{-0.86}$ \\ [1ex]
AzTEC5 & $11.10^{+0.18}_{-0.14}$ & $13.32^{+0.18}_{-0.14}$ & $2\,089^{+1073}_{-575}$ & $16.6^{+18.1}_{-8.6}$ & $6.3^{+3.3}_{-1.7}$ & $48.7^{+3.3}_{-4.1}$ & $9.04^{+0.00}_{-0.28}$ \\ [1ex]
AzTEC7 & $11.72^{+0.05}_{-0.09}$ & $13.17^{+0.08}_{-0.02}$ & $1\,479^{+299}_{-66}$ & $2.8^{+1.3}_{-0.4}$ & $2.4^{+0.5}_{-0.1}$ & $47.4^{+3.5}_{-1.3}$ & $8.82^{+0.21}_{-0.02}$ \\ [1ex]
AzTEC8 & $10.93^{+0.00}_{-0.02}$ & $13.44^{+0.00}_{-0.02}$ & $2\,754^{+0}_{-124}$ & $32.4^{+1.5}_{-1.5}$ & $13.0^{+0.0}_{-0.6}$ & $44.1^{+3.9}_{-0.0}$ & $9.18^{+0.08}_{-0.00}$ \\ [1ex]
AzTEC9 & $10.67^{+0.06}_{-0.12}$ & $13.17^{+0.07}_{-0.11}$ & $1\,479^{+259}_{-331}$ & $31.6^{+17.4}_{-10.2}$ & $8.5^{+1.5}_{-1.9}$ & $42.0^{+1.9}_{-4.0}$ & $9.19^{+0.10}_{-0.06}$ \\ [1ex]
AzTEC10 & $10.24^{+0.00}_{-0.41}$ & $12.42^{+0.00}_{-0.41}$ & $263^{+0}_{-161}$ & $15.1^{+23.8}_{-9.3}$ & $4.8^{+0.0}_{-3.0}$ & $30.6^{+2.4}_{-2.2}$ & $9.20^{+0.15}_{-0.46}$ \\ [1ex]
AzTEC11-S & $11.01^{+0.00}_{-0.00}$ & $12.68^{+0.00}_{-0.00}$ & $479^{+0}_{-0}$ & $4.7^{+0.0}_{-0.0}$ & $4.4^{+0.0}_{-0.0}$ & $34.1^{+0.0}_{-0.0}$ & $9.15^{+0.00}_{-0.00}$ \\ [1ex]
AzTEC12 & $11.62^{+0.04}_{-0.07}$ & $13.21^{+0.44}_{-0.05}$ & $1\,622^{+2847}_{-177}$ & $3.9^{+8.7}_{-0.7}$ & $2.4^{+4.2}_{-0.3}$ & $42.2^{+6.1}_{-1.6}$ & $9.20^{+0.04}_{-0.08}$ \\ [1ex]
AzTEC15 & $11.41^{+0.28}_{-0.06}$ & $12.63^{+0.58}_{-0.36}$ & $427^{+1195}_{-241}$ & $1.7^{+5.6}_{-1.3}$ & $1.0^{+2.7}_{-0.5}$ & $37.6^{+16.8}_{-5.3}$ & $9.01^{+0.19}_{-0.40}$ \\ [1ex]
AzTEC17a & $10.24^{+0.02}_{-0.01}$ & $12.77^{+0.01}_{-0.02}$ & $589^{+14}_{-27}$ & $33.9^{+1.6}_{-3.0}$ & $10.3^{+0.2}_{-0.5}$ & $39.9^{+0.0}_{-2.5}$ & $8.86^{+0.12}_{-0.00}$ \\ [1ex]
AzTEC19a & $10.86^{+0.10}_{-0.20}$ & $13.05^{+0.07}_{-0.16}$ & $1\,122^{+196}_{-346}$ & $15.5^{+13.3}_{-7.0}$ & $6.0^{+1.0}_{-1.8}$ & $39.1^{+2.1}_{-2.1}$ & $9.13^{+0.08}_{-0.06}$ \\ [1ex]
AzTEC21a & $11.49^{+0.03}_{-0.04}$ & $12.77^{+0.06}_{-0.06}$ & $589^{+87}_{-76}$ & $1.9^{+0.5}_{-0.4}$ & $1.2^{+0.2}_{-0.2}$ & $49.0^{+6.8}_{-8.4}$ & $8.72^{+0.26}_{-0.26}$  \\ [1ex]
AzTEC21b & $10.34^{+0.13}_{-0.03}$ & $11.33^{+0.13}_{-0.46}$ & $21^{+8}_{-14}$ & $1.0^{+0.5}_{-0.7}$ & $0.3^{+0.1}_{-0.2}$ & $36.4^{+11.0}_{-7.0}$ & $8.03^{+0.49}_{-0.90}$ \\ [1ex]
AzTEC24b & $11.31^{+0.03}_{-0.15}$ & $12.25^{+0.05}_{-0.11}$ & $178^{+22}_{-40}$ & $0.9^{+0.5}_{-0.2}$ & $0.8^{+0.1}_{-0.2}$ & $29.9^{+2.2}_{-0.0}$ & $9.20^{+0.00}_{-0.15}$ \\ [1ex]
\hline 
Median & $10.96^{+0.34}_{-0.19}$ & $12.93^{+0.09}_{-0.19}$ & $856^{+191}_{-310}$ & $9.9^{+21.4}_{-8.1}$ & $4.6^{+4.0}_{-3.5}$ & $40.6^{+7.5}_{-8.1}$ & $9.17^{+0.03}_{-0.33}$ \\ [1ex]
\hline 
\end{tabular} 
\tablefoot{The columns are as follows: (1) the name of the SMG; (2) stellar mass; (3) IR luminosity calculated by integrating the SED over the rest-frame wavelength range of 
$\lambda_{\rm rest}=8-1\,000$~$\mu$m; (4) SFR calculated using the $L_{\rm IR}-{\rm SFR}$ relationship of Kennicutt (1998); (5) specific SFR ($={\rm SFR}/M_{\star}$); 
(6) ratio of SFR to that of a main-sequence galaxy of the same stellar mass (i.e. offset from the main sequence; see Sect.~3.3); (7) luminosity-weighted dust temperature (see Eq.~(8) in \cite{dacunha2015}); (8) dust mass. The quoted values and their uncertainties represent the median of the likelihood distribution, and its 68\% confidence interval (corresponding to the 16th--84th percentile range). The uncertainties of the photometric redshifts (see Table~\ref{table:sample}) have also been propagated to the derived parameters (see text for details). The last row tabulates the median value of the parameters, where the indicated $\pm$ range corresponds to the 16th--84th percentile range.}
\end{minipage} }
\end{table*}



\subsection{Pure radio spectral energy distributions, and spectral indices}

\begin{figure*}[!htb]
\begin{center}
\includegraphics[width=0.2465\textwidth]{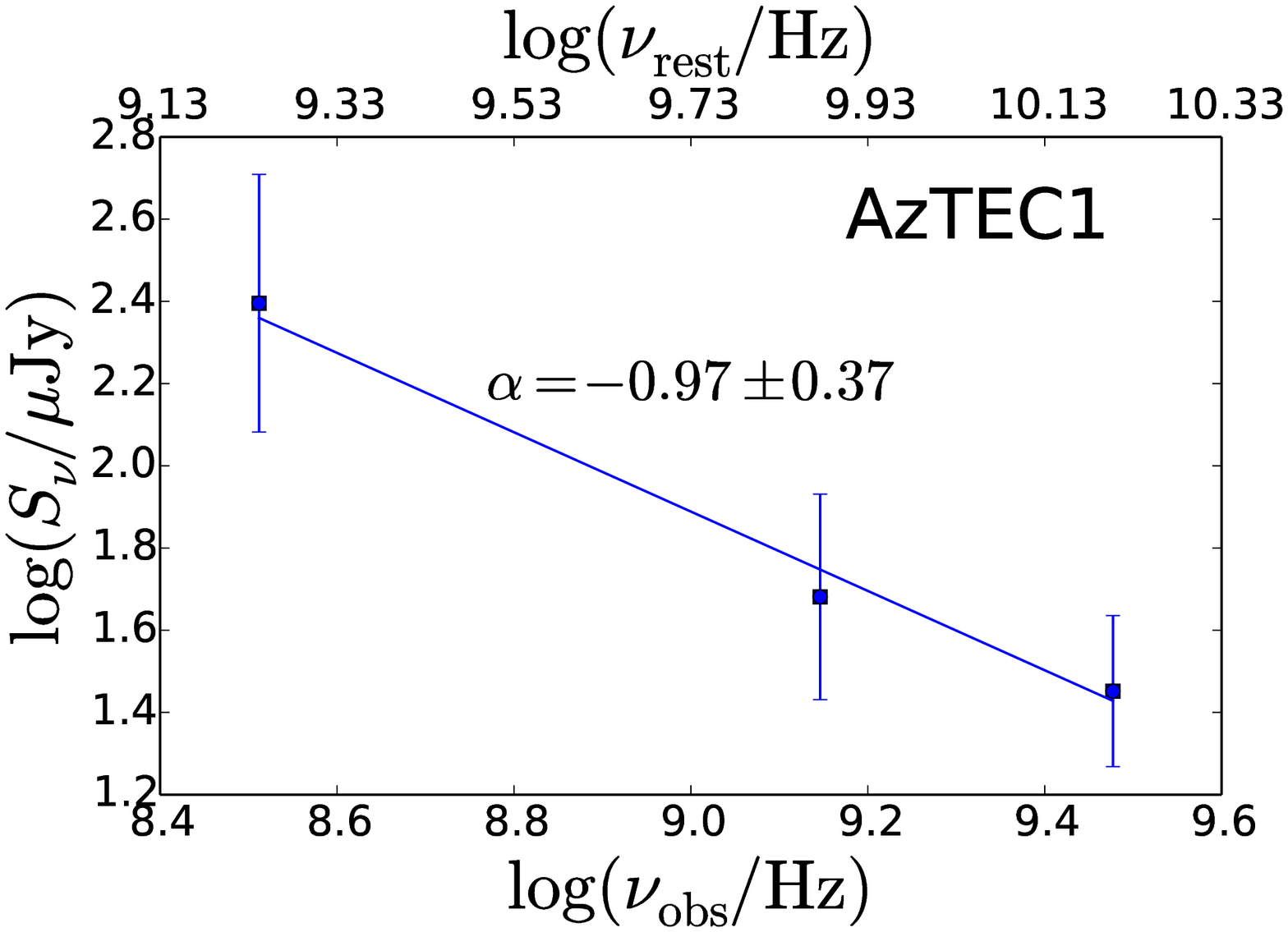}
\includegraphics[width=0.2465\textwidth]{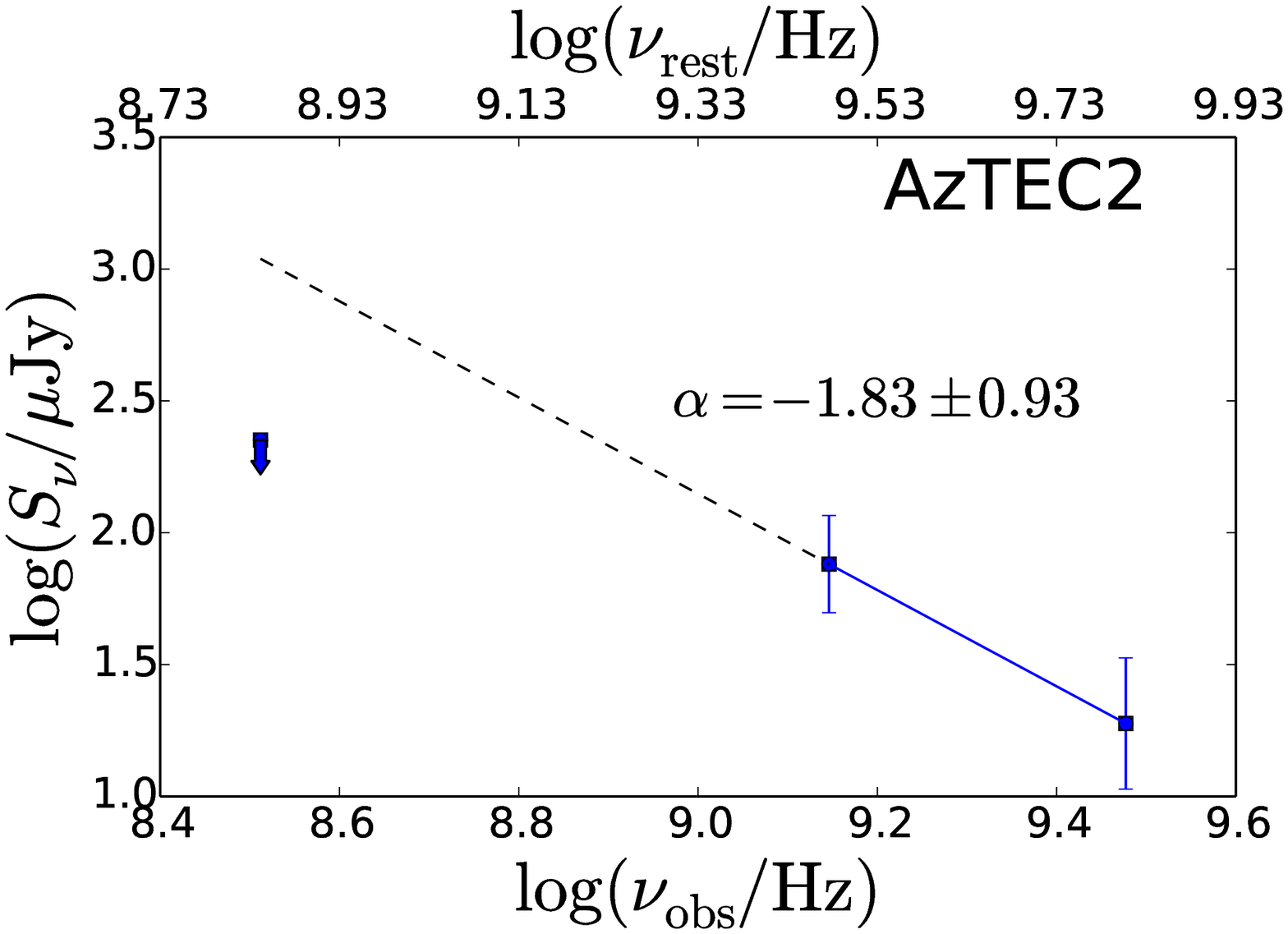}
\includegraphics[width=0.2465\textwidth]{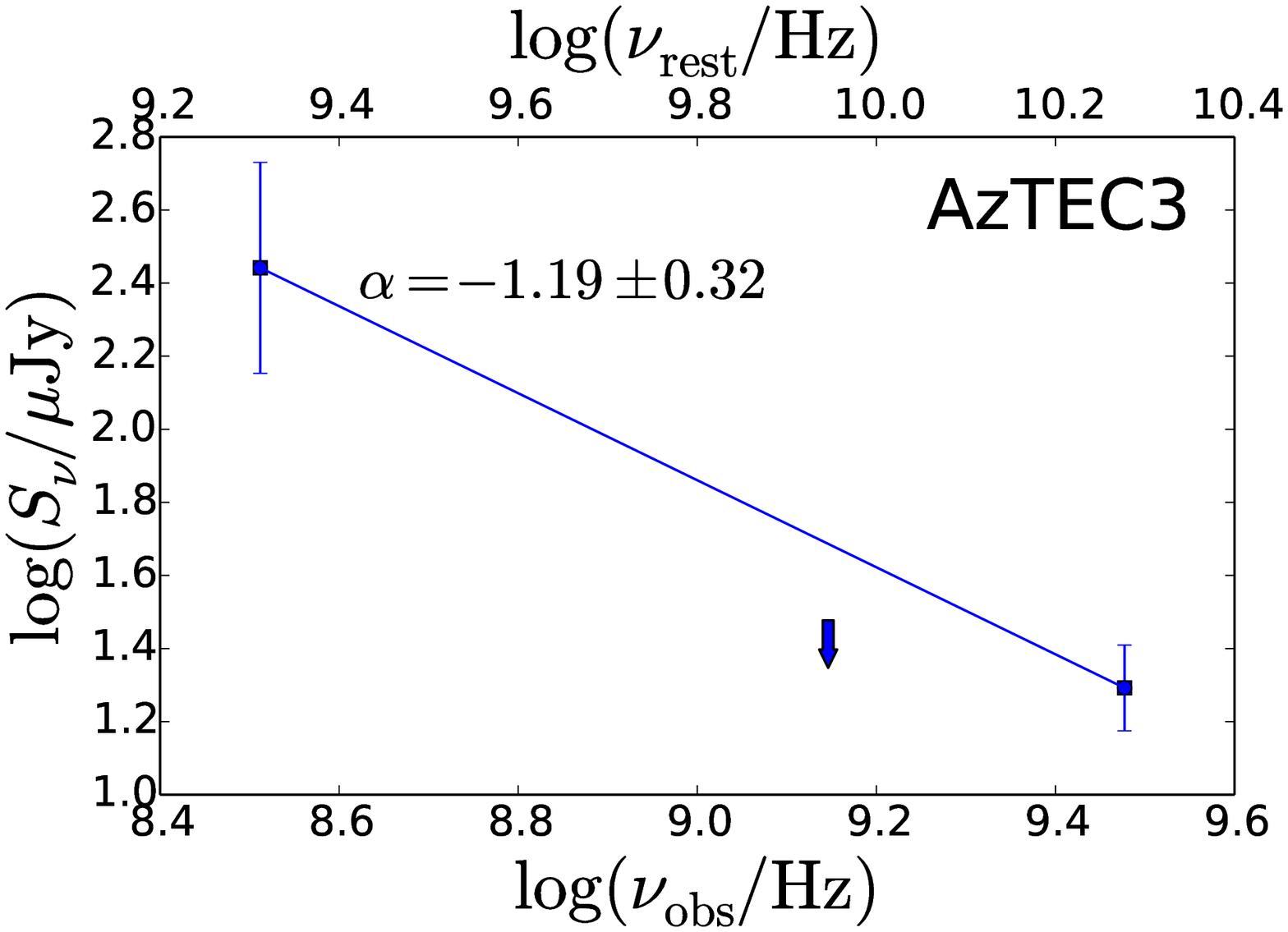}
\includegraphics[width=0.2465\textwidth]{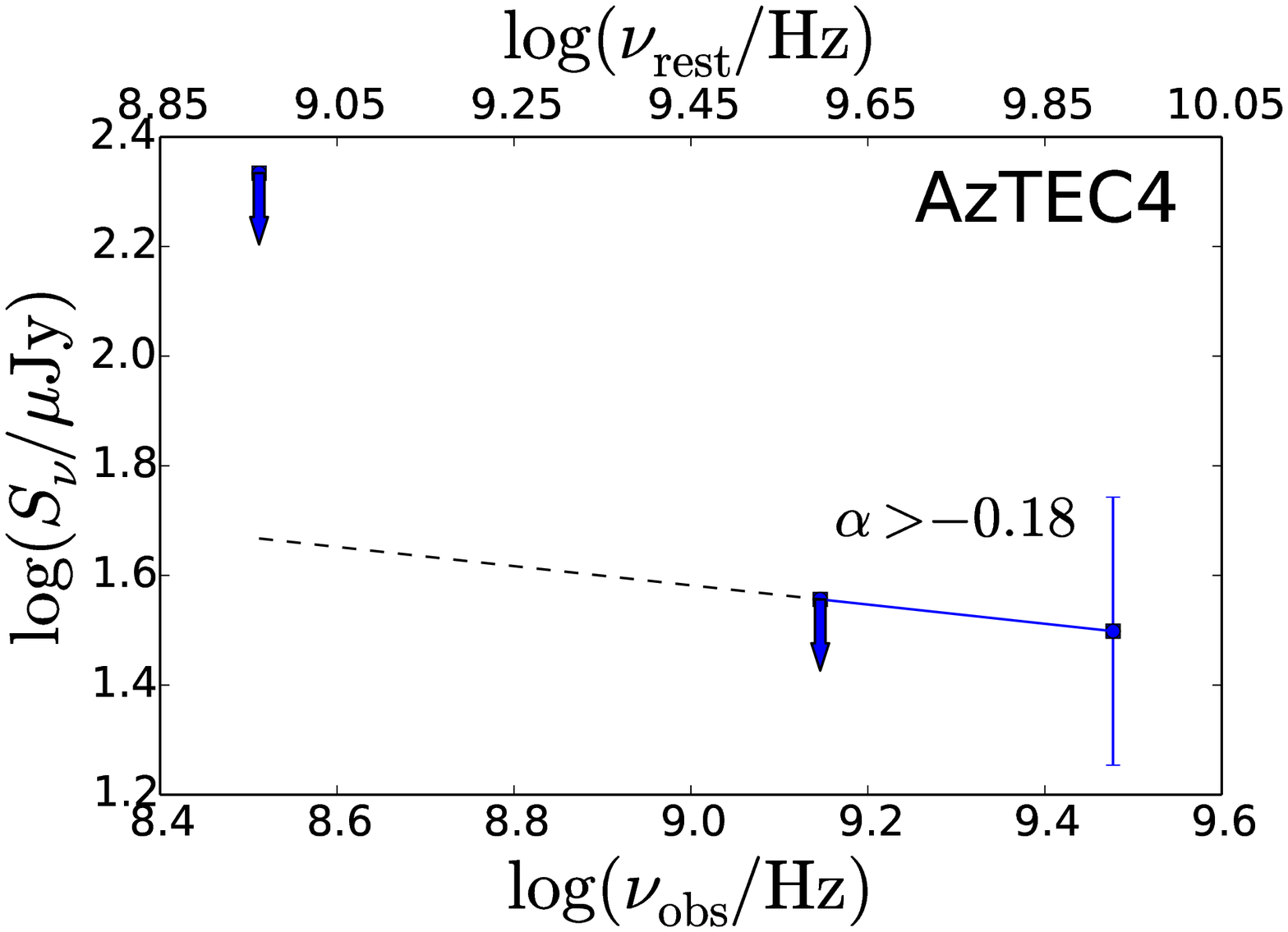}
\includegraphics[width=0.2465\textwidth]{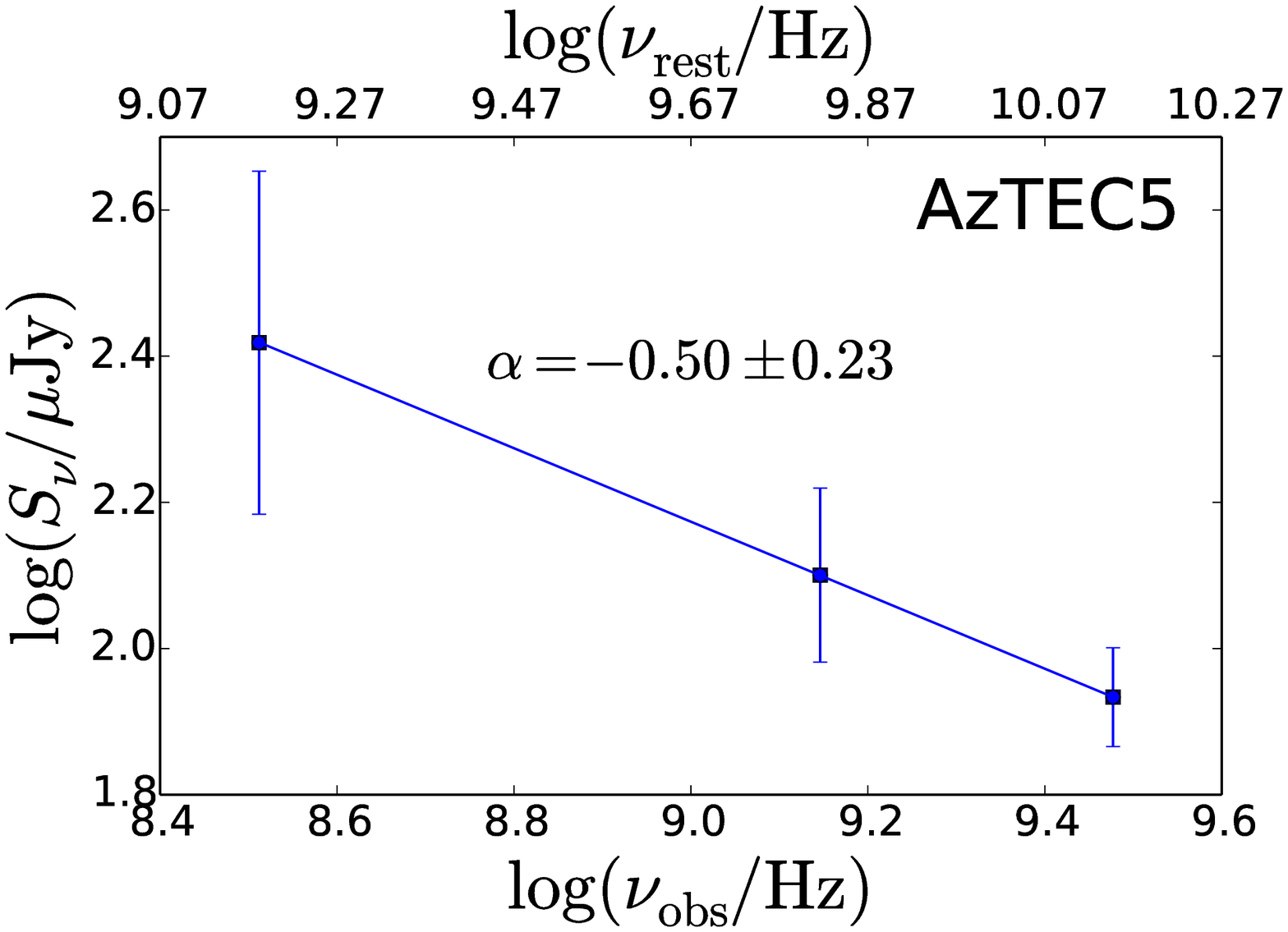}
\includegraphics[width=0.2465\textwidth]{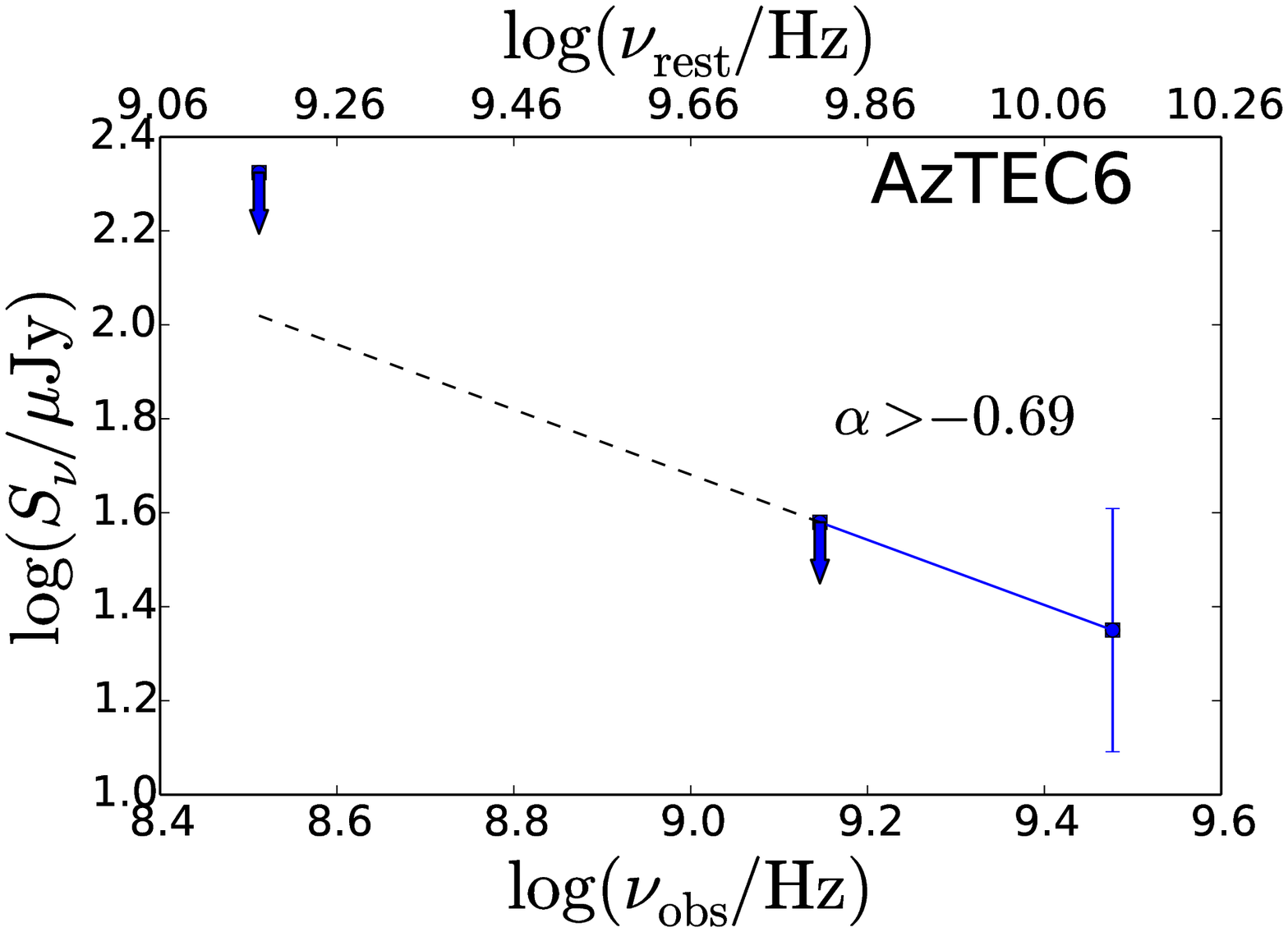}
\includegraphics[width=0.2465\textwidth]{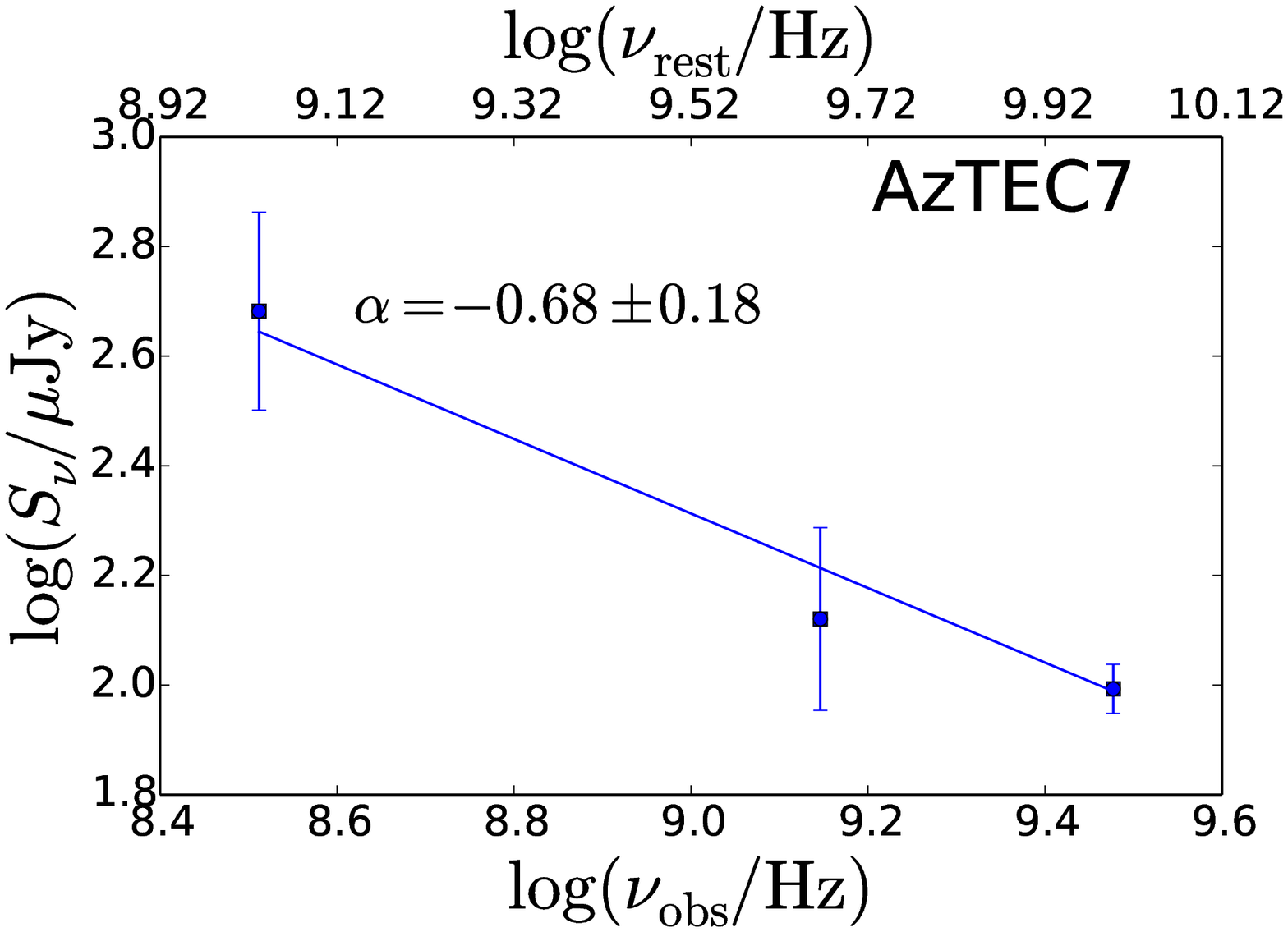}
\includegraphics[width=0.2465\textwidth]{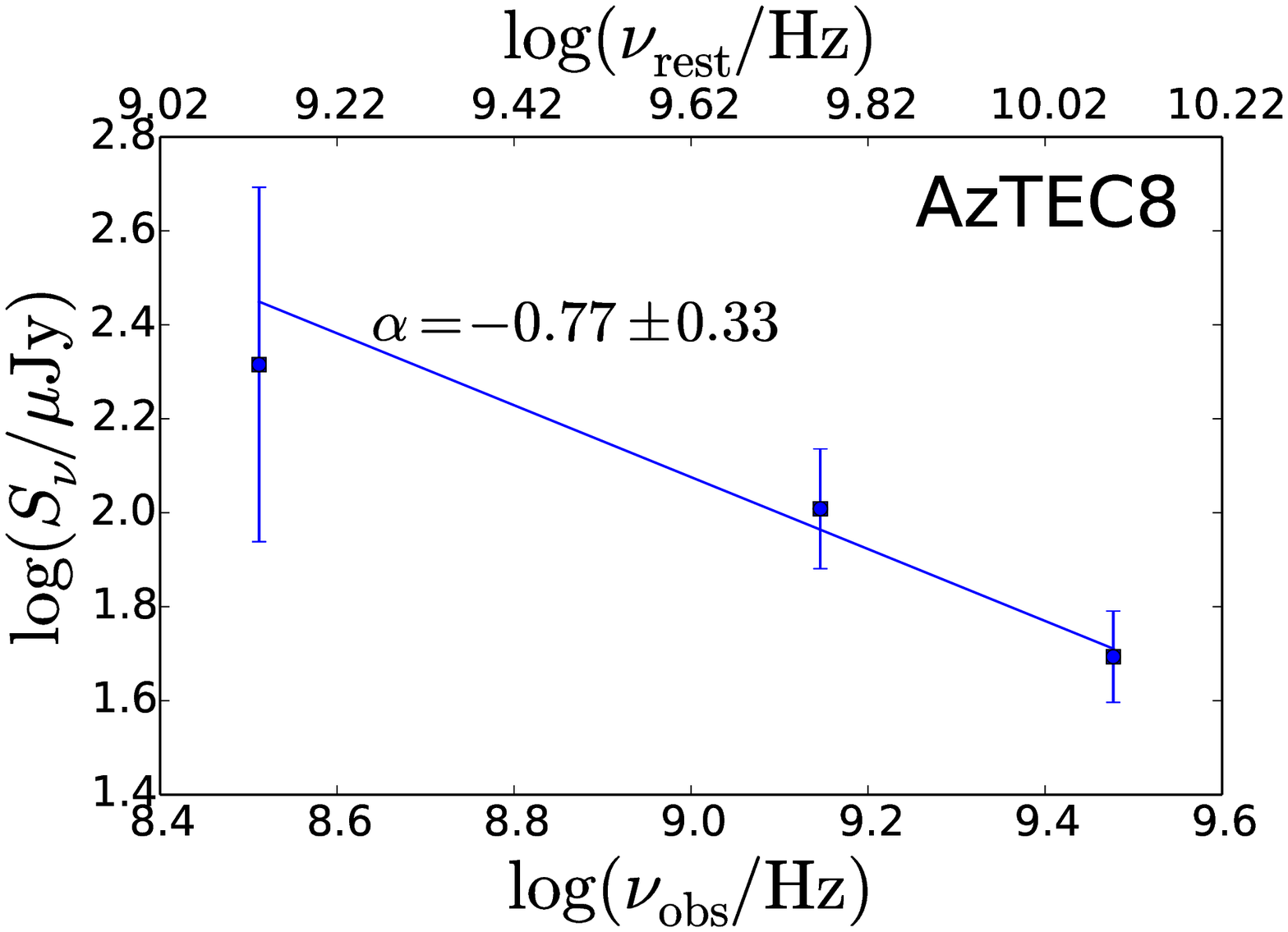}
\includegraphics[width=0.2465\textwidth]{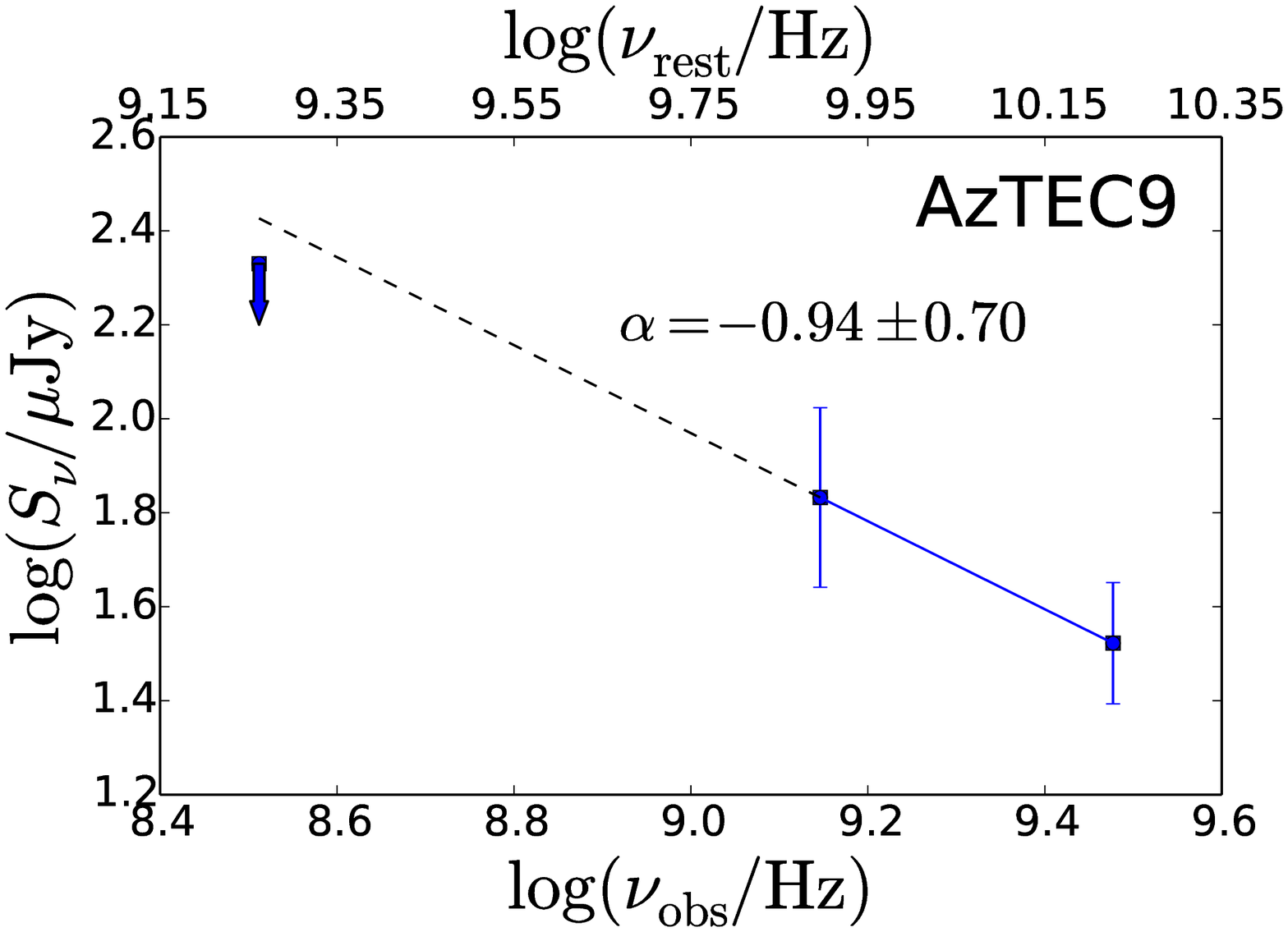}
\includegraphics[width=0.2465\textwidth]{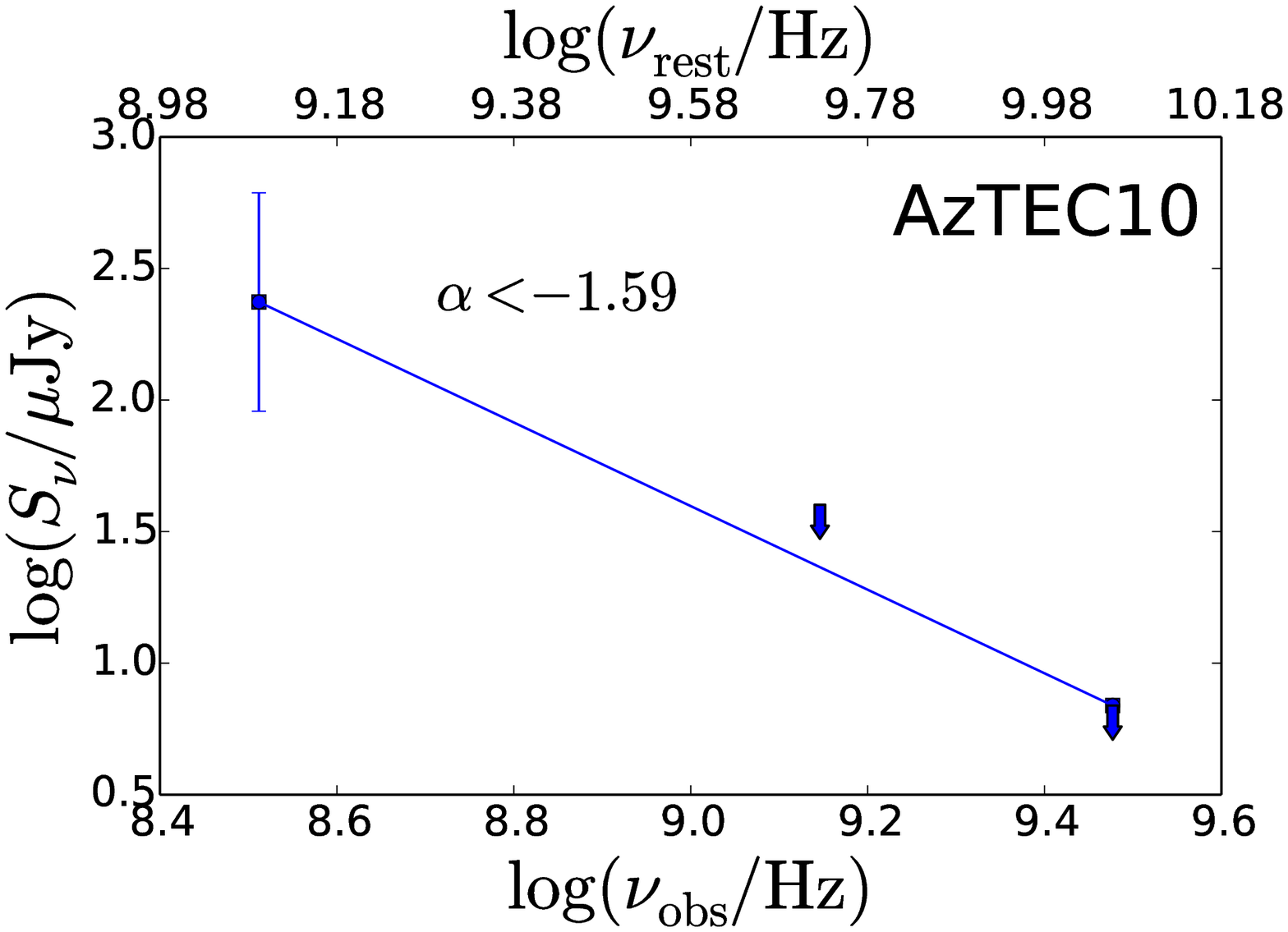}
\includegraphics[width=0.2465\textwidth]{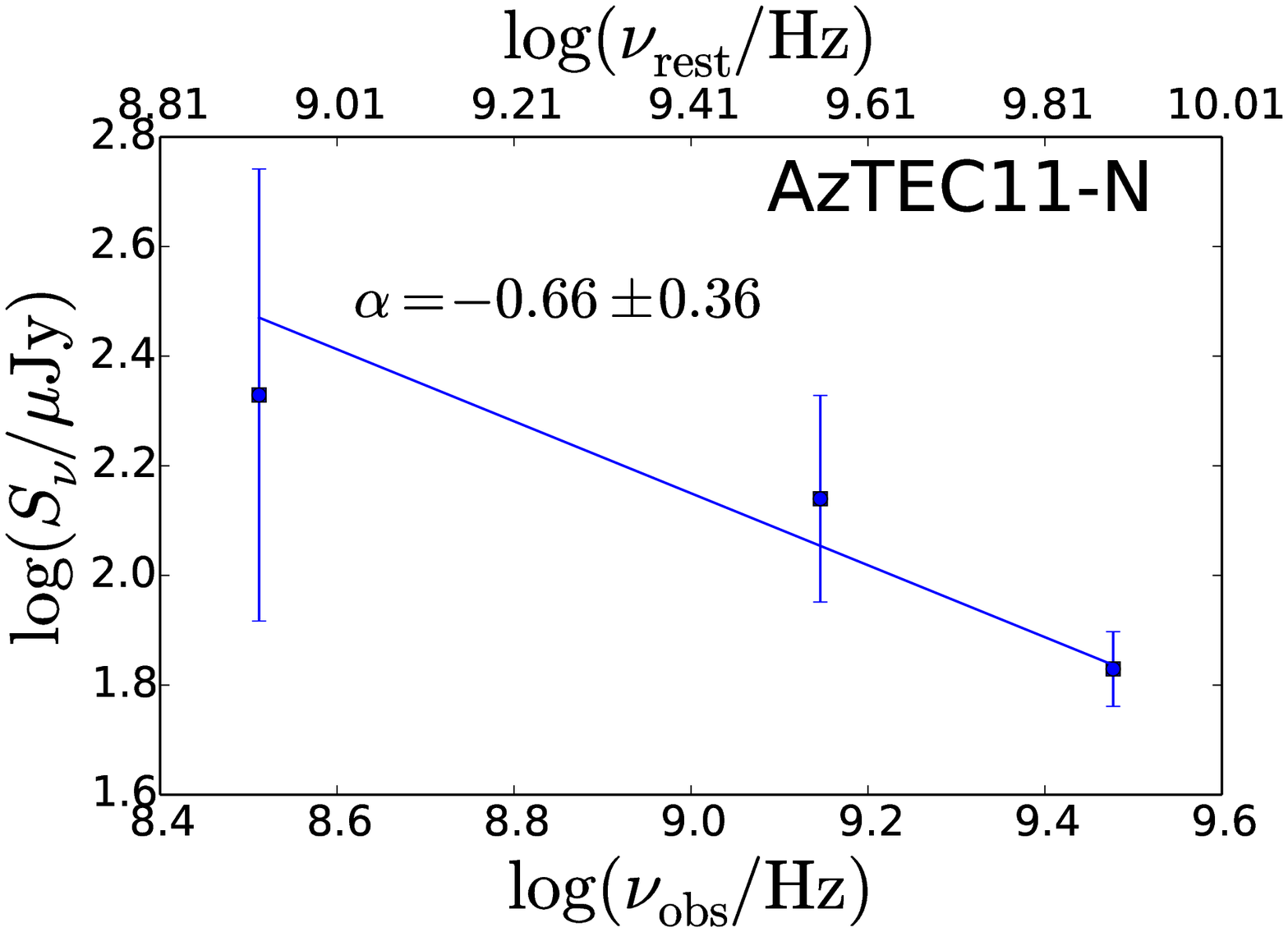}
\includegraphics[width=0.2465\textwidth]{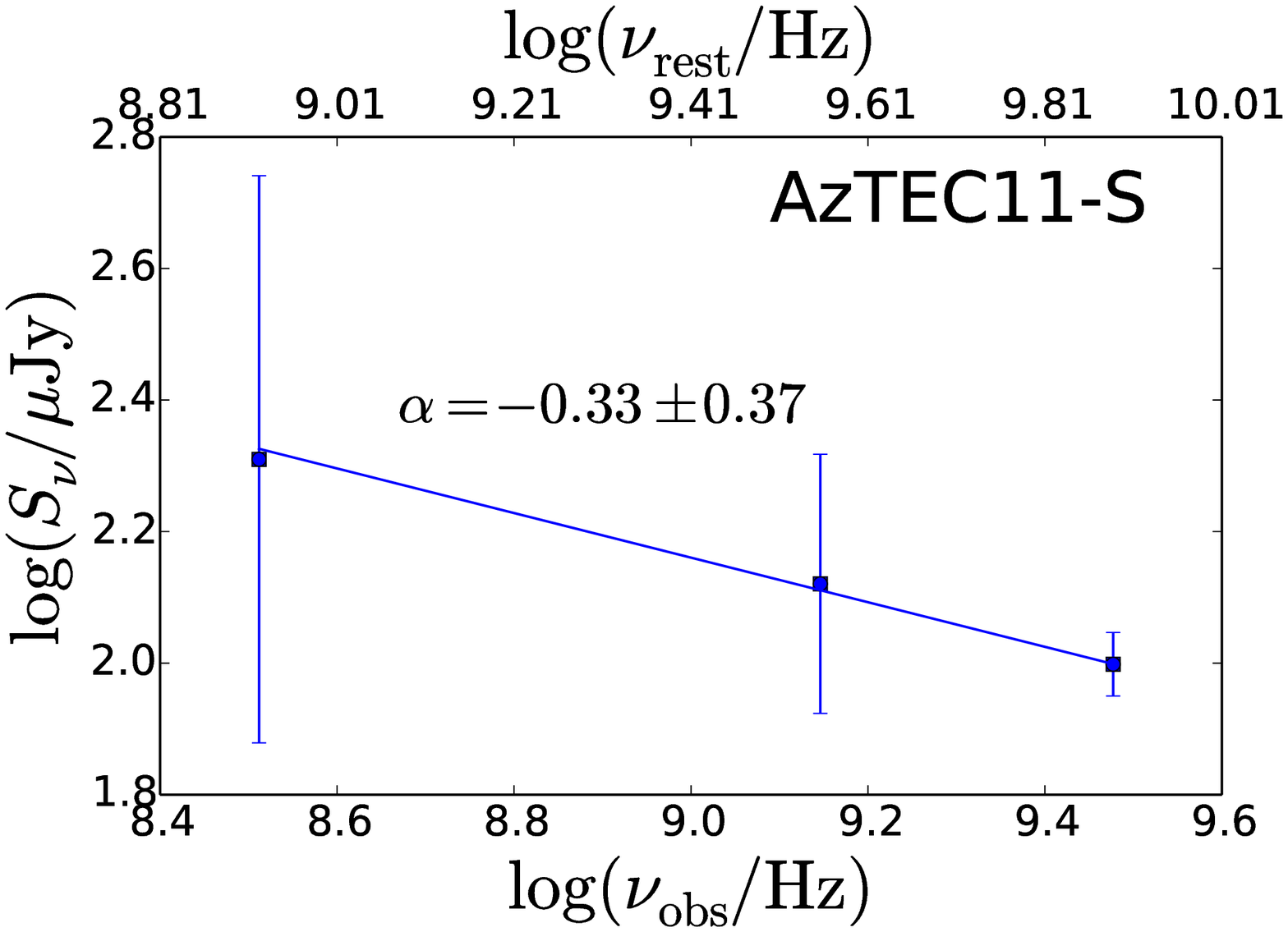}
\includegraphics[width=0.2465\textwidth]{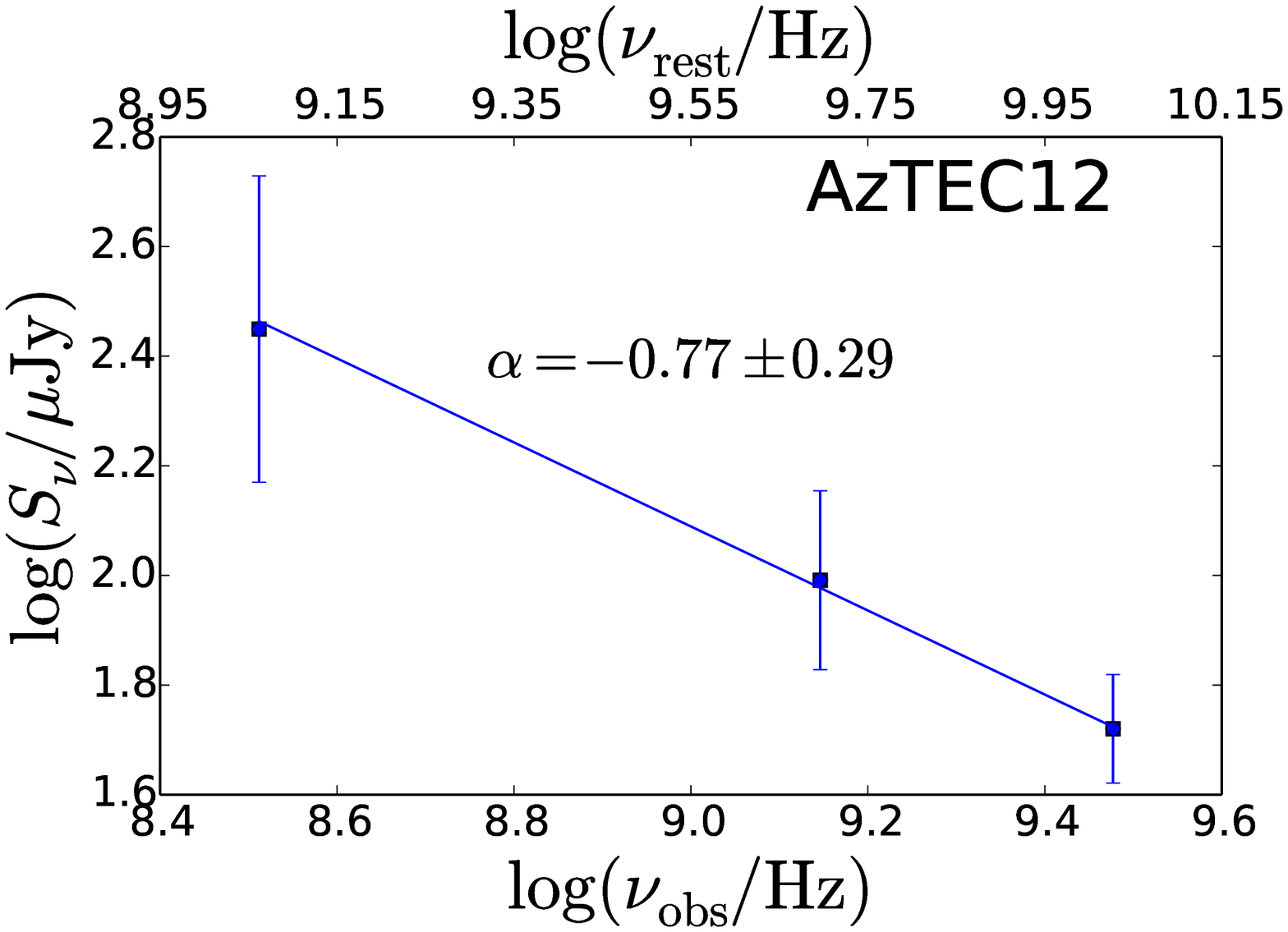}
\includegraphics[width=0.2465\textwidth]{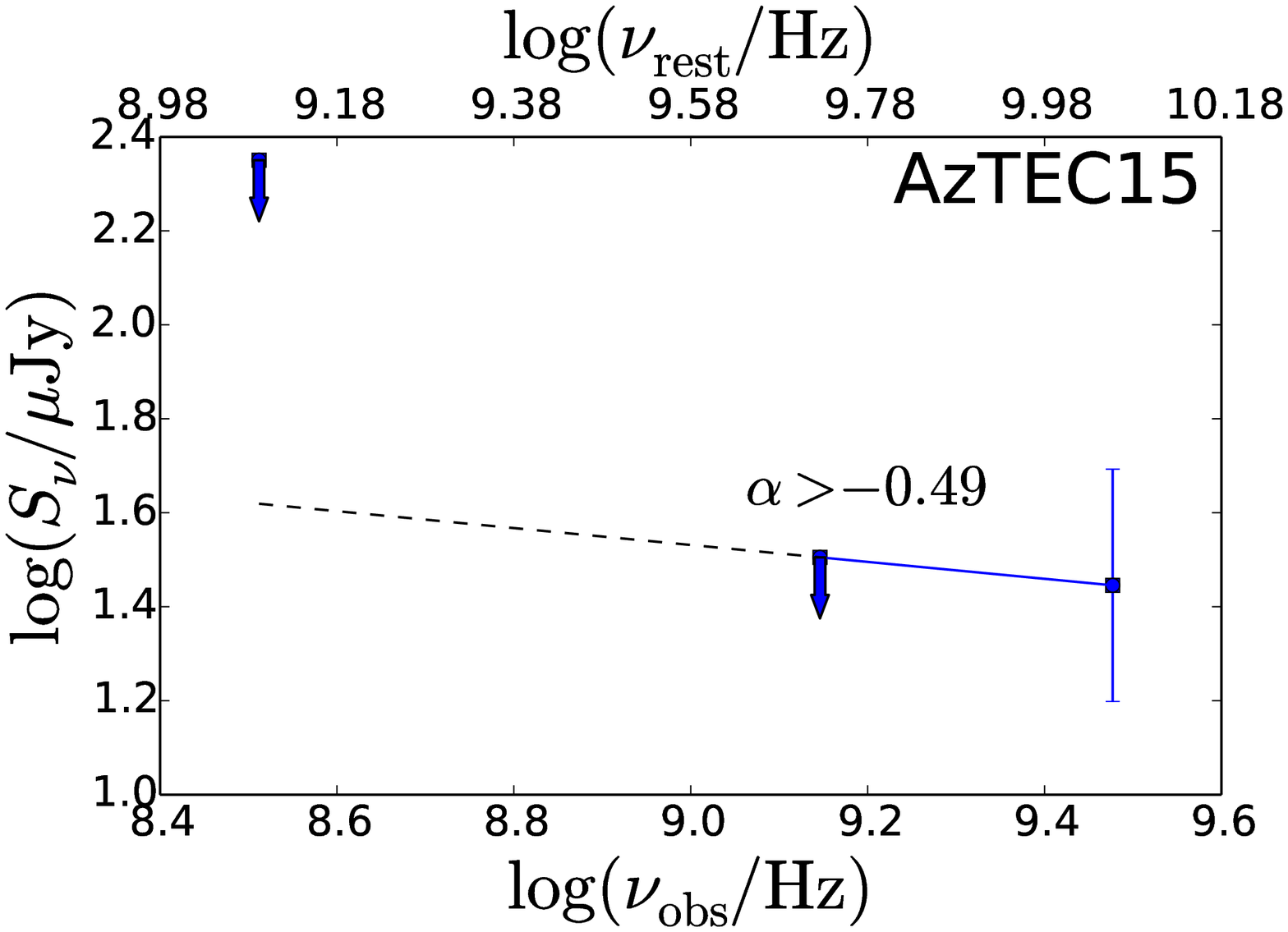}
\includegraphics[width=0.2465\textwidth]{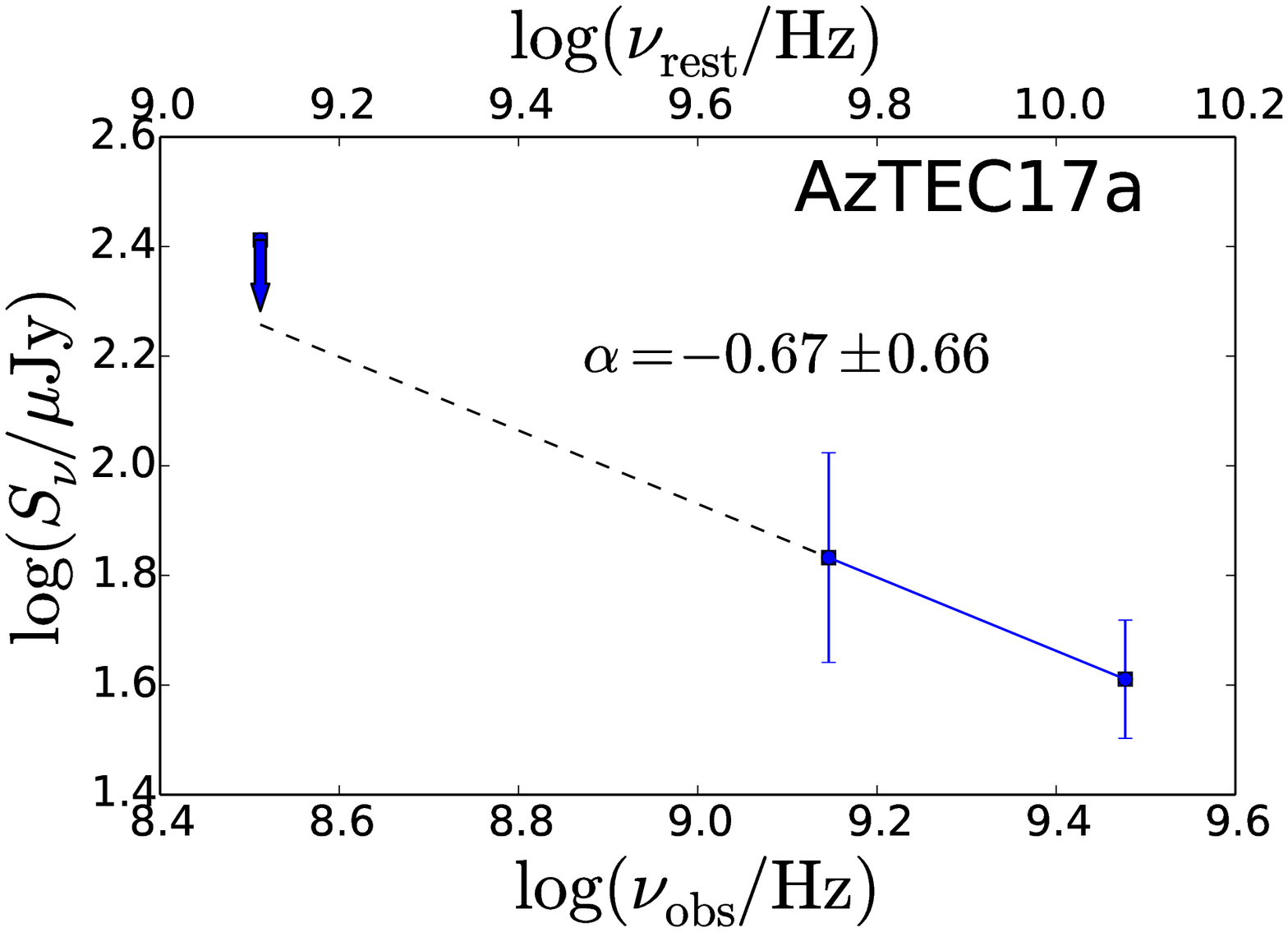}
\includegraphics[width=0.2465\textwidth]{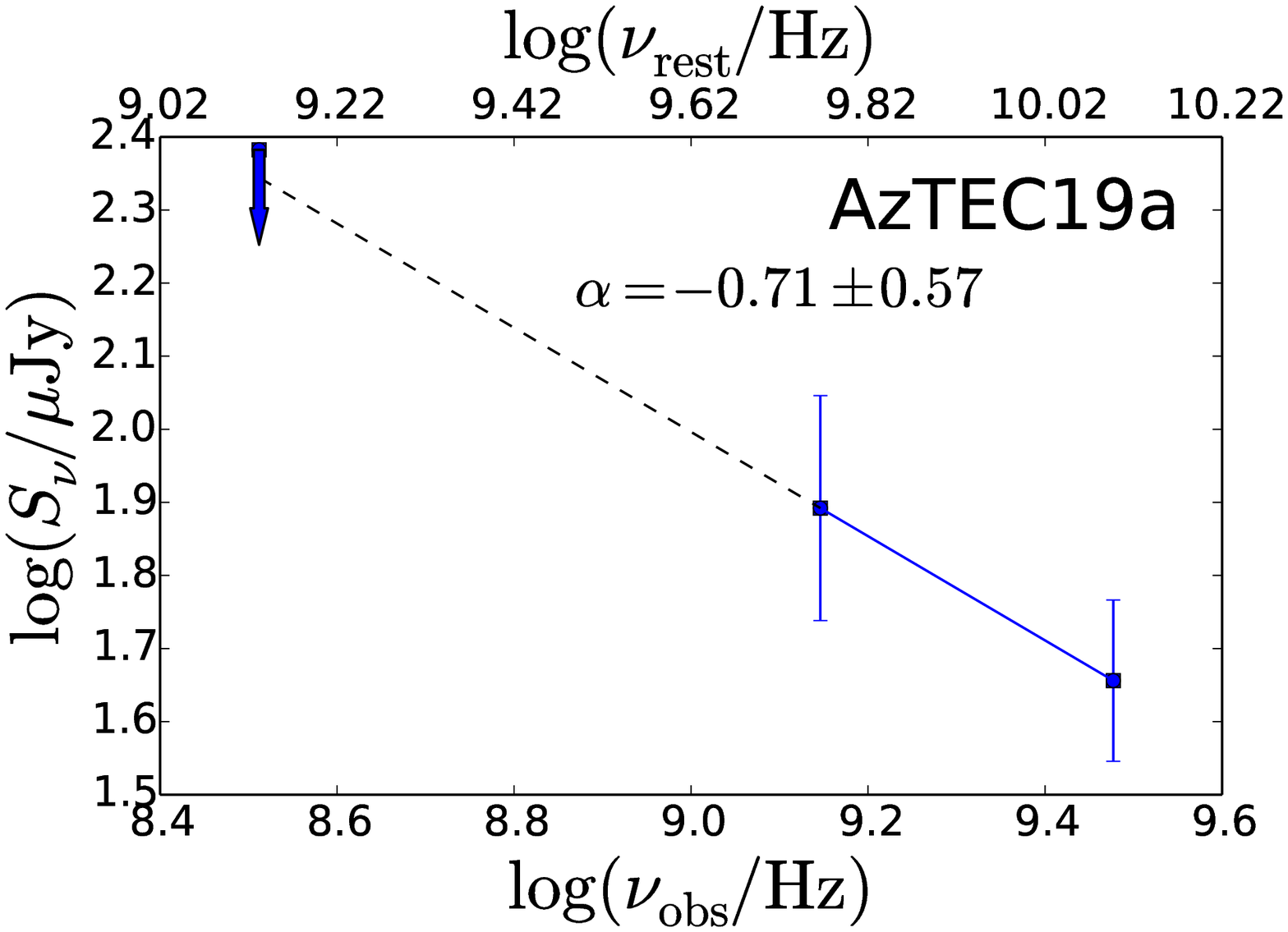}
\includegraphics[width=0.2465\textwidth]{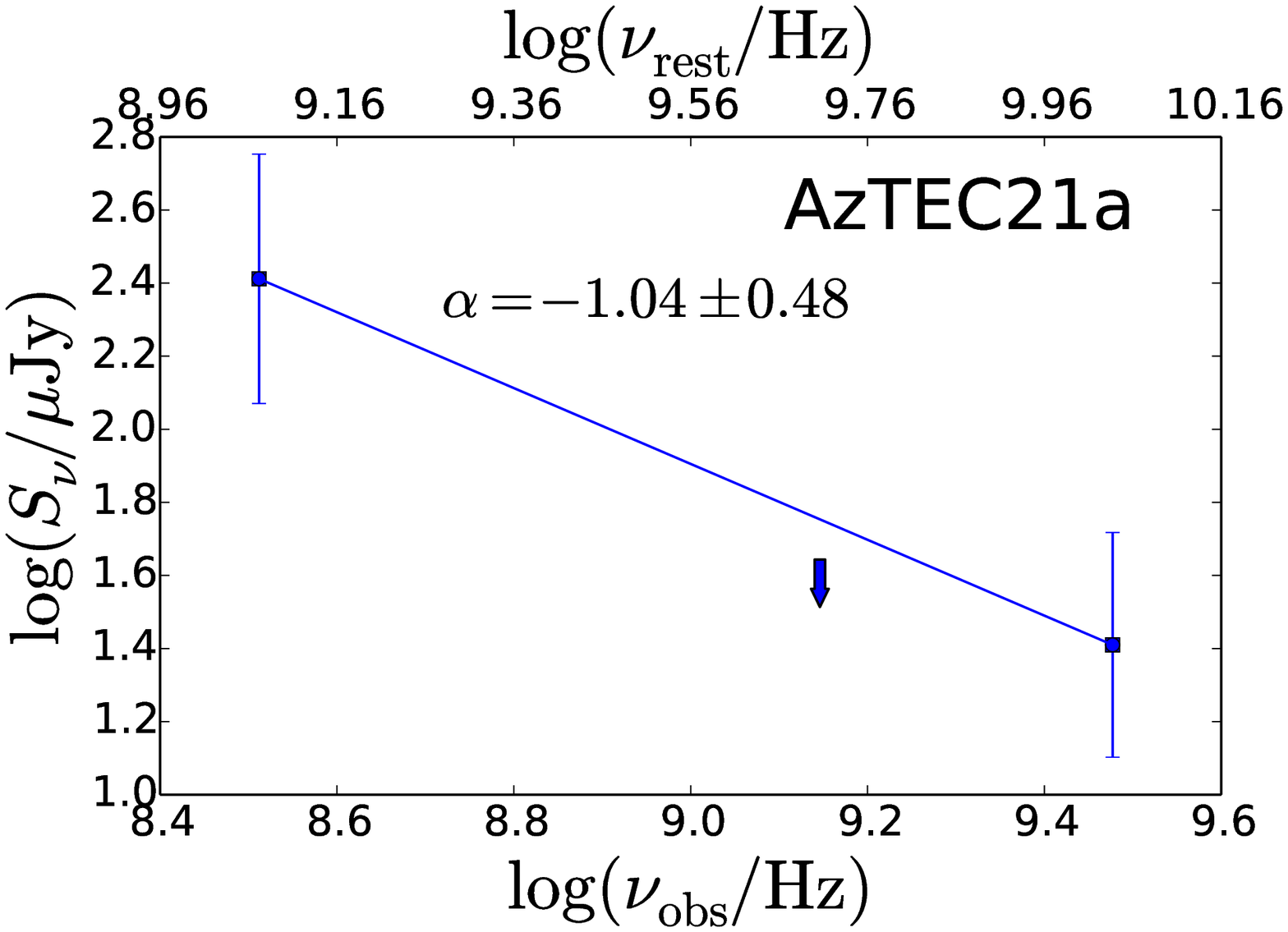}
\includegraphics[width=0.2465\textwidth]{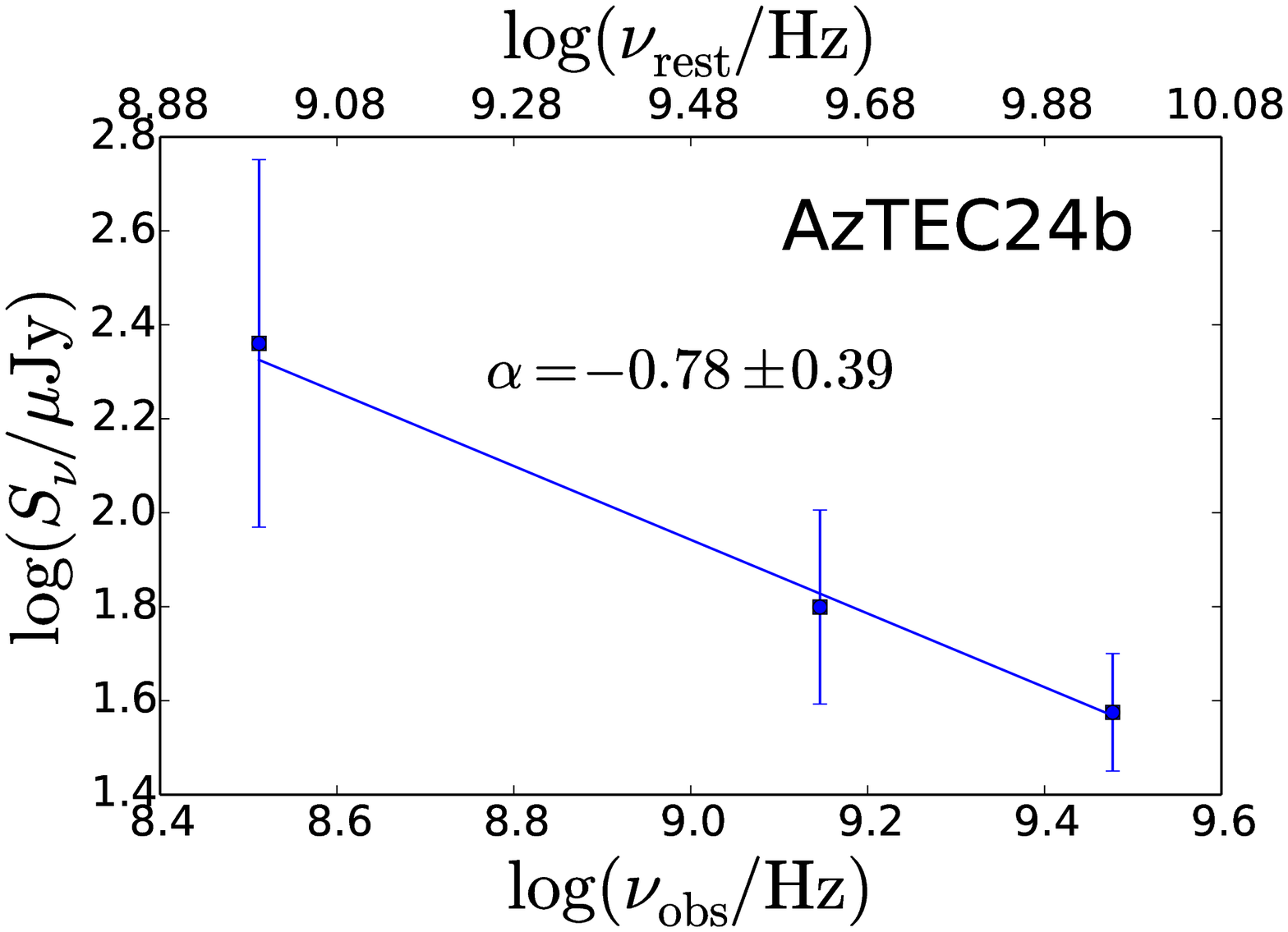}
\includegraphics[width=0.2465\textwidth]{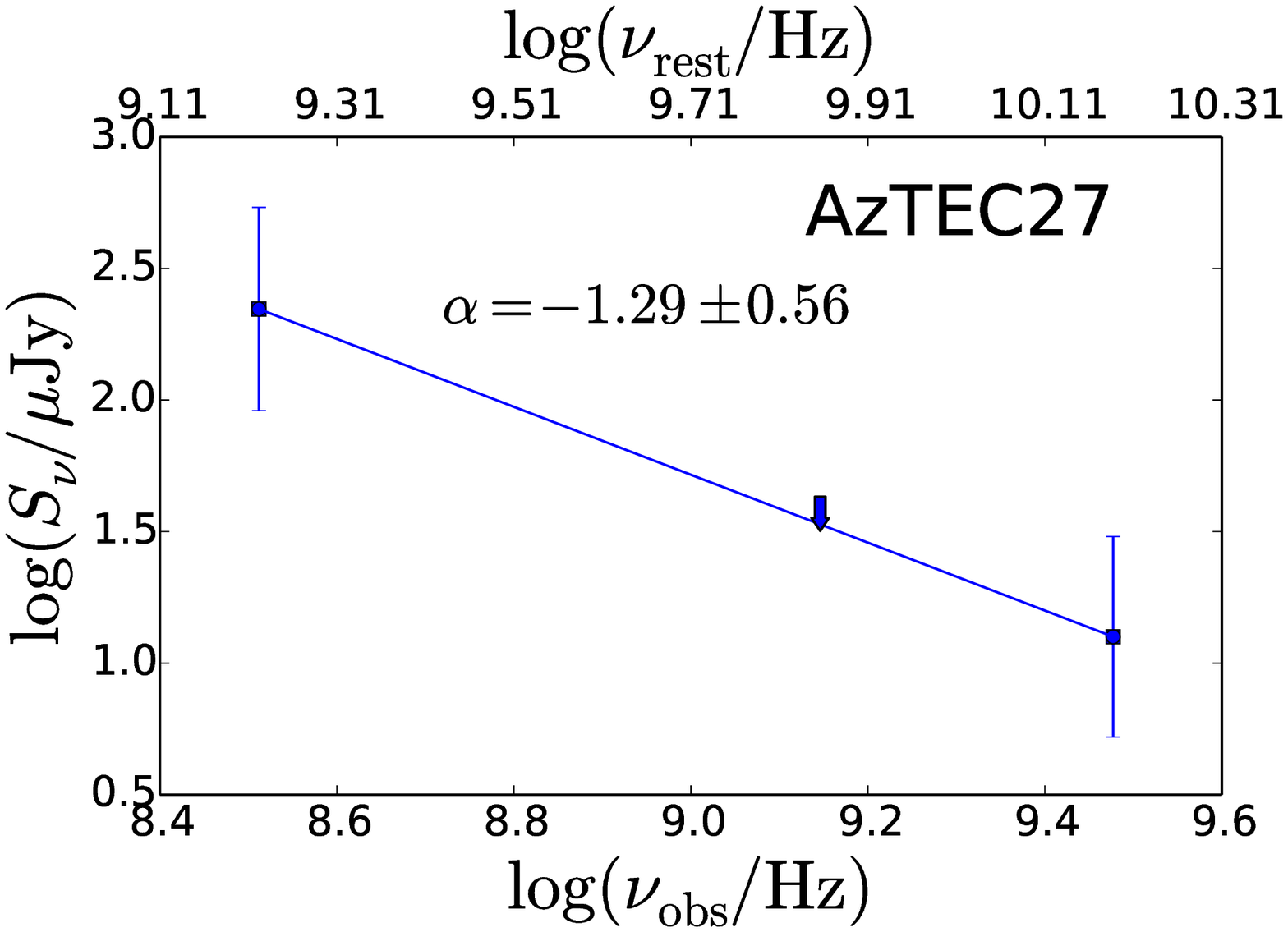}
\caption{Radio SEDs (on a log-log scale) of 19 of our target SMGs that were detected in at least one of the three observed frequencies of 
325~MHz, 1.4~GHz, and 3~GHz (the three data points in each panel with vertical error bars). The downward pointing arrows show $3\sigma$ upper 
limits. The solid lines show the least squares fits to the data points, and the continuation of the fit is illustrated by the dashed 
line. The derived spectral index values are shown in each panel. In each panel, the lower $x$-axis shows the observed frequency, while the upper $x$-axis gives the corresponding rest-frame frequency (for AzTEC6 and 27 only a lower $z$ limit is available, and hence the $\nu_{\rm rest}$ shown is only a lower limit).} 
\label{figure:radio}
\end{center}
\end{figure*}

To study the radio SEDs of our SMGs, we used the 325~MHz GMRT data, and 1.4~GHz and 3~GHz VLA data as described in Sect.~2.2. 
The radio SEDs for the SMGs detected in at least one of the three radio frequencies are shown in Fig.~\ref{figure:radio}. A linear least squares regression was used to fit the data points on a log--log scale to derive the radio spectral indices. In most cases (the following 15 sources: AzTEC1, 4, 5, 6, 7, 8, 10, 11-N, 11-S, 12, 15, 17a, 19a, 24b, and 27) the observed data are consistent with a single power-law spectrum. However, for AzTEC2 and AzTEC9 the $3\sigma$ upper limit to the 325~MHz flux density lies below a value suggested by the 1.4--3~GHz part of the SED, while in the AzTEC3 and AzTEC21a SEDs the 1.4~GHz upper flux density limit lies below the fitted power-law function. We note that because we only have three data points in our galaxy-integrated radio SEDs, we did not try to fit them with anything more complex than a single power law, like a broken or curved power law. The derived spectral indices are tabulated in column~(2) in Table~\ref{table:radio2}.\footnote{In Paper~II, we derived the values of $\alpha_{\rm 1.4\, GHz}^{\rm 3\, GHz}$ for our SMGs (see Table~4 therein). Those are almost identical to the corresponding present values, but the quoted uncertainties differ in some cases because of the difference in the method they were derived. In the present study, the uncertainties represent the standard deviation errors weighted by the flux density uncertainty, while in Paper~II the spectral index uncertainty was directly propagated from that associated with the flux densities.} For AzTEC2, the reported spectral index refers to a frequency range between 1.4 and 3~GHz. We also note that for AzTEC21b we could not constrain the radio spectral index needed in the IR-radio correlation analysis (Sect.~3.5). Hence, for this source we assumed a value of $\alpha_{\rm 325\, MHz}^{\rm 3\, GHz}=-0.75\pm0.05$ to be consistent with the canonical non-thermal synchrotron spectral index range of $\alpha_{\rm synch}\in [-0.8, \, -0.7]$ (e.g. \cite{niklas1997}; \cite{lisenfeld2000}; \cite{marvil2015}), but it is not included in the subsequent radio analysis.

\begin{figure}[!h]
\centering
\resizebox{0.9\hsize}{!}{\includegraphics{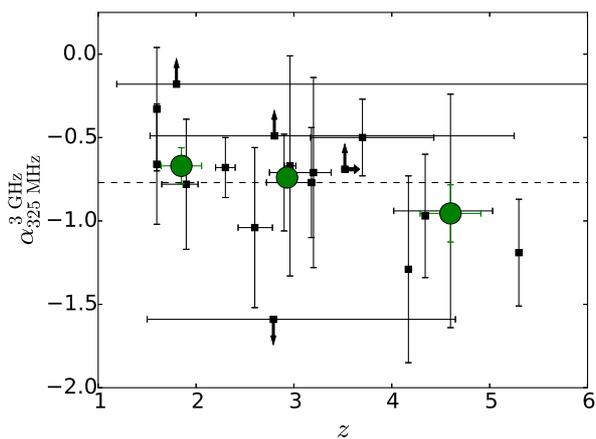}}
\caption{Radio spectral index between 325~MHz and 3~GHz as a function of redshift. The arrows indicate lower and upper limits to $\alpha_{\rm 325\, MHz}^{\rm 3\, GHz}$ and $z$. The horizontal dashed line shows the median spectral index value of $\alpha_{\rm 325\, MHz}^{\rm 3\, GHz}=-0.77$. The green filled circles represent the median values of the binned data computed using survival analysis (each bin contains six SMGs), with the error bars showing the standard errors of the median values.}
\label{figure:alphavsz}
\end{figure}

The derived $\alpha_{\rm 325\, MHz}^{\rm 3\, GHz}$ values contain both lower and upper limits. To estimate the median of these doubly-censored data, we applied survival analysis. First, we used the {\tt dblcens} package for {\tt R}\footnote{\tt https://cran.r-project.org/web/packages/dblcens/}, which computes the non-parametric maximum likelihood estimation of the cumulative distribution function from doubly-censored data via an expectation-maximization algorithm. This was used to estimate the contribution of the right-censored data. We then used the Kaplan-Meier (K-M) method to construct a model of the input data by assuming that the left-censored data follow the same distribution as the non-censored values (for this purpose, we used the Nondetects And Data Analysis for environmental data (NADA; \cite{helsel2005}) package for {\tt R}). Using this method the median of $\alpha_{\rm 325\, MHz}^{\rm 3\, GHz}$ was found to be $-0.77^{+0.28}_{-0.42}$, where the $\pm$ uncertainty represents the 16th--84th percentile range.
 Although the derived median spectral index is fully consistent with the value of $\alpha_{\rm synch}=-0.8$ assumed in the {\tt MAGPHYS} analysis, some of the individual nominal values are significantly different, which is reflected as poor fits in the radio regime as shown in Fig.~\ref{figure:seds}. The distribution of the derived $\alpha_{\rm 325\, MHz}^{\rm 3\, GHz}$ values as a function of redshift is plotted in Fig.~\ref{figure:alphavsz}. We note that the large number of censored data points allowed us to calculate the median values of 
the binned data showed in Fig.~\ref{figure:alphavsz}, but not the corresponding mean values. 
The binned median data points suggest that there is a hint of decreasing (i.e. steepening) spectral index with increasing 
redshift. To quantify this dependence, we fit the binned data points using a linear regression line, and derived a relationship of the form 
$\alpha_{\rm 325\, MHz}^{\rm 3\, GHz} \propto -(0.098 \pm 0.072)\times z$, where the uncertainty in the slope is based on the one standard deviation errors of the spectral index data points. 
The Pearson correlation coefficient of the binned data was found to be $r=-0.99$. Hence, a negative relationship is present, but it is not statistically significant.

\subsection{Stellar mass-SFR correlation: comparison with the main sequence of star-forming galaxies}

\begin{figure}[!htb]
\centering
\resizebox{0.9\hsize}{!}{\includegraphics{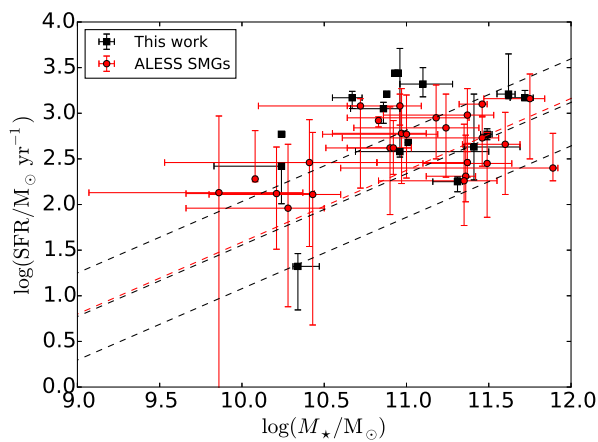}}
\caption{A log-log plot of the SFR versus stellar mass. The black squares show our AzTEC SMG data points, while the red filled circles show 
the ALESS SMG data from da Cunha et al. (2015), where the ALESS sample was limited to sources having similar flux densities to our sample 
(see Sect.~4.2 for details). The dashed lines show the position of the star-forming main sequence at the median redshift of the analysed AzTEC SMGs ($z=2.85$; black line) and the flux-limited ALESS sample ($z=3.12$; red line) as given by Speagle et al. (2014); the lower and upper black dashed lines indicate a factor of three below and above the main sequence at $z=2.85$.}
\label{figure:ms}
\end{figure}

In Fig.~\ref{figure:ms}, we plot our SFR values as a function of stellar mass. A tight relationship found between these 
two quantities is known as the main sequence of star-forming galaxies (e.g. \cite{brinchmann2004}; \cite{noeske2007}; 
\cite{elbaz2007}; \cite{daddi2007}; \cite{karim2011}; \cite{whitaker2012}; \cite{speagle2014}; \cite{salmon2015}). 
For comparison, in Fig.~\ref{figure:ms} we also plot the ${\rm SFR}-M_{\star}$ values for ALMA 870~$\mu$m-detected SMGs 
from da Cunha et al. (2015; the so-called ALESS SMGs) who used the same high-$z$ extension of {\tt MAGPHYS} in 
their analysis as we have used here. Here, we have limited the da Cunha et al. (2015) sample to those SMGs that 
are equally bright to our target sources (AzTEC1--30; see Sect.~4.2 for a detailed description). 

\begin{figure}[!htb]
\centering
\resizebox{0.9\hsize}{!}{\includegraphics{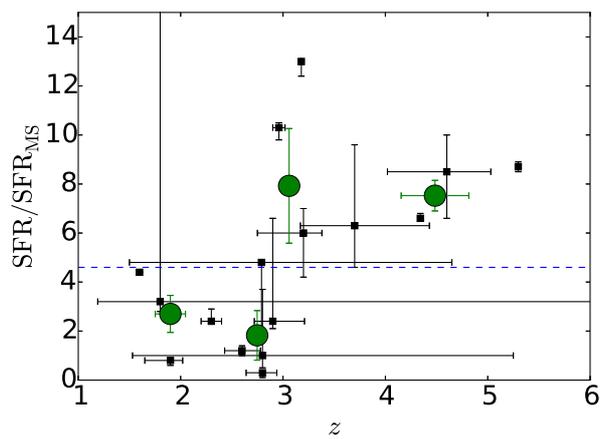}}
\caption{Starburstiness or the distance from the main sequence (parameterised as ${\rm SFR}/{\rm SFR}_{\rm MS}$) as a function of redshift. The blue horizontal dashed line marks the sample median of 4.6, while the green filled circles represent the mean values of the binned data (each bin contains four SMGs), with the error bars showing the standard errors of the mean values.}
\label{figure:burst}
\end{figure}

To illustrate how our data compare with the star-forming galaxy main sequence, we overlay 
the best fit from Speagle et al. (2014), which is based on a compilation of 25 studies, and is given by 
$\log({\rm SFR}/{\rm M}_{\sun}~{\rm yr}^{-1})=(0.84-0.026\times \tau_{\rm univ})\log(M_{\star}/{\rm M}_{\sun})-(6.51-0.11\times \tau_{\rm univ})$, 
where $\tau_{\rm univ}$ is the age of the universe in Gyr, i.e. the normalisation rises with increasing redshift. 
We show the main sequence position at the median redshift of our analysed SMGs ($z=2.85$) and that of the aforementioned 
ALESS SMG sample ($z=3.12$). We also plot the factor of 3 lines below and above the main sequence at $z=2.85$; 
this illustrates the accepted thickness, or scatter of the main sequence (see e.g. \cite{magdis2012}; \cite{dessauges2015}).
To further quantify the offset from the main sequence, we calculated the ratio of the derived SFR to that expected for 
a main sequence galaxy of the same stellar mass, i.e. ${\rm SFR}/{\rm SFR}_{\rm MS}$. The values of this ratio are given in column~(6) 
in Table~\ref{table:sed}, and they range from $0.3^{+0.1}_{-0.2}$ to $13.0^{+0}_{-0.6}$ with a median of $4.6^{+4.0}_{-3.5}$. 
The values of ${\rm SFR}/{\rm SFR}_{\rm MS}$ are plotted as a function of redshift in Fig.~\ref{figure:burst}. The binned data suggest 
a bimodal behaviour of our SMGs in the sense that the sources at $z<3$ are consistent with the main sequence, while those at $z>3$ lie above 
the main sequence. The $M_{\star}$-SFR plane of our SMGs will be discussed further in Sect.~4.1.

\subsection{Stellar mass-size relationship}

\begin{figure}[!htb]
\centering
\resizebox{0.9\hsize}{!}{\includegraphics{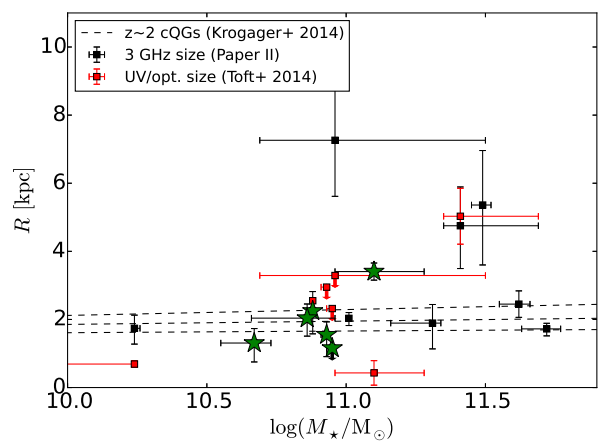}}
\caption{3~GHz radio continuum sizes (radii defined as half the major axis FWHM) derived in Paper~II (and scaled to the revised redshifts and cosmology adopted here) plotted against the stellar masses derived in the present work (black squares). The SMGs at $z>3$ are highlighted by green star symbols. For comparison, the red squares show the rest-frame UV/optical radii from Toft et al. (2014; also scaled to the present redshifts and cosmology). The upper size limits are indicated by arrows pointing down. The three dashed lines show the mass-size relationship of $z\sim2$ compact, quiescent galaxies from Krogager et al. (2014), where the lower and upper lines represent the dispersion in the parameters (see text for details). }
\label{figure:size}
\end{figure}

In Fig.~\ref{figure:size}, we plot the 3~GHz radio continuum sizes of our SMGs derived in Paper~II against their stellar masses derived in the present paper. We also show the rest-frame UV/optical radii for AzTEC1, 3, 4, 5, 8, 10, and 15 derived by Toft et al. (2014), but which were scaled to our adopted cosmology, and we used the revised redshifts for AzTEC1, 4, 5, and 15. The radio size data points of the SMGs lying at $z>3$ are highlighted by green star symbols in Fig.~\ref{figure:size}, while the UV/optical sizes are for $z\simeq 1.8-5.3$ SMGs, out of which 6/7 (86\%) lie at $z\gtrsim2.8$. As shown in the figure, with a few exceptions the largest spatial scales of both stellar and radio emission are seen among the highest stellar mass sources ($\log (M_{\star}/{\rm M_{\sun}})\geq 11.41$). We also note that most of the data points ($52\%$ ($64\%$) of all the plotted data (radio sizes)) are clustered within the dispersion of the mass-size relationship of $z\sim2$ cQGs derived by Krogager et al. (2014), namely $r_{\rm e}=\gamma (M_{\star}/10^{11}~{\rm M}_{\sun})^{\beta}$, where $\log(\gamma/{\rm kpc})=0.29\pm0.07$ and $\beta=0.53^{+0.29}_{-0.21}$ for their galaxies having spectroscopic redshifts. The $M_{\star}$-size plane analysed here will be discussed further in Sect.~4.1.

\subsection{Infrared-radio correlation}

The quantities derived in the present study allow us to examine the IR-radio correlation among our SMGs (e.g. \cite{vanderkruit1971}; \cite{dejong1985}; 
\cite{helou1985}; \cite{condon1991}; \cite{yun2001}). As usually done, we quantify this analysis by calculating the $q$ parameter, which can be defined as 
(see e.g. \cite{sargent2010}; \cite{magnelli2015}) 

\begin{equation}
q = \log \left(\frac{L_{\rm IR}}{3.75 \times 10^{12}\,{\rm W}} \right )-\log \left(\frac{L_{\rm 1.4\, GHz}}{{\rm W}~{\rm Hz}^{-1}}    \right )\, ,
\end{equation}
where the value $3.75 \times 10^{12}$ is the normalising frequency (in Hz) corresponding to $\lambda=80$~$\mu$m, and $L_{\rm 1.4\, GHz}$ is 
the rest-frame monochromatic 1.4~GHz radio luminosity density. Since our $L_{\rm IR}$ is calculated over the wavelength range of $\lambda_{\rm rest}=8-1\,000$~$\mu$m, 
our $q$ value refers to the total-IR value, i.e. $q\equiv q_{\rm TIR}$.\footnote{A FIR luminosity ($L_{\rm FIR}$) calculated by integrating over $\lambda_{\rm rest}=42.5-122.5$~$\mu$m is sometimes used to define $q\equiv q_{\rm FIR}$. We note that $q_{\rm TIR}=q_{\rm FIR}+\log (L_{\rm IR}/L_{\rm FIR})$.} The 1.4~GHz luminosity density is given by 

\begin{equation}
L_{\rm 1.4\, GHz}=\frac{4\pi d_{\rm L}^2}{(1+z)^{1+\alpha}}S_{\rm 1.4\, GHz}\,,
\end{equation}
where $d_{\rm L}$ is the luminosity distance. Following Smol{\v c}i{\'c} et al. (2015), the 1.4~GHz flux density was calculated as 
$S_{\rm 1.4\, GHz}=S_{\rm 325\, MHz}\times (1.4\,{\rm GHz}/325\,{\rm MHz})^{\alpha}$, because the observed-frame frequence of 
$\nu_{\rm obs}=325$~MHz corresponds to the rest-frame frequence of $\nu_{\rm rest}=1.4$~GHz at $z=3.3$, which is only a 16\% higher redshift than the median redshift of the SMGs analysed here ($z=2.85$). Hence, the smallest and least uncertain $K$-correction 
$(1+z)^{\alpha}$ to rest-frame 1.4~GHz is needed to derive the value of $L_{\rm 1.4\, GHz}$. The derived values of $q$ are listed in 
column~(3) in Table~\ref{table:radio2}. 

\begin{figure}[!h]
\centering
\resizebox{0.9\hsize}{!}{\includegraphics{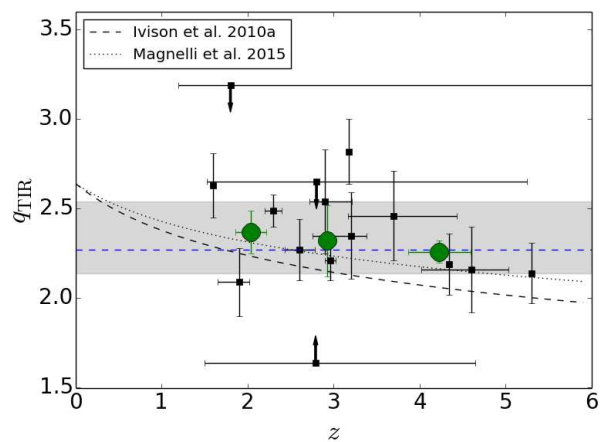}}
\caption{
Infrared-radio correlation $q\equiv q_{\rm TIR}$ parameter as a function of redshift. The arrows pointing up and down show the lower and upper limits, respectively. The green filled circles represent the survival analysis-based mean values of the binned data (each bin contains five SMGs), with the error bars showing the standard errors of the mean values. The blue dashed line marks the median value of $q=2.27^{+0.27}_{-0.13}$, and the quoted 16th--84th percentile range is illustrated by the grey shaded region. The black dashed curve shows the $q(z)\propto(1+z)^{-0.15\pm0.03}$ relationship from Ivison et al. (2010a), while the dotted curve represents the $q(z)\propto(1+z)^{-0.12\pm0.04}$ relationship from Magnelli et al. (2015). The latter relationships are normalised here to give a value of $q=2.64$ at $z=0$ (see Sect.~4.4). }
\label{figure:q}
\end{figure}

As in the case of $\alpha_{\rm 325\, MHz}^{\rm 3\, GHz}$ (Sect.~3.2), our $q$ values contain both lower and upper limits. Hence, to estimate the sample median, we employed the K-M survival analysis as described in Sect.~3.2. The median $q$ value and the 16th--84th percentile range is found to be $2.27^{+0.27}_{-0.13}$. In Fig.~\ref{figure:q}, we show the derived $q$ values as a function of redshift, and discuss the results further in Sect.~4.4.


\begin{table}[H]
\renewcommand{\footnoterule}{}
\caption{The radio spectral indices and IR-radio correlation $q$ parameter values.}
{\normalsize
\begin{minipage}{1\columnwidth}
\centering
\label{table:radio2}
\begin{tabular}{c c c}
\hline\hline 
Source ID & $\alpha_{\rm 325\, MHz}^{\rm 3\, GHz}$ & $q$\tablefootmark{a} \\[1ex]
\hline 
AzTEC1 & $-0.97\pm0.37$ & $2.19\pm0.17$ \\ 
AzTEC2 & $-1.83\pm0.93$\tablefootmark{b} & \ldots \\
AzTEC3 & $-1.19\pm0.32$ & $2.14\pm0.17$ \\
AzTEC4 & $>-0.18$ & $<3.19$ \\
AzTEC5 & $-0.50\pm0.23$ & $2.46\pm0.25$ \\ 
AzTEC6 & $>-0.69$ & \ldots \\ 
AzTEC7 & $-0.68\pm0.18$ & $2.49\pm0.09$ \\ 
AzTEC8 & $-0.77\pm0.33$ & $2.82\pm0.18$ \\ 
AzTEC9 & $-0.94\pm0.70$ & $2.16\pm0.24$ \\
AzTEC10 & $<-1.59$ & $>1.64$ \\ 
AzTEC11-N & $-0.66\pm0.36$ & \ldots \\ 
AzTEC11-S & $-0.33\pm0.37$ & $2.63\pm0.18$ \\
AzTEC12 & $-0.77\pm0.29$ & $2.54\pm0.29$ \\ 
AzTEC15 & $>-0.49$ & $<2.65$ \\
AzTEC17a & $-0.67\pm0.66$ & $2.21\pm0.11$ \\
AzTEC19a & $-0.71\pm0.57$ & $2.35\pm0.24$ \\
AzTEC21a & $-1.04\pm0.48$ & $2.27\pm0.17$ \\
AzTEC21b & $-0.75\pm0.05$\tablefootmark{c} & $>0.51$ \\
AzTEC24b & $-0.78\pm0.39$ & $2.09\pm0.19$ \\
AzTEC27 & $-1.29\pm0.56$ & \ldots \\
\hline
Median\tablefootmark{d} & $-0.77^{+0.28}_{-0.42}$  & $2.27^{+0.27}_{-0.13}$\\[1ex]
\hline 
\end{tabular} 
\tablefoot{\tablefoottext{a}{The $q$ parameter refers to the total-IR 
($\lambda_{\rm rest}=8-1\,000$~$\mu$m) value, i.e. $q \equiv q_{\rm TIR}$.}\tablefoottext{b}{The spectral index for AzTEC2 refers to a frequency interval between 1.4 and 3~GHz, and is not considered in the statistical analysis.}\tablefoottext{c}{The radio spectral index for AzTEC21b could not be constrained, and hence it was assumed to be $\alpha=-0.75\pm0.05$.}\tablefoottext{d}{The sample median value was derived using all the tabulated values except those for AzTEC2 and 21b. The quoted uncertainties of the sample medians represent the 16th--84th percentile range.} }
\end{minipage} }
\end{table}

\section{Discussion}

The SEDs of some of our COSMOS/AzTEC SMGs have already been analysed in previous studies. We discuss those studies and compare their results with ours 
in Appendix~C. AzTEC1 and 3 both have a gas mass estimate available 
in the literature (\cite{yun2015} and \cite{riechers2010}, respectively), which
allows us to examine their ISM physical properties in more detail; these two high-redshift SMGs are discussed in more detail in Appendix~D. After discussing 
the stellar mass-SFR and mass-size relationships of our SMGs in Sect.~4.1, we compare the physical properties of our SMGs with those of the ALESS SMGs derived by da Cunha et al. (2015), and discuss the similarities and differences between the two SMG samples in Sect.~4.2. The radio SEDs and IR-radio correlation are discussed in Sects.~4.3 and 4.4, while the present results are discussed in the context of evolution of massive galaxies in Sect.~4.5. 

\subsection{How do the COSMOS/AzTEC SMGs populate the $M_{\star}$-SFR and $M_{\star}$-size planes ?}

\subsubsection{Comparison with the galaxy main sequence}

We have found that 10 out of the 16 SMGs ($62.5\%$ with a Poisson counting error of 
$19.8\%$) analysed here lie above the main sequence with ${\rm SFR}/{\rm SFR}_{\rm MS}>3$. 
AzTEC8 is found to be the most significant outlier with a ${\rm SFR}/{\rm SFR}_{\rm MS}$ ratio of 
about 13. The remaining six SMGs have ${\rm SFR}/{\rm SFR}_{\rm MS}\simeq 0.3-2.4$, 
and hence lie on or within the main sequence. Our result is consistent with previous studies 
where some of the SMGs are found to be located on or close to the main sequence 
(at the high-mass end), while a fair fraction of SMGs -- 
especially the most luminous objects -- are found to lie above the main sequence (e.g. \cite{magnelli2012}; \cite{michalowski2012}; \cite{roseboom2013}; \cite{dacunha2015}; \cite{koprowski2016}). This suggests that SMGs are a mix of populations of two star formation modes, namely ``normal-type'' 
star-forming galaxies and starbursts. Hydrodynamic simulations have also suggested that the SMG population 
can be divided into two subpopulations consisting of major-merger driven starbursts and disk galaxies where star formation, while possibly driven by mergers or smooth gas accretion, is occuring more quiescently (e.g. \cite{hayward2011}, 2012, and references therein).

\subsubsection{Stellar mass-size relationship}

As we discussed in Paper~II, the rest-frame UV/optical sizes (i.e. the stellar emission size scales), derived by Toft et al. (2014) for a subsample of seven of our SMGs, are smaller than the radio size for AzTEC4 and 5, in agreement for AzTEC15, and the upper stellar emission size limits for AzTEC1, 3, and 8 are larger than their radio sizes, and hence formally consistent with each other 
(AzTEC10 was also analysed by Toft et al. (2014), but it is not detected at 3~GHz). A difference in the radio and stellar emission sizes can be partly caused by the fact that rest-frame UV/optical emission can 
be subject to a strong, and possibly differential dust extinction, and stellar population effects. 

In the present work, we have revised the stellar masses of the aforementioned seven SMGs, and for those sources where the redshifts used in the analysis were similar, our values are found to be 0.3 to 1.6 (median 0.6) times those from Toft et al. (2014) (see Appendix~C). Despite these discrepancies, the way our SMGs populate the $M_{\star}$-size plane shown in Fig.~\ref{figure:size} is consistent with the finding of Toft et al. (2014), i.e. the $M_{\star}-R$ distribution of $3 < z < 6$ SMGs is comparable to that of cQGs at $z\sim2$. Hence, our result supports the authors' conclusion of high-$z$ SMGs being potential precursors of the $z\sim2$ cQGs, where the quenching of the starburst phase in the former type of galaxies leads to the formation of the latter population (Sect.~4.5). 

It is worth noting that five of our $z<3$ SMGs plotted in Fig.~\ref{figure:size} also exhibit stellar masses and radio sizes that place them in the $z\sim2$ cQGs' mass-size relationship from Krogager et al. (2014). While these authors found that the slope and scatter of this 
mass-size relationship are consistent with the local ($z=0$) values, the galaxies grow in size towards lower redshifts (e.g. via mergers). However, as can be seen in Fig.~7 of Krogager et al. (2014; see also references therein), the median size of the quiescent galaxy population at a fixed stellar mass of $M_{\star}=10^{11}$~M$_{\sun}$ is $R\sim2$~kpc in the redshift range $z\sim1.3-2$. In the context of massive galaxy evolution, the $z<3$ SMGs could potentially evolve into lower-redshift ($z<2$) cQGs, and then grow in size at later epochs (see Sect.~4.5). As mentioned in Sect.~3.4, there is a hint that the largest 3~GHz radio sizes are found among the most massive of our analysed SMGs (see the ouliers at $M_{\star}\gtrsim 2.6\times10^{11}$~M$_{\sun}$ in Fig.~\ref{figure:size}). While this is not statistically significant, also the most extended stellar emission spatial scale is found for the most massive SMG with available rest-frame UV-optical size measurement, namely AzTEC15. Although these largest radio sizes have large error bars, it is possible that the spatially extended radio emission in these massive SMGs is caused by processes not related to star formation, like galaxy mergers leading to magnetic fields being pulled out from the interacting disks (see Paper~II and references therein; O.~Miettinen et al., in prep.). Interestingly, all the three SMGs that exhibit the largest radio sizes in our sample, namely AzTEC4, 15, and 21a, are found to lie on the main sequence or only slightly above it (factor of 3.2 for AzTEC4), but they have high SFRs of $\sim 380-590$~${\rm M}_{\sun}~{\rm yr}^{-1}$, which could be induced by gravitational interaction of merging galaxies. This is further supported by the fact that many of our target SMGs exhibit clumpy or disturbed morphologies, or show evidence of close companions at different observed wavelengths, e.g. in the UltraVISTA NIR images (\cite{younger2007}, 2009; \cite{toft2014}; Paper~I). 

\subsection{Comparison with the physical properties of the ALESS 870~$\mu$m-selected SMGs}

Our main comparison sample of SMGs from the literature is the ALESS SMGs studied by da Cunha et al. (2015). The ALESS SMGs were uncovered in the LABOCA (Large APEX BOlometer CAmera) 870~$\mu$m survey of the Extended \textit{Chandra} Deep Field South (ECDFS) or LESS survey by Wei{\ss} et al. (2009), and followed up with $1\farcs5$ resolution Cycle 0 ALMA observations (\cite{hodge2013}; \cite{karim2013}). The reasons why we compare with the da Cunha et al. (2015) study are that \textit{i)} we have also used the new, high-$z$ version of {\tt MAGPHYS} as first presented and used by da Cunha et al. (2015) to derive the SMG physical properties, which allows for a direct comparison, and \textit{ii)} the SMG sample from da Cunha et al. (2015) is relatively large: they analysed the 99 most reliable SMGs detected in the ALESS survey (\cite{hodge2013}). 

\subsubsection{Description of the basic physical properties of the ALESS SMGs}

For their full sample of 99 ALESS SMGs, da Cunha et al. (2015) derived the following median properties (see their Table~1): 
$\log(M_{\star}/{\rm M}_{\sun})=10.95^{+0.6}_{-0.8}$, $\log(L_{\rm dust}/{\rm L}_{\sun})=12.55^{+0.3}_{-0.5}$, ${\rm SFR}=282^{+27}_{-31}$~M$_{\sun}$~yr$^{-1}$, 
${\rm sSFR}=2.8^{+8.4}_{-2.1}$~Gyr$^{-1}$, $T_{\rm dust}=43^{+10}_{-10}$~K, and $\log(M_{\rm dust}/{\rm M}_{\sun})=8.75^{+0.3}_{-0.4}$. The quoted uncertainties represent the 16th--84th percentile of the likelihood distribution. The value of $L_{\rm dust}$ reported by da Cunha et al. (2015) refers to the total dust IR ($3-1\,000$~$\mu$m) luminosity, which we have found to be almost equal to $L_{\rm IR}$ with the $L_{\rm IR}/L_{\rm dust}$ ratio ranging from 0.91 to 0.99 (both the mean and median being 0.95). 
We also note that da Cunha et al. (2015) defined the current SFR over the last 10~Myr, while the corresponding timescale in the present study is 100~Myr. When comparing the {\tt MAGPHYS} output SFRs, we found that the 10~Myr-averaged values are 1.0 to 4.7 times higher than those averaged over the past 100~Myr (the mean and median being 2.1 and 1.6, respectively). The authors concluded that the physical properties of the ALESS SMGs are very similar to those of local ultraluminous IR galaxies or ULIRGs (see \cite{dacunha2010}).

For a more quantitative comparison, we limit the da Cunha et al. (2015) sample to those SMGs that have 870~$\mu$m flux densities corresponding to our AzTEC 1.1~mm flux density range in the parent sample (AzTEC1--30), i.e. $3.3~{\rm mJy}\leq S_{\rm 1.1\, mm} \leq9.3$~mJy. Assuming that the dust emissivity index is $\beta=1.5$, this flux density range corresponds to $7.5~{\rm mJy}\leq S_{\rm 870\, \mu m} \leq21.1$~mJy.\footnote{We note that in the high-$z$ model libraries of {\tt MAGPHYS}, the value of $\beta$ is fixed at 1.5 for the warm dust component (30--80~K), while that for the colder (20--40~K) dust is $\beta=2$. A manifestation of this $T_{\rm dust}$-dependent $\beta$ is that scaling the 1.1~mm flux densities to those at 870~$\mu$m with a simple assumption of $\beta=1.5$ yields slightly different values than suggested by the {\tt MAGPHYS} SEDs shown in Fig.~\ref{figure:seds}.} The LESS SMGs that fall in this flux density range are LESS1--18, 21--23, 30, 35, and 41, where LESS1, 2, 3, 7, 15, 17, 22, 23, 35, and 41 were resolved into multiple components with ALMA (\cite{hodge2013}; \cite{karim2013}). The photometric redshifts of these SMGs, as derived by da Cunha et al. (2015), lie in the range of $z_{\rm phot}=1.42-5.22$ with a median and its standard error of $z_{\rm phot}=3.12\pm0.24$. We note that this median redshift is only 9.5\% higher than that of our analysed SMGs ($z=2.85\pm0.32$). As mentioned in Sect.~3.1.2, the error bars for the ALESS SMGs' parameters were propagated from the photo-$z$ uncertainties by da Cunha et al. (2015), and they are typically much larger than those of our parameters (see e.g. Fig.~\ref{figure:ms} herein).

For the aforementioned flux-limited sample, the median values of the physical parameters are $\log(M_{\star}/{\rm M}_{\sun})=11.0^{+0.46}_{-0.61}$, $\log(L_{\rm dust}/{\rm L}_{\sun})=12.69^{+0.31}_{-0.43}$, ${\rm SFR}=417^{+578}_{-243}$~M$_{\sun}$~yr$^{-1}$, ${\rm sSFR}=4.8^{+7.6}_{-3.7}$~Gyr$^{-1}$, $T_{\rm dust}=42.0^{+4.3}_{-5.3}$~K, and $\log(M_{\rm dust}/{\rm M}_{\sun})=8.98^{+0.20}_{-0.48}$, where we quote the 16th--84th percentile range. All the other quantities except $T_{\rm dust}$ are higher than for the aforementioned full sample, up to a factor of 1.7 for sSFR and $M_{\rm dust}$, which is not surprising because the subsample in question is composed of the brightest ALESS SMGs.

\subsubsection{Comparison of the AzTEC and ALESS SMGs}

In what follows, we compare the physical properties of our SMGs with those of the aforementioned flux-limited ALESS 
sample composed of equally bright sources. The ratios between our median $M_{\star}$, $L_{\rm dust}$, SFR, sSFR, $T_{\rm dust}$, and $M_{\rm dust}$ values and those of the ALESS SMGs are given in Table~\ref{table:comparison}. We note that for a proper comparison, the comparison of the SFR and sSFR values were done using the {\tt MAGPHYS} output values averaged over the past 10~Myr. As can be seen in Table~\ref{table:comparison}, the median values of $M_{\star}$ and $T_{\rm dust}$ are similar, and our median $M_{\rm dust}$ value is a factor of 1.5 times higher than for the equally bright ALESS SMGs. On the other hand, our dust luminosities and (s)SFR values appear to be about two times higher on average.


We also performed a two-sided Kolmogorov–Smirnov (K-S) test between the aforementioned physical parameter values to check if our SMGs and the flux-limited ALESS SMG sample could be drawn from a common underlying parent distribution. The null hypothesis was that these two samples are drawn from the same parent distribution. The K-S test statistics and $p$-values for the comparisons of the $M_{\star}$, $L_{\rm dust}$, ${\rm SFR}_{\rm MAGPHYS}$, $T_{\rm dust}$, and $M_{\rm dust}$ values are also given in Table~\ref{table:comparison}. The K-S test results suggest that the underlying stellar mass distribution is the same ($p= 0.92$), while those of the remaining properties might differ. We note that for $T_{\rm dust}$, for which the median value between our AzTEC SMGs and the ALESS SMGs was found to be very similar, the K-S test $p$-value is 0.29, the second highest after the stellar mass comparison. It should also be noted that the comparison samples are small (16 AzTEC and 25 ALESS sources, respectively), and hence the K-S tests presented here are subject to small number statistics. Nevertheless, we cannot exclude the possibility that at least part of the differences found here is caused by the different selection wavelength ($\lambda_{\rm obs}=1.1$~mm versus $\lambda_{\rm obs}=870$~$\mu$m), and different depths of the optical to IR observations available in COSMOS and the ECDFS, although the stellar mass estimates based on the optical regime of the galaxy SED are found to be similar.

da Cunha et al. (2015) found that, at $z \simeq 2$, about half of the ALESS SMGs (49\%) lie above the star-forming main sequence 
(i.e. ${\rm SFR}/{\rm SFR}_{\rm MS}>3$), while the other half (51\%) are consistent with being at the high-mass end of the main sequence, 
where the main sequence definition was also adopted from Speagle et al. (2014). For the ALESS SMG sample flux-limited to match our sample limit, which has a median redshift of $z=3.12$, only $24\pm10\%$ of the sources are found to have ${\rm SFR}/{\rm SFR}_{\rm MS}>3$, while the remaining $76\pm17\%$ lie within a factor of 3 of the main sequence (see our Fig.~\ref{figure:ms}). It should be noted, however, that the ALESS sources have significant error bars in their SFR and $M_{\star}$ values (propagated from the photo-$z$ uncertainties; \cite{dacunha2015}). The fractions we have found for the analysed AzTEC SMGs are nominally more extreme, i.e. $62.5\%\pm19.8\%$ are above the main sequence, and $37.5\%\pm15.3\%$ are consistent with the main sequence. If we base our analysis on the 10~Myr-averaged {\tt MAGPHYS} output SFRs as da Cunha et al. (2015) did, we find that the fraction of the AzTEC SMGs having ${\rm SFR}/{\rm SFR}_{\rm MS}>3$ is the same $62.5\%\pm19.8\%$ as derived above from the Kennicutt (1998) $L_{\rm IR}-{\rm SFR}$ calibration.

\begin{figure}[!htb]
\centering
\resizebox{0.9\hsize}{!}{\includegraphics{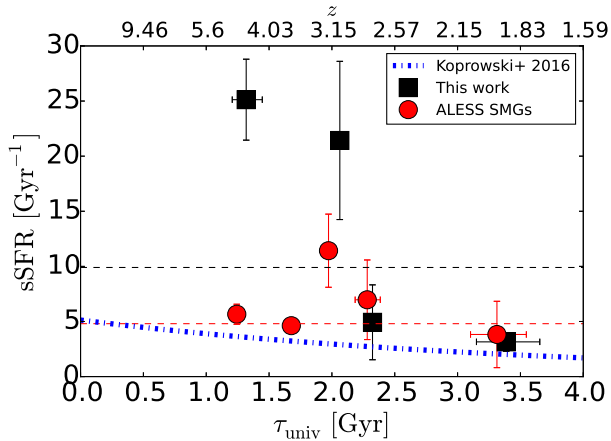}}
\caption{Specific SFR (sSFR) as a function of cosmic time (lower $x$-axis) and redshift (upper $x$-axis) for our SMGs and the flux-limited sample of ALESS SMGs. The plotted data points represent the mean values of our data and the ALESS SMG data binned in redshift and sSFR (each of our bin contains four sources, while the ALESS data have five values in each bin). The error bars represent the standard error of the mean. The horizontal dashed lines mark the full sample median sSFRs of $9.9^{+21.4}_{-8.1}$~Gyr$^{-1}$ and $4.8^{+7.6}_{-3.7}$~Gyr$^{-1}$, respectively. The blue dash-dot line corresponds to the relationship $\log({\rm sSFR}/{\rm Gyr^{-1}})=-0.12\times (\tau_{\rm univ}/{\rm Gyr})+0.71$ derived by Koprowski et al. (2016).}
\label{figure:sSFR}
\end{figure}

In Fig.~\ref{figure:sSFR}, we plot the sSFR as a function of cosmic time (lower $x$-axis) and redshift (upper $x$-axis). For legibility purposes, we only show the binned version of the data; the plotted data points represent the mean values of the full data with four SMGs per bin in our AzTEC sample, and five SMGs per bin in the ALESS sample. The sSFR of the ALESS comparison sample appears to be relatively constant as a function of cosmic time, while our individual SMGs, lying in a similar redshift range, show more scatter with a factor of $2.1^{+26.4}_{-2.0}$ higher median sSFR (our median sSFR is higher by the same nominal factor of $2.1^{+35.4}_{-2.0}$ when the comparison is done between the 10~Myr-averaged {\tt MAGPHYS} outputs; Table~\ref{table:comparison}). The blue dash-dot line overplotted in the figure represents the sSFR-cosmic time relationship derived by Koprowski et al. (2016). As can be seen, our lowest redshift data point (corresponding to the latest time) lies slightly (by a factor of 1.57) above the normalisation of this relationship, and the second lowest redshift data point, though having a factor of 1.82 higher sSFR than suggested by the Koprowski et al. (2016) relationship, is still consistent with it within the standard error. However, our two highest redshift bins lie at much higher sSFRs than computed from the Koprowski et al. (2016) relationship (by factors of $\sim7$), but as mentioned by the authors, their study could not set tight constraints on the sSFR beyond redshift of $z\simeq3$ ($\tau_{\rm univ}=2.1$~Gyr), where many of our SMGs are found. Indeed, as can be seen in Fig.~10 of Koprowski et al. (2016), the scatter of data increases at $z>3$, and many data points lie above their derived relationship. It should also be pointed out that our SFRs and stellar masses were derived using a different method than those in Koprowski et al. (2016) and their reference studies, and this can be part of the reason why our values lie above the Koprowski et al. (2016) sSFR($\tau_{\rm univ}$) function. On the other hand, while the three lowest redshift ALESS data points shown in Fig.~\ref{figure:sSFR} show a trend similar to ours, with a jump in sSFR near $z\simeq3$, the two highest redshift ALESS bins are only by factors of 1.43--1.56 above the Koprowski et al. (2016) relationship. Larger, multi-field samples of SMGs are required to limit cosmic variance, and examine the evolution of SMGs' sSFR as a function of cosmic time further, particularly at $z\gtrsim3$.

\begin{figure}[!h]
\centering
\resizebox{0.9\hsize}{!}{\includegraphics{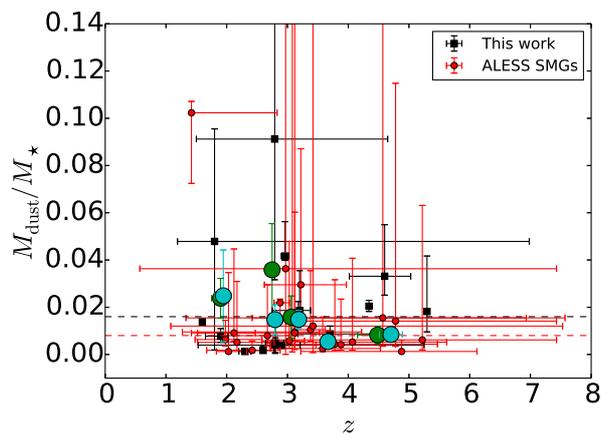}}
\caption{Dust-to-stellar mass ratio as a function of redshift for our SMGs and the flux-limited ALESS SMGs. The green filled circles 
represent the mean values of the binned AzTEC data (each bin contains four SMGs), while the cyan filled circles represent the mean values of 
the binned ALESS data (each bin contains five SMGs). The error bars of the binned data points represent the standard errors of the mean 
values. The horizontal dashed lines mark the corresponding median values of $0.016^{+0.022}_{-0.012}$ and $0.008^{+0.009}_{-0.004}$, 
respectively.}
\label{figure:corr}
\end{figure}

Figure~\ref{figure:corr} plots the dust-to-stellar mass ratio as a function of redshift for our SMGs and the comparison ALESS SMG sample. 
The median values for these samples are $M_{\rm dust}/M_{\star}=0.016^{+0.022}_{-0.012}$ and 
$M_{\rm dust}/M_{\star}=0.008^{+0.009}_{-0.004}$, respectively, where the quoted uncertainties represent the 16th--84th percentile range. The binned data points shown in Fig.~\ref{figure:corr} show a hint of decreasing dust-to-stellar mass ratio towards higher redshifts. The AzTEC (ALESS) data points suggest a linear regression of the form $M_{\rm dust}/M_{\star} \propto -(0.006 \pm 0.003)\times z$ ($M_{\rm dust}/M_{\star} \propto -(0.002 \pm 0.002)\times z$), with a Pearson $r$ of $-0.68$ ($-0.86$). These trends are not statistically significant, but the fact that both the samples show a comparable behaviour is indicative of a star formation and dust production history being fairly similar between the AzTEC and ALESS SMGs, and hence suggesting a similar level of metallicity, which is not surprising given that these SMGs lie at the same cosmic epoch.

Thomson et al. (2014) studied the radio properties of the ALESS SMGs. They used 610~MHz GMRT and 1.4~GHz VLA data, 
and derived a median$\pm$standard error radio spectral index of $\alpha_{\rm 610\, MHz}^{\rm 1.4\, GHz}=-0.79\pm0.06$ 
for a sample of 52 SMGs. Again, if we limit this comparison sample to those sources that are equally bright to ours, 
we derive a median spectral index of $\alpha_{\rm 610\, MHz}^{\rm 1.4\, GHz}=-0.79\pm0.19$ from the values reported in their Table~3 
(survival analysis was used to take the lower limits into account). This is the same as for their full sample, 
and also consistent with our median $\alpha_{\rm 325\, MHz}^{\rm 3\, GHz}$ value of $-0.77^{+0.28}_{-0.42}$ although our observed frequency range is broader.


\begin{table}[H]
\renewcommand{\footnoterule}{}
\caption{Comparison of the physical properties between the target AzTEC SMGs and the equally bright ALESS SMGs.}
{\normalsize
\begin{minipage}{1\columnwidth}
\centering
\label{table:comparison}
\begin{tabular}{c c}
\hline\hline 
Parameter\tablefootmark{a} & Value\\
\hline 
$M_{\star}^{\rm AzTEC}/M_{\star}^{\rm ALESS}$ & $0.9^{+7.2}_{-0.7}$ \\ [1ex]
$L_{\rm dust}^{\rm AzTEC}/L_{\rm dust}^{\rm ALESS}$ & $1.9^{+9.3}_{-1.6}$ \\ [1ex]
${\rm SFR}^{\rm AzTEC}/{\rm SFR}^{\rm ALESS}$ & $2.3^{+11.2}_{-2.0}$\tablefootmark{b} \\ [1ex]
${\rm sSFR}^{\rm AzTEC}/{\rm sSFR}^{\rm ALESS}$ & $2.1^{+35.4}_{-2.0}$\tablefootmark{b} \\[1ex]
$T_{\rm dust}^{\rm AzTEC}/T_{\rm dust}^{\rm ALESS}$ & $1.0^{+0.3}_{-0.3}$ \\ [1ex]
$M_{\rm dust}^{\rm AzTEC}/M_{\rm dust}^{\rm ALESS}$ & $1.5^{+3.5}_{-1.0}$ \\[1ex]
\hline 
\multicolumn{2}{c}{K-S test results\tablefootmark{c}}\\ [1ex]
\hline
$M_{\star}$ & $D=0.17$, $p=0.92$ \\ [1ex]
$L_{\rm dust}$ & $D=0.46$, $p=0.02$ \\ [1ex]
${\rm SFR}$ & $D=0.38$\tablefootmark{b}, $p=0.10$\tablefootmark{b} \\ [1ex]
$T_{\rm dust}$ & $D=0.30$, $p=0.29$ \\ [1ex]
$M_{\rm dust}$ & $D=0.39$, $p=0.08$ \\ [1ex]
\hline
\end{tabular} 
\tablefoot{\tablefoottext{a}{A ratio between the median values.}\tablefoottext{b}{The comparison was done between the {\tt MAGPHYS} output values averaged over 10~Myr.}\tablefoottext{c}{Results from a two-sided K-S test between the two sets of physical properties. The maximum distance between the two cumulative distribution functions is given by the K-S test statistic $D$, while the corresponding $p$-value describes the probability that the two datasets are drawn from the same underlying parent distribution. }   }
\end{minipage} }
\end{table}

\subsection{Properties of the radio SEDs}

\subsubsection{Radio continuum spectra of star-forming galaxies}

The radio continuum emission arising from a star-forming galaxy can be divided into two main components: (1) the thermal free-free emission (bremsstrahlung) from 
\ion{H}{ii} regions, and (2) non-thermal synchrotron emission from relativistic cosmic-ray electrons. Both of these emission mechanisms are directly linked 
to the evolution of high-mass stars: a newly forming high-mass star photoionises its surrounding medium to create an \ion{H}{ii} region, while the explosive 
deaths of high-mass ($>8$~M$_{\sun}$) stars -- supernovae and their remnants -- are associated with shock fronts that can accelerate cosmic-ray particles to relativistic energies. 

While a typical synchrotron spectral index is $\alpha_{\rm synch}\in [-0.8,\, -0.7]$ (Sect.~3.2, and references therein), that of thermal free-free emission is about seven to eight times flatter, i.e. $\alpha_{\rm ff}=-0.1$. At low frequencies of $\nu_{\rm rest}<30$~GHz the total radio emission is expected to be dominated by the non-thermal synchrotron component, while at higher frequencies (up to about 100~GHz) the thermal emission component makes a larger contribution of the total observed radio wave emission (e.g. \cite{condon1992}; \cite{niklas1997}).\footnote{We note that the radio flux density at 33~GHz might be affected by the anomalous dust emission, but its significance in the galaxy-integrated measurements is unclear (\cite{murphy2010}).} Because of the thermal emission, the radio spectrum exhibits flattening towards higher frequencies. Besides this effect, the low-frequency regime of a radio SED ($\nu_{\rm rest}\lesssim1$~GHz) can also exhibit flattening because the free-free optical thickness increases ($\tau \propto \nu^{-2.1}$), and hence the free-free absorption due to ionised gas becomes important. 

On the other hand, cosmic-ray electrons suffer from a number of cooling or energy-loss mechanisms. These include synchrotron radiation, inverse-Compton (IC) scattering, ionisation, and bremsstrahlung losses (e.g. \cite{murphyonly2009}; Paper~II and references therein), all of which can shape the radio SED, i.e. modify the spectral index. For example, besides the free-free absorption, the ionisation energy-losses can flatten the radio spectrum towards lower frequencies ($<1$~GHz; e.g. \cite{thompson2006}). These energy-loss mechanisms are dependent on the physical properties of the galaxy, like the magnetic field strength, neutral gas density, and the energy density of the radiation field (e.g. \cite{basu2015}; Paper~II and references therein). Moreover, the IC losses at high redshifts can be aided by IC losses off the cosmic microwave background (CMB) photons, because the CMB energy density goes as $u_{\rm CMB}\propto(1+z)^4$. 
 
Because the aforementioned physical properties (magnetic field strength, density, etc.) have spatial gradients within the galaxy, the interpretation of the observed galaxy-integrated radio SEDs can become difficult (\cite{basu2015}). However, measurements of the radio spectral index over different frequency ranges can provide useful information about the energy-loss and gain mechanisms of the leptonic (e$^{\pm}$) cosmic-ray population. In the next subsection, we attempt to classify our SMGs into different categories on the basis of the observed radio spectral indices.

\subsubsection{Radio SED classification}

To classify our SMGs into different radio SED categories, we define the following three classes: I$=$classic synchrotron spectrum ($\alpha_{\rm synch}\simeq -0.8\ldots-0.7$), 
II$=$flattened radio spectrum, and III$=$steepened radio spectrum. The classification in the observed-frame frequency interval of 325~MHz--3~GHz 
is provided in Table~\ref{table:categories}. Seven of our sources agree with a canonical synchrotron spectrum (our category~I), while the remaining 11
fall into a category~II or III with a 5:6 proportion. We note that AzTEC3, 10, 21a, and 27 exhibit a nominal spectral index of $\alpha<-1.0$ 
(and AzTEC2 has $\alpha_{\rm 1.4\, GHz}^{\rm 3\, GHz}=-1.83\pm0.93$), and these sources could also be classified as ultra-steep spectrum sources (\cite{thomson2014}, 2015). 

In the next subsection, we will examine the contribution of the thermal free-free emission to the flattest radio spectrum found in the present study.

\subsubsection{Thermal fraction}

Considering the contribution of the thermal free-free emission to the derived spectral indices, we note that the highest rest-frame frequency probed in the present study 
is 18.9~GHz, which corresponds to $\nu_{\rm obs}=3$~GHz at the redshift of AzTEC3 ($z_{\rm spec}\simeq5.3$). However, AzTEC3 has one of the steepest spectral index derived 
here ($\alpha_{\rm 325\, MHz}^{\rm 3\, GHz}=-1.19\pm0.32$), although the non-detection at 1.4~GHz could be an indication of a flattening towards higher frequencies 
($\alpha_{\rm 1.4\, GHz}^{\rm 3\, GHz}>-0.56$). On the other hand, the flattest index found here is $\alpha_{\rm 325\, MHz}^{\rm 3\, GHz}>-0.18$ for AzTEC4, which might be an indication that 
the observed-frame 3~GHz flux density has a contribution from thermal free-free emission. To quantify the level of this free-free contribution, we can translate the 
$L_{\rm IR}$-based SFR into the thermal luminosity using Eq.~(11) of Murphy et al. (2011), which for a Chabrier (2003) IMF scaling can be written as

\begin{equation}
\frac{{\rm SFR}_{\rm ff}}{{\rm M}_{\sun}~{\rm yr}^{-1}}=3.1\times10^{-21}\, \left(\frac{T_{\rm e}}{10^4\,{\rm K}}\right)^{-0.45}\left(\frac{\nu}{{\rm GHz}}\right)^{0.1}\left(\frac{L_{\nu}^{\rm ff}}{{\rm W\, Hz^{-1}}}\right)\,,
\end{equation}
where $T_{\rm e}$ is the electron temperature, which is here set to $10^4$~K. For AzTEC4, for which the nominal SFR is $380$~${\rm M}_{\sun}~{\rm yr}^{-1}$, we obtain a luminosity density of $\sim1\times10^{23}$~W~Hz$^{-1}$ at the rest-frame frequency $\nu_{\rm rest}=8.4$~GHz (corresponding to $\nu_{\rm obs}=3$~GHz). By comparing this with the radio luminosity density calculated using the observed 3~GHz flux density (see our Eq.~(2)), we estimate the thermal fraction to be $ >33\%$ for AzTEC4 at $\nu_{\rm obs}=3$~GHz (the lower limit results from the lower $\alpha_{\rm 325\, MHz}^{\rm 3\, GHz}$ limit used in the calculation). Although the thermal contribution makes the derivation of the non-thermal spectral index more uncertain, our result demonstrates that it also offers a possibility to estimate the SFR of high-$z$ galaxies more directly than relying on the IR-radio correlation. The high sensitivity of the future facilities, like the Square Kilometre Array\footnote{{\tt https://www.skatelescope.org}}, has the potential to use thermal free-free emission as a reliable, dust-unbiased SFR tracer.

\subsubsection{Radio characteristics and the offset from the main sequence}

\begin{figure}[!h]
\centering
\resizebox{0.999\hsize}{!}{\includegraphics{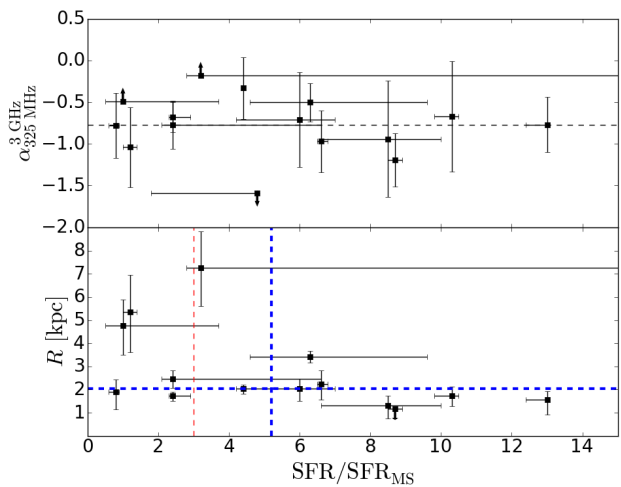}}
\caption{\textbf{Top:} Radio spectral index between 325~MHz and 3~GHz as a function of 
starburstiness or the distance from the main sequence (parameterised as 
${\rm SFR}/{\rm SFR}_{\rm MS}$). For reference, the horizontal dashed line marks the median 
spectral index value of $-0.77^{+0.28}_{-0.29}$ (see Table~\ref{table:radio2}). \textbf{Bottom:} 3~GHz radio size (radius calculated as $\theta_{\rm maj}({\rm FWHM})/2$) from Paper~II as a function of the distance above the main sequence. The blue vertical dashed line marks 
the median value of ${\rm SFR}/{\rm SFR}_{\rm MS}=5.2$ derived for the plotted sample, while the horizontal line marks the survival 
analysis-based median radius of 2.03~kpc for the plotted sample. The thin, red vertical dashed line shows the upper boundary limit of the main sequence, i.e. ${\rm SFR}/{\rm SFR}_{\rm MS}=3$.}
\label{figure:alpha}
\end{figure}

Finally, because most ($63\%$) of our SMGs are found to be starbursts (Sect.~4.1.1), we compare our results with Murphy et al. (2013) who found 
that the radio spectral indices of galaxies flatten with increasing distance above the galaxy main sequence. The authors interpreted this as 
compact starbursts being more optically thick in the radio, i.e. free-free absorption being more important in these systems ($\tau_{\nu}\propto n_{\rm e}^2$, 
where $n_{\rm e}$ is the electron density). To test whether our SMGs exhibit 
such a trend, in the top panel in Fig.~\ref{figure:alpha} we plot our 
$\alpha_{\rm 325\, MHz}^{\rm 3\, GHz}$ values as a function of the starburstiness parameter ${\rm SFR}/{\rm SFR}_{\rm MS}$. 
As can be seen, no trend is discernible in Fig.~\ref{figure:alpha}. We also note that in Paper~II we did not find a correlation between the 3~GHz radio sizes 
of our SMGs and $\alpha_{\rm 1.4\, GHz}^{\rm 3\, GHz}$ (see Fig.~4 therein)

If the scenario proposed by Murphy et al. (2013) plays out, then one might expect to see a correlation between the size of a galaxy and its distance from the main sequence. To see if this is the case for our SMGs, in the bottom panel in Fig.~\ref{figure:alpha} we plot the 3~GHz radio sizes from Paper~II as a function of ${\rm SFR}/{\rm SFR}_{\rm MS}$. Indeed, we see a possible hint of more compact sources being those that lie the furthest away from the main sequence. Interestingly, the largest radio-emitting sizes are seen among SMGs within or near the main sequence (${\rm SFR}/{\rm SFR}_{\rm MS}\lesssim3$), although we note that SMGs with a radio size close to the sample median size are also found within the main sequence envelope. Hence, no firm conclusions can be drawn on the basis of current data.


\begin{table}[H]
\renewcommand{\footnoterule}{}
\caption{Classification of the target SMGs into different radio SED categories in the observed frequency interval 0.325--3~GHz.}
{\normalsize
\begin{minipage}{1\columnwidth}
\centering
\label{table:categories}
\begin{tabular}{c c}
\hline\hline 
Category\tablefootmark{a} & AzTEC ID\\
\hline 
I & 7, 8, 11-N, 12, 17a, 19a, 24b \\
II & 4, 5, 6, 11-S, 15\\        
III & 1, 3, 9, 10, 21a, 27\\ 
\hline
\end{tabular} 
\tablefoot{\tablefoottext{a}{We define our categories as follows: I$=$classic synchrotron radio SED; II$=$flattened radio SED; and III$=$steepened radio SED.}}
\end{minipage} }
\end{table}

\subsection{The IR-radio correlation}

\subsubsection{Comparison with literature data}

The IR-radio $q$ parameters we have derived, ranging from $q >1.64$ to $q= 2.82\pm0.18$ with a median of $q=2.27^{+0.27}_{-0.13}$, exhibit a hint of negative correlation with redshift (see Sect.~3.5, and the binned data in Fig.~\ref{figure:q}). 

Thomson et al. (2014) reported a median $q_{\rm TIR}$ of $2.56\pm0.05$ for their ALESS SMGs. However, to compute $q$ Thomson et al. (2014) used the IR luminosities derived by Swinbank et al. (2014). Hence, we recalculated the $q$ values for the flux-limited comparison sample using the $L_{\rm dust}$ values (and photo-$z$'s) derived by da Cunha et al. (2015), and the 610~MHz GMRT flux densities from Thomson et al. (2014) (following our Eqs.~(1) and (2) modified for $\nu_{\rm obs}=610$~MHz).\footnote{We note that for the radio non-detected (at both 610~MHz and 1.4~GHz) ALESS SMGs, we adopted a spectral index of $\alpha=-1.57\pm0.08$ derived through stacking by Thomson et al. (2014). The authors reported upper $q$ limits for the radio non-detected SMGs in their Table~1, but in the case of an upper limit to $S_{\rm 610 \, MHz}$, the corresponding $q$ value becomes a lower limit.} The median $q$ we derived this way is $2.29\pm0.09$, where we quote the standard error of the median. Because this value refers to the total dust luminosity ($3-1\,000$~$\mu$m), the total-IR $q$ is slightly lower. Given that the ratio of our median $q$ value to the recalculated ALESS $q$ is 0.99, we conclude that the IR-radio correlation parameter is not significantly different between our SMGs and the equally bright ALESS SMGs. 
 
Similarly to Thomson et al. (2014), Barger et al. (2012) reported an average $q_{\rm TIR}$ value of $2.51\pm0.01$ for their sample of five SCUBA 850~$\mu$m-identified SMGs at $z=2-4$ in the Great Observatories Origins Deep Survey-North (GOODS-N). The AzTEC 1.1~mm flux densities of these SMGs lie in the range of $S_{\rm 1.1\, mm}=(1.9\pm1.4)-(5.4\pm1.1)$~mJy (see Table~6 in \cite{barger2012}), and they are generally fainter at $\lambda_{\rm obs}=1.1$~mm than our target SMGs with $S_{\rm 1.1\, mm}=3.3^{+1.4}_{-1.6}-9.3^{+1.3}_{-1.3}$~mJy (\cite{scott2008}). Additionally, because Barger et al. (2012) assumed a radio spectral index of $\alpha=-0.8$, and used a Arp 220 SED-based relationship of $L_{\rm IR}=1.42 \times L_{\rm FIR}$ in their calculation, a direct comparison with our $q_{\rm TIR}$ calculation is not feasible.

The average FIR-based $q_{\rm FIR}$ values derived for SMGs lying at median redshifts of $z \sim 2-2.5$ and without AGN signatures are usually found to be about 2.0--2.2, which are comparable with those of other types of star-forming galaxies at similar redshifts (e.g. \cite{kovacs2006}; \cite{magnelli2010}; \cite{alaghband2012}). For our SMGs, 36\% to 90\% of the total-IR luminosity is found to emerge in the FIR, with a mean (median) fraction of 75\% (80\%). Hence, our median $q_{\rm FIR}$ can be derived to be about 2.17, in excellent agreement with previous results.

\subsubsection{Evolution of the IR-radio correlation}

The question then arises as to what is causing the relatively low $q_{\rm TIR}$ in our (and ALESS) SMGs with respect to the local universe median value of $q_{\rm TIR}=2.64$ (see \cite{bell2003}; \cite{sargent2010}). One possible explanation is related to the dynamical/evolutionary stage of the SMG system. The models by Bressan et al. (2002) predict that the $q$ parameter reaches its minimum value during the post-starburst phase. The authors suggested that this evolutionary effect is due to radio emission fading more slowly after an intense burst of star formation than the (F)IR emission, which is manifested as an apparent radio excess or lower $q$ value. Because SMGs are potentially driven by mergers (e.g. \cite{tacconi2008}; \cite{engel2010}), some of the low $q$ values we have observed suggest that the target SMG is being observed during the post-starburst phase. 

Schleicher \& Beck (2013) also discussed the possibility of accretion/merger-driven turbulence enhancing the magnetic field strength. If an SMG is driven by a major merger, this could be one mechanism to yield a low $q$ value. It is also possible that environmental effects increase the galactic magnetic field strength and/or reaccelerate cosmic rays, hence causing an excess in radio emission, which lowers the $q$ value (\cite{miller2001}; \cite{reddy2004}; \cite{murphy2009}). On the other hand, the value of $q_{\rm TIR}$ we have derived for AzTEC8, i.e. $ 2.82\pm0.18$, is relatively high compared to the local median value. An elevated $q_{\rm TIR}$ can result from a recent starburst phenomenon, where the synchrotron emission is generated with a time delay (due to the generation of relativistic electrons and/or magnetic field) with respect to the stellar radiation field that heats the dust (\cite{roussel2003}).

Some previous studies have found a moderate evolution of $q$ as a function of redshift (\cite{ivison2010a},b; \cite{magnelli2010}; \cite{casey2012}; \cite{magnelli2015}; see also \cite{delhaize2016}). For instance, Magnelli et al. (2015) found a statistically significant redshift evolution of the form $q_{\rm FIR}=(2.35\pm0.08)\times(1+z)^{-0.12 \pm 0.04}$ for their $M_{\star}$-selected, multi-field sample of $0<z<2.3$ galaxies (including COSMOS), in agreement with that found earlier by Ivison et al. (2010a) for 250~$\mu$m-selected galaxies at $z=0.038-2.732$, i.e. $q_{\rm TIR}\propto(1+z)^{-0.15 \pm 0.03}$. The latter relationships are overplotted in Fig.~\ref{figure:q}, normalised to yield the local universe median value of $q_{\rm TIR}(z=0)=2.64$. We note that the slope of the $q_{\rm TIR}(z)$ curves becoming shallower with increasing $z$ is consistent with a turbulent dynamo theory prediction by Schober et al. (2016).

As can be seen in Fig.~\ref{figure:q}, our binned SMG data show a weak hint of decreasing evolution in $q_{\rm TIR}$ as a function of cosmic time. A linear regression yields a relationship of the form $q_{\rm TIR} \propto -(0.050 \pm 0.060)\times z$, with a Pearson correlation coefficient of $r=-1$. Hence, the decreasing trend seen here is not statistically significant. The two lowest redshift bins at $z=2.04$ and $z=2.93$ are consistent within the standard errors with the nominal $q=q(z)$ relationships of Ivison et al. (2010a) and Magnelli et al. (2015). Our highest redshift binned data point at $z=4.23$ lies above those relationships. We note that the ALESS SMGs do not exhibit a (negative) correlation in the $q_{\rm TIR}$-$z$ plane (\cite{thomson2014}; see Fig.~3 therein). 

It is not clear what would cause a trend of decreasing $q$ with redshift. One possibility is an erroneous radio $K$-correction for higher redshift sources due to an increasing thermal free-free contribution to the observed emission (\cite{delhaize2016}). In the present work, we have minimised the $K$-correction by using the observed-frame 325~MHz radio flux densities (Sect.~3.5). If, however, the trend as a function of cosmic time is physical, the question arises whether it is driven by an enhanced radio emission or a depressed IR emission from dust at higher redshifts. If the ISM density of galaxies increases towards higher redshifts, the corresponding galactic magnetic field could be stronger, which in turn would lead to a stronger radio synchrotron emission implying a lower $q$ value (\cite{schleicher2013}). Given the enhanced IC-CMB cooling of cosmic-ray electrons at high redshifts (Sect.~4.3), the associated suppression of synchrotron radiation could even lead to an increase of $q$ towards higher redshifts, but the magnetic field strength in SMGs might be strong enough to compensate this effect (\cite{thomson2014}). One possible interpretation of the weak or even a complete lack of any obvious $q(z)$ trend among SMGs compared to the negative $q(z)$ relationship among less-extreme star-forming galaxies is that the latter type of galaxies exhibit a lower dust content (and hence a lower $L_{\rm IR}$ and $q$) at higher redshifts (e.g. \cite{capak2015}), while SMGs are rich in dust at all cosmological epochs they are being observed (and hence $q$ does not decrease, or does so only weakly).

\subsection{The 1.1~mm-selected COSMOS/AzTEC SMGs in the context of massive galaxy evolution}

Putting the results from our previous studies and the present one together, we are in a position to discuss our SMGs in the context of the evolution of massive galaxies. We start by noting that although the SMGs studied in the present work appear to preferentially lie above the galaxy main sequence ($63\%$ have a ${\rm SFR}/{\rm SFR}_{\rm MS}$ ratio $>3$), and can be considered starbursts, a fair fraction of about $38\%$ are consistent with being main-sequence star-forming galaxies. Nevertheless, the high SFRs of $\sim20-1\,500~{\rm M}_{\sun}~{\rm yr}^{-1}$ with a median of $508~{\rm M}_{\sun}~{\rm yr}^{-1}$ derived for these main-sequence SMGs show that a phase of intense star formation is taking place in these galaxies. We also find an indication for an anti-correlation between distance from the main sequence and radio size (Fig.~\ref{figure:alpha}, bottom panel). Moreover, as shown in Fig.~\ref{figure:burst}, there is an average trend of $z>3$ SMGs having a higher level of starburstiness (${\rm SFR}/{\rm SFR}_{\rm MS}$ ratio) than those SMGs that lie at $z<3$. On the other hand, in Paper~II we did not find any trend between radio size and redshift of our target SMGs (Fig.~7 therein; see also O.~Miettinen et al., in prep.). 

As further shown for two of our high-$z$ SMGs AzTEC1 and AzTEC3 in Appendix~D, an early starburst phase can be characterised by a SFR occuring near or at the Eddington limit within a spatially compact region. Such high SFRs of luminous SMGs are plausibly triggered by a wet merger between gas-rich galaxies, or, perhaps in the case of less luminous SMGs, by gravitational instabilities fuelled by more continuous, cold gas accretion (\cite{narayanan2009}; \cite{dekel2009}; \cite{engel2010}; \cite{wiklind2014}; \cite{narayanan2015}). For example, the AzTEC11 system, which is composed of two very nearby ($1\farcs46$ or $12.4$~kpc in projection) dust and radio-emitting components, is possibly in the process of merging, in a phase before the parent nuclei have fully coalesced (see Paper~II; Fig.~2 therein, and O.~Miettinen et al., in prep.). 

\subsubsection{$z>3$ SMGs}

\begin{figure}[!h]
\centering
\resizebox{0.99\hsize}{!}{\includegraphics{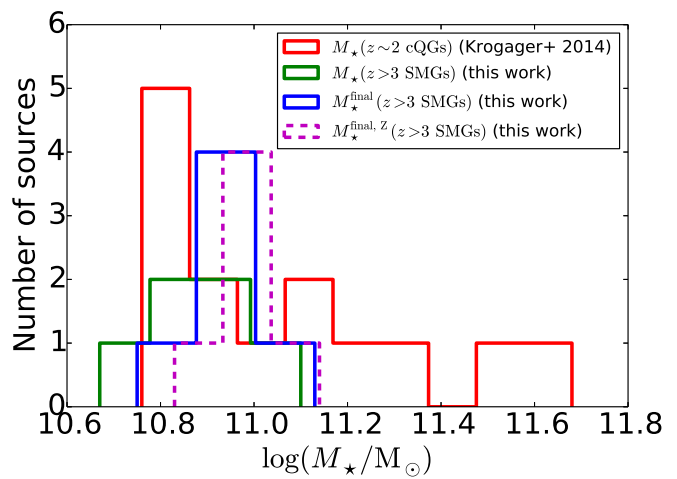}}
\caption{Stellar mass distribution of our $z>3$ SMGs (green histogram), and $z\sim2$ cQGs (\cite{krogager2014}; red histogram). The blue histogram shows the distribution of the 
final stellar masses of the $z>3$ SMGs by assuming that $10\%$ of their putative gas mass content is converted into stars by the end of the ongoing starburst event, while the magenta dashed histogram represents the final stellar mass distribution where $M_{\rm gas}$ is based on the mass-metallicity relationship (see text for details). The CO-based gas mass estimate was adopted for AzTEC1 and AzTEC3 (Appendix~D).}
\label{figure:masshist}
\end{figure}

Our parent sample contains 12 SMGs that lie at $z>3$ (six of which were analysed in the present work), i.e. at the formation epoch of $z\sim2$ cQGs (\cite{toft2014}). As shown in Sect.~3.4, a comparison with the stellar mass-size relationship of $z\sim2$ cQGs supports an evolutionary link between these high stellar-density systems and our $z>3$ SMGs, in accordance with the results of Toft et al. (2014). To further quantify this evolutionary path, we follow the approach of Toft et al. (2014), and compare the stellar mass distributions of our $z>3$ SMGs and $z\sim2$ cQGs. In Fig.~\ref{figure:masshist}, we plot 
the mass distribution of our $z>3$ SMGs, and the stellar masses of the spectroscopically confirmed $z\sim2$ cQGs from Krogager et al. (2014; their Table~2). To estimate the final stellar masses of 
the $z>3$ SMGs, we adopt the CO-based molecular gas masses available for AzTEC1 and AzTEC3 (see Appendix~D), and assume a value of $M_{\rm gas}\sim10^{11}$~M$_{\sun}$ -- the average of the gas masses of 
AzTEC1 and 3 -- for the remaining four sources. Following Toft et al. (2014), we assumed that $10\%$ of the gas mass is converted into stars by the end of the starburst (this star formation efficiency is based on the hydrodynamic simulations by Hayward et al. (2011) as described in \cite{toft2014}). We note that the time interval between $z>3$ and $z\sim2$ is $\gtrsim1.1$~Gyr, which is much longer than the gas depletion timescale in SMGs; for comparison, the depletion timescale we derived for AzTEC1 and AzTEC3 is only $86^{+15}_{-14}$~Myr and $19^{+1}_{-0}$~Myr, respectively (Appendix~D). 

The derived final stellar mass distribution is shown by the blue histogram in Fig.~\ref{figure:masshist}. A two-sided K-S test between this mass distribution and the Krogager et al. (2014) cQG $M_{\star}$ distribution yielded a K-S test statistic of $D= 0.36$ and a $p$-value of $p= 0.55$, suggesting that the two distributions might share a common parent distribution, which would further support the evolutionary link between $z>3$ SMGs and $z\sim2$ cQGs. However, given the small sample sizes, this comparison should be considered with caution in mind.

As another gas mass estimate (for the sources other than AzTEC1 and 3), we used the method adopted by Toft et al. (2014), namely the gas-to-dust mass ratio-metallicity ($\delta_{\rm gdr}-Z$) relationship derived by Leroy et al. (2011) in combination with the stellar mass-metallicity ($M_{\star}-Z$) relationship from Genzel et al. (2012 and references therein). Using the {\tt MAGPHYS}-based $M_{\star}$ and $M_{\rm dust}$ values, we could then estimate the gas masses, and hence the final stellar masses under the assumption of a $10\%$ star formation efficiency. The resulting distribution is shown by the magenta dashed histogram in Fig.~\ref{figure:masshist}. The median values of the two estimated final stellar mass distributions (the blue and magenta dashed histograms) are the same ($\log(M_{\star}/{\rm M}_{\sun})=10.97$), and a K-S test performed between the final stellar mass distribution shown by the magenta dashed line and the $z\sim2$ cQG $M_{\star}$ distribution yielded the same results as above ($D= 0.36$ and $p= 0.55$). We emphasise, though, that larger sample sizes are required to better test the validity of the aforementioned stellar mass-metallicity relationship, because it is based on the analysis of $z\sim2$ galaxies, and hence its applicability for $z>3$ SMGs is questionable (see also \cite{toft2014}). Indeed, the gas-to-dust mass ratios derived through the empirical relationships for AzTEC1 and 3 are 123 and 119, respectively, which are about 1.4 and 3.6 times higher than the observed values (Appendix~D). 

At this point, we conclude that on the basis of the mass-size relationship and the stellar mass distribution analyses presented here, the evolutionary connection between $z>3$ SMGs and $z\sim2$ cQGs seems possible, as proposed by Toft et al. (2014). The subsequent growth of $z\sim2$ cQGs through minor dry (gas-poor) mergers can then turn them into the giant ellipticals seen in the present-day universe (\cite{toft2014}). 

\subsubsection{$z\leq3$ SMGs}

While the above discussion concerns the $z>3$ SMG population, most of the SMGs in our parent sample lie at $z\leq3$ (18 sources with either a spectroscopic or photometric redshift, plus seven lower redshift limits at $z<3$; 10 sources at $z\simeq1.6-3$ were analysed here). Because there are also cQGs at $z<2$ (see e.g. the sample of $0.9<z<1.6$ cQGs in \cite{belli2014}), one could think of a scenario where $z\lesssim3$ SMGs represent their progenitors. However, our $z\leq3$ SMGs already have a considerable stellar mass content of $M_{\star}\sim 1.8\times10^{10}-4.9\times10^{11}$~M$_{\sun}$ with a median of $ 2.3\times10^{11}$~M$_{\sun}$, while for example the Chabrier (2003) IMF-based stellar masses of the $0.9<z<1.6$ cQGs from Belli et al. (2014) are $M_{\star}\sim1.9\times10^{10}-2.2\times10^{11}$~M$_{\sun}$ with a median of $7\times10^{10}$~M$_{\sun}$. 

Since the stellar masses of our $z\leq3$ SMGs are already higher on average than those of the Belli et al. (2014) cQGs, an evolutionary link between these two populations seems very unlikely, although we note that our SMG subsample here consists of only 10 sources, which prevents us from making any firm conclusions. On the other hand, it has been shown that also lower-redshift ($z\leq3$) SMGs have the potential to evolve into the present-day passive ellipticals (\cite{swinbank2006}; \cite{simpson2014}). For instance, based on the $H$-band magnitude and space density comparisons, and the stellar age analysis of local ellipticals, Simpson et al. (2014) showed that the properties of their sample of 77 ALESS SMGs with a median redshift of $z_{\rm phot}=2.3$ and local ellipticals agree well with each other, which strongly supports an evolutionary connection between them. Based on the stellar population ages of quiescent, red $z\sim1.5-2$ galaxies (e.g. \cite{whitaker2013}), Simpson et al. (2014) concluded that those stellar populations were formed at a cosmic epoch consistent with the median redshift of their SMGs, and hence the passive $z\sim1.5-2$ galaxies could descend from the $z\lesssim3$ SMGs, and ultimately turn into today's ellipticals. It is unclear, however, how our $z\leq3$ SMGs fit into this scenario because their median stellar mass is about three times higher than those of the $z\simeq1.6-2.8$ ALESS SMGs from da Cunha et al. (2015). Perhaps they have the potential to evolve into very massive red galaxies at $z\sim1-2$, and then undergo mergers and grow to the most massive ellipticals seen today.

Overall, our study of the COSMOS/AzTEC SMGs supports a scenario where the present-day massive ellipticals formed the bulk of their stars in a vigorous starburst in the early universe. This in concert with the finding that many of our SMGs populate galaxy overdensities or even protoclusters (\cite{capak2011}; Paper~III), suggest we are witnessing the formation process of the massive, red-and-dead ellipticals that are found to reside in rich galaxy clusters today.

\section{Summary and conclusions}

We have studied the physical properties of a flux-limited sample of SMGs in the COSMOS field. The target SMGs were originally 
uncovered in a 1.1~mm continuum survey carried out with the AzTEC bolometer, and followed up with higher-resolution interferometric 
(sub)millimetre continuum observations. Our main results are summarised as follows:

\begin{enumerate}
\item We have used the new version of the {\tt MAGPHYS} code of da Cunha et al. (2008, 2015) to interpret the observed panchromatic SEDs of a sample of 16 of our SMGs, which lie at redshifts of $z \simeq 1.6-5.3$. Based on this analysis, we derived the following median values and 16th and 84th percentiles for the stellar mass, total ($8-1\,000~\mu$m) IR luminosity, SFR, sSFR, dust temperature, and dust mass: $\log(M_{\star}/{\rm M}_{\sun})= 10.96^{+0.34}_{-0.19}$, $\log(L_{\rm IR}/{\rm L}_{\sun})= 12.93^{+0.09}_{-0.19}$, ${\rm SFR}= 856^{+191}_{-310}$~${\rm M}_{\sun}~{\rm yr}^{-1}$, ${\rm sSFR}=9.9^{+21.4}_{-8.1}$~${\rm Gyr}^{-1}$, $T_{\rm dust}=40.6^{+7.5}_{-8.1}$~K, and $\log(M_{\rm dust}/{\rm M}_{\sun})=9.17^{+0.03}_{-0.33}$, respectively. Our stellar masses and dust temperatures are similar to those of the 870~$\mu$m-selected ALESS SMGs that are equally bright to our AzTEC SMGs (\cite{dacunha2015}), while the other parameter values for our SMGs are $\sim1.5-2$ times higher on average. However, given the spread of the values (the 16th--84th percentile ranges), the latter discrepancies are not statistically signifcant. Nevertheless, part of this discrepancy is potentially caused by the different observed wavelengths of the two SMG samples (i.e. $\lambda_{\rm obs}=1.1$~mm versus $\lambda_{\rm obs}=870$~$\mu$m).
\item When compared with the galaxy main-sequence as defined by Speagle et al. (2014), our SMGs lie by a factor of $ 0.3^{+0.1}_{-0.2}$ to $13.0^{+0.0}_{-0.6}$ above the main sequence. The median ${\rm SFR}/{\rm SFR}_{\rm MS}$ ratio is found to be $4.6^{+4.0}_{-3.5}$, and most ($63\%$) of the target SMGs can be considered starbursts.
\item The 3~GHz radio sizes of our SMGs measured in Paper~II were used in conjunction with the present stellar mass values to examine their possible 
stellar mass-size relationship. In particular, the radio sizes of the $z>3$ SMGs appear fairly consistent with the $z\sim2$ cQGs' mass--size relation, 
supporting a scenario where the high-redshift SMGs experience a cQG phase at $z\sim2$ (\cite{toft2014}).
\item We used 325~MHz GMRT data in conjuction with the 1.4~GHz and 3~GHz VLA data to investigate the radio SED properties of our SMGs. A radio SED could be constructed for 19 SMGs in total, 
where each source was detected at least in one of the aforementioned radio bands. The median radio spectral index measured between the observed frame 325~MHz and 3~GHz was found to be $ -0.77^{+0.28}_{-0.42}$, which is consistent with the canonical radio synchrotron value ($\alpha \simeq -0.8 \ldots-0.7 $).
\item We found evidence of both spectral flattening and steepening in the radio SEDs of our SMGs, which are indicative of different cosmic-ray energy gain and loss mechanisms taking place 
in these sources. In the case of the $z_{\rm phot}\simeq 1.8$ SMG AzTEC4, which shows the flattest radio spectral index derived here ($\alpha > -0.18$), the flattening could be due to a non-negligible contribution of thermal free-free emission, which is estimated to be at least one-third of the observed-frame 3~GHz flux density for this source.
\item We found an indication for an anti-correlation between distance from the galaxy main sequence and 3~GHz radio size. 
This suggests that starburst SMGs are more compact, and hence more susceptible to free-free absorption.
\item The present data allowed us to study the IR-radio correlation among our SMGs. This was quantified by calculating the total-IR $q$ parameter, and its median value was determined to be $q=2.27^{+0.27}_{-0.13}$. This is in very good agreement with the flux-limited (i.e. equally bright) ALESS SMGs' median $q$ parameter of $2.29\pm0.09$ (recalculated from \cite{thomson2014}). We found a statistically insignificant hint of negative correlation between the $q$ parameter and the source redshift in the binned average data. The weakness of this $q$ evolution might be partially caused by the selection effect of our SMGs being essentially IR-selected, but not necessarily detected at the rest-frame 1.4~GHz used to calculate $q$. This can bias the $q$ parameter towards higher values, and flatten the $q(z)$ distribution. Some of the observed low and high $q$ values compared to the local median value could be linked to the source evolutionary phase.
\item The two high redshift SMGs in our sample that benefit from previous CO line data, AzTEC1 and AzTEC3, were analysed in further details (see Appendix~D). These SMGs are found to form 
stars near or at the Eddington limit, where the process of star formation can exhaust the gas reservoir in only $86^{+15}_{-14}$ and $19^{+1}_{-0}$~Myr, respectively. The current stellar and gas mass estimates for AzTEC1 and 3 suggest they can both evolve to have a final stellar mass of $>10^{11}$~${\rm M}_{\sun}$. Overall, the results of our study support the general view of high-$z$ ($z \gtrsim3$) SMGs being the progenitors of the present-day giant, gas-poor ellipticals, mostly sitting in the cores of galaxy clusters. 

\end{enumerate}

\begin{acknowledgements}

We would like to thank the anonymous referee for providing us with comments and 
suggestions. This research was funded by the European Union's Seventh 
Framework programme under grant agreement 337595 (ERC Starting Grant, 'CoSMass'). This work
was completed at the Aspen Center for Physics, which is supported by National Science Foundation grant 
PHY-1066293. This work was partially supported by a grant from the Simons Foundation. 
M.~A. acknowledges partial support from FONDE-CYT through grant 1140099. 
A.~K. acknowledges support by the Collaborative Research Council 956, sub-project A1, funded by the Deutsche Forschungsgemeinschaft (DFG). 
This paper makes use of the following ALMA data: ADS/JAO.ALMA\#2012.1.00978.S and ADS/JAO.ALMA\#2013.1.00118.S. 
ALMA is a partnership of ESO (representing its member states), NSF (USA) and NINS (Japan), together with NRC (Canada), 
NSC and ASIAA (Taiwan), and KASI (Republic of Korea), in cooperation with the Republic of Chile. 
The Joint ALMA Observatory is operated by ESO, AUI/NRAO and NAOJ. This paper is also partly based
on data products from observations made with ESO Telescopes at the La Silla Paranal Observatory under 
ESO programme ID 179.A-2005 and on data products produced by TERAPIX and the Cambridge Astronomy Survey Unit 
on behalf of the UltraVISTA consortium. This research has made use of NASA's Astrophysics Data System, 
and the NASA/IPAC Infrared Science Archive, which is operated by the JPL, California Institute of Technology, 
under contract with the NASA. We greatfully acknowledge the contributions of the entire COSMOS collaboration 
consisting of more than 100 scientists. More information on the COSMOS survey is available at 
{\tt http://cosmos.astro.caltech.edu}. 

PACS has been developed by a consortium of institutes led by MPE (Germany) and including UVIE (Austria); 
KU Leuven, CSL, IMEC (Belgium); CEA, LAM (France); MPIA (Germany); INAF-IFSI/OAA/OAP/OAT, LENS, SISSA (Italy); 
IAC (Spain). This development has been supported by the funding agencies BMVIT (Austria), ESA-PRODEX (Belgium), 
CEA/CNES (France), DLR (Germany), ASI/INAF (Italy), and CICYT/MCYT (Spain). 

SPIRE has been developed by a consortium of institutes led by Cardiff University (UK) and including Univ. 
Lethbridge (Canada); NAOC (China); CEA, LAM (France); IFSI, Univ. Padua (Italy); IAC (Spain); 
Stockholm Observatory (Sweden); Imperial College London, RAL, UCL-MSSL, UKATC, Univ. Sussex (UK); 
and Caltech, JPL, NHSC, Univ. Colorado (USA). This development has been supported by national funding agencies: CSA (Canada); NAOC (China); CEA, CNES, CNRS (France); 
ASI (Italy); MCINN (Spain); SNSB (Sweden); STFC, UKSA (UK); and NASA (USA).

\end{acknowledgements}

\appendix

\section{Photometric tables}

A selection of mid-IR to mm flux densities of our SMGs are listed in Table~\ref{table:fluxes}. The GMRT (325~MHz) and VLA (1.4 and 3~GHz) radio flux densities are tabulated in 
Table~\ref{table:radio}.

\begin{landscape}
\begin{table}
\caption{Flux densities at mid-IR to mm wavelengths in mJy.}
{\small
\centering
\renewcommand{\footnoterule}{}
\label{table:fluxes}
\begin{tabular}{c c c c c c c c c c c c}
\hline\hline 
Source ID & $S_{\rm 24\, \mu m}$ & $S_{\rm 100\, \mu m}$ & $S_{\rm 160\, \mu m}$ & $S_{\rm 250\, \mu m}$ & $S_{\rm 350\, \mu m}$ & $S_{\rm 450\, \mu m}$ & $S_{\rm 500\, \mu m}$ & $S_{\rm 850\, \mu m}$ & $S_{\rm 890\, \mu m}$ & $S_{\rm 1.1\, mm}$ & $S_{\rm 1.3\, mm}$ \\
          & MIPS & PACS & PACS & SPIRE & SPIRE & SCUBA-2\tablefootmark{a} & SPIRE & SCUBA-2\tablefootmark{a} & SMA\tablefootmark{b} & AzTEC\tablefootmark{c} & ALMA/PdBI\tablefootmark{d} \\
\hline
AzTEC1\tablefootmark{e} & $<0.054$ & $<5.0$ & $<10.2$ & $17.54\pm1.60$ & $30.57\pm1.62$ & \ldots & $27.72\pm1.90$ & \ldots & $15.6\pm1.1$ & $9.3\pm1.3$ & $4.41\pm0.11$(A) \\
AzTEC2 & $0.29\pm0.02$ & $<5.0$ & $<10.2$ & $23.60\pm1.97$ & $31.82\pm2.99$ & $20.46\pm4.78$ & $35.68\pm3.58$ & $10.86\pm0.54$ & $12.4\pm1.0$ & $8.3\pm1.3$ & $4.64\pm0.11$(A) \\
AzTEC3\tablefootmark{f} & $<0.054$ & $<5.0$ & $<10.2$ & $16.50\pm1.64$ & $19.29\pm1.66$ & \ldots & $11.64\pm1.89$ & \ldots & $8.7\pm1.5$ & $5.9\pm1.3$ & \ldots \\
AzTEC4 & $<0.054$ & $<5.0$ & $<10.2$ & $30.57\pm1.73$ & $36.02\pm1.58$ & \ldots & $34.78\pm1.87$ & \ldots & $14.4\pm1.9$ & $5.2^{+1.3}_{-1.4}$ & $3.88\pm0.10$(A)\\
AzTEC5 & $0.19\pm0.01$ & $<5.0$ & $<10.2$ & $33.54\pm1.97$ & $36.31\pm2.94$ & $25.35\pm6.04$ & $24.74\pm4.92$ & $10.54\pm0.90$ & $9.3\pm1.3$ & $6.5^{+1.2}_{-1.4}$ & $2.40\pm0.10$(A) \\
AzTEC6 & $<0.054$ & $<5.0$ & $<10.2$ & $<8.1$ & $<10.7$ & \ldots & $<15.4$ & \ldots & $8.6\pm1.3$ & $6.3^{+1.3}_{-1.2}$ & $3.68\pm0.10$(A) \\
AzTEC7 & $0.55\pm0.05$ & $22.50\pm2.73$ & $71.90\pm4.51$ & $77.63\pm1.97$ & $50.11\pm3.62$ & \ldots & $<15.4$ & \ldots & $12.0\pm1.5$ & $7.1\pm1.4$ & \ldots\\
AzTEC8 & $<0.054$ & $9.89\pm1.60$ & $35.56\pm 3.57$ & $62.32\pm1.65$ & $71.89\pm 1.59$ & \ldots & $68.61\pm1.88$ & \ldots & $21.6\pm2.3$ & $5.5\pm1.3$ & $4.10\pm0.15$(A) \\
AzTEC9 & $<0.054$ & $<5.0$ & $<10.2$ & $<8.1$ & $27.27\pm1.57$ & \ldots & $25.75\pm1.95$ & $11.49\pm1.10$ & $7.4\pm3.0$ & $5.8^{+1.3}_{-1.5}$ & $5.02\pm0.10$(A) \\
AzTEC10 & $<0.054$ & $<5.0$ & $<10.2$ & $8.56\pm1.97$ & $14.07\pm2.63$ & \ldots & $19.86\pm3.90$ & \ldots & $4.7\pm1.7$ & $4.7\pm1.3$ & \ldots \\
AzTEC11-N & $<0.054$ & $<5.0$ & $<10.2$ & $<8.1$ & $<10.7$ & \ldots & $<15.4$ & \ldots & $10.0\pm2.1$ & $3.3\pm0.9$ & $0.81\pm0.11$(A)  \\
AzTEC11-S & $0.66\pm0.02$ & $10.50\pm1.71$ & $45.11\pm3.95$ & $74.29\pm1.97$ & $70.12\pm2.44$ & \ldots & $45.46\pm3.67$ & \ldots & $4.4\pm2.1$ & $1.4\pm0.4$ & $2.23\pm0.10$(A)   \\
AzTEC12 & $0.36\pm0.02$ & $<5.0$ & $25.02\pm3.09$ & $51.10\pm1.97$ & $55.17\pm2.49$ & \ldots & $30.36\pm3.78$ & \ldots & $12.8\pm2.9$ & $4.5^{+1.3}_{-1.5}$ & $3.70\pm0.10$(A)    \\
AzTEC15 & $0.09\pm0.02$ & $<5.0$ & $<10.2$ & $11.93\pm1.97$ & $18.87\pm2.56$ &\ldots & $14.36\pm3.91$ & \ldots & $5.8\pm1.7$ & $4.2^{+1.3}_{-1.4}$ & $1.73\pm0.15$(A)  \\
AzTEC17a & $<0.054$ & $<5.0$ & $<10.2$ & $29.21\pm2.71$ & $28.33\pm3.33$ & \ldots & $20.85\pm5.18$ & \ldots & \ldots & $2.0\pm0.7$ & $1.58\pm0.43$(P) \\
AzTEC19a & $0.20\pm0.01$ & $<5.0$ & $<10.2$ & $33.30\pm1.97$ & $37.95\pm2.45$ & $37.54\pm6.58$ & $29.15\pm3.71$ & $9.21\pm1.45$ & \ldots & $3.8^{+1.3}_{-1.6}$ & $3.98\pm0.91$(P) \\
AzTEC19b & $<0.054$ & $<5.0$ & $<10.2$ & $<8.1$ & $<10.7$ & \ldots & $<15.4$ & \ldots & \ldots & \ldots & $5.21\pm1.30$(P) \\
AzTEC21a & $0.56\pm0.02$ & $<5.0$ & $<10.2$ & $18.49\pm1.97$ & $15.41\pm2.54$ & \ldots & $<15.4$ & \ldots & \ldots & $1.9^{+0.7}_{-0.8}$ & $3.37\pm1.03$(P)  \\
AzTEC21b & $<0.054$ & $<5.0$ & $<10.2$ & $<8.1$ & $<10.7$ & \ldots & $<15.4$ & \ldots & \ldots & $0.8\pm0.3$ & $1.34\pm0.38$(P) \\
AzTEC24b & $0.15\pm0.02$ & $<5.0$ & $<10.2$ & $18.25\pm1.97$ & $19.75\pm2.46$ & \ldots & $19.14\pm3.76$ & \ldots & \ldots & $4.9^{+1.1}_{-1.2}$ & $1.39\pm0.10$(A)\\  
AzTEC26a & $<0.054$ & $<5.0$ & $<10.2$ & $<8.1$ & $<10.7$ & \ldots & $<15.4$ & \ldots & \ldots & $1.7^{+0.7}_{-0.8}$ & $0.98\pm0.28$(P) \\
\hline
\end{tabular} }
\tablefoot{The second row of the table gives the name of the instrument used to measure the flux density. The quoted \textit{Herschel} flux density uncertainties refer to the total error, i.e. instrumental plus confusion noise. We place a $3\sigma$ flux density upper limit for the non-detections. The \textit{Herschel} flux density upper limits include the confusion noise.\tablefoottext{a}{From \cite{casey2013}.}\tablefoottext{b}{From \cite{younger2007}, 2009.}\tablefoottext{c}{From \cite{scott2008}, except for AzTEC24b the AzTEC 1.1~mm flux density is from \cite{aretxaga2011} (see Appendix~B.1). For sources resolved into multiple components at $\sim2\arcsec$ resolution (AzTEC11, 17, 21, 26), a component's $S_{\rm 1.1\, mm}$ value was estimated using the relative flux densities measured interferometrically. The single-dish 1.1~mm flux density of AzTCE19 can be fully assigned to the component AzTEC19a because the secondary component AzTEC19b lies $10\farcs6$ away from the phase centre, and could be spurious (see Paper~I and Appendix~B.2 herein).}\tablefoottext{d}{The 1.3~mm flux density given in the last column was determined with either ALMA (M.~Aravena et al., in prep.) or PdBI (Paper~I); these are marked with ``(A)'' or ``(P)'', respectively.}\tablefoottext{e}{For AzTEC1 we also have a 870~$\mu$m flux density of $S_{\rm 870\, \mu m}=14.12\pm0.25$~mJy measured with ALMA (Cycle 2 project, PI: A.~Karim), and a PdBI 3~mm flux density of $S_{\rm 3\, mm}=0.30\pm0.04$~mJy (\cite{smolcic2011}).}\tablefoottext{f}{For AzTEC3 we additionally have a 1~mm flux density of $S_{\rm 1\, mm}=6.20\pm0.25$~mJy from the ALMA observations by Riechers et al. (2014).}}
\end{table}
\end{landscape}

\begin{table}
\renewcommand{\footnoterule}{}
\caption{Flux densities of our SMGs at the observed radio frequencies of 325~MHz, 1.4~GHz, and 3~GHz. Apart from AzTEC21b, only sources that are detected in at least one of these three frequency bands are listed.}
{\small
\begin{minipage}{1\columnwidth}
\centering
\label{table:radio}
\begin{tabular}{c c c c}
\hline\hline 
Source ID & $S_{\rm 325\, MHz}$\tablefootmark{a} & $S_{\rm 1.4\, GHz}$\tablefootmark{b} & $S_{\rm 3\, GHz}$\tablefootmark{c} \\
          & [$\mu$Jy] & [$\mu$Jy] & [$\mu$Jy] \\
\hline
AzTEC1 & $248.6\pm77.9$ & $48\pm12$ & $28.3\pm5.2$ \\
AzTEC2 & $<224.7$ & $76\pm14$ & $18.9\pm4.7$ \\
AzTEC3 & $276.5\pm79.8$ & $<30$ & $19.6\pm2.3$ \\
AzTEC4 & $<215.7$ & $<36$ & $31.5\pm7.7$ \\
AzTEC5 & $438.9\pm61.5$ & $126\pm15$ & $85.8\pm5.8$ \\
AzTEC6 & $<210.9$ & $<38$ & $22.4\pm5.8$ \\
AzTEC7 & $481.0\pm86.8$ & $132\pm22$ & $98.4\pm4.4$ \\
AzTEC8 & $531.1\pm78.0$ & $102\pm13$ & $49.4\pm4.8$ \\
AzTEC9 & $<213.9$ & $68\pm13$ & $33.3\pm4.3$ \\
AzTEC10 & $235.8\pm88.8$ & $<40$ & $<6.9$ \\
AzTEC11-N & $213.4\pm88.0$ & $138\pm26$ & $67.5\pm4.6$ \\
AzTEC11-S & $204.1\pm88.0$ & $132\pm26$ & $99.6\pm4.8$ \\
AzTEC12 & $281.5\pm78.7$ & $98\pm16$ & $52.5\pm5.2$ \\
AzTEC15 & $<224.1$ & $<32$ & $27.9\pm6.9$ \\
AzTEC17a & $<258.3$ & $68\pm13$ & $40.8\pm4.4$ \\
AzTEC19a & $<241.2$ & $78\pm12$ & $45.3\pm5.0$ \\
AzTEC21a & $258.1\pm88.1$ & $<44$ & $25.7\pm7.9$ \\
AzTEC21b & $<264.6$ & $<44$ & $<6.9$ \\
AzTEC24b & $229.4\pm89.7$ & $63\pm13$ & $37.6\pm4.7$ \\
AzTEC27 & $221.7\pm85.7$ & $<43$ & $12.6\pm4.8$ \\
\hline 
\end{tabular} 
\tablefoot{A $3\sigma$ upper limit is reported for non-detections.\tablefoottext{a}{The value of $S_{\rm 325\, MHz}$ refers to the peak surface brightness value of the source in the 325~MHz GMRT-COSMOS mosaic (angular resolution $10\farcs76 \times 9\farcs49$; A.~Karim et al., in prep.). AzTEC5, 8, and 11 each have two radio components at 1.4 and 3~GHz (see Paper~II), which are blended in the large GMRT beam FWHM. To estimate the $S_{\rm 325\, MHz}$ values of these components, we divided up the observed flux density in the same proportion as that seen in the 1.4~GHz data.}\tablefoottext{b}{The values of $S_{\rm 1.4\, GHz}$ were taken from the COSMOS VLA Deep Catalogue May 2010 (\cite{schinnerer2010}) except for AzTEC1, 8, and 11, for which we use the values from Paper~II (Table~4 therein).}\tablefoottext{c}{The 3~GHz flux densities are adopted from Paper~II (Table~2 therein).}  }
\end{minipage} 
}
\end{table}

\section{Revised redshifts, unsatisfying SED model fits, and revisit of the sample statistics}

\subsection{Revised redshifts}

AzTEC17a was previously thought to lie at a spectroscopic 
redshift of $z_{\rm spec}=0.834$ (see Paper~I, and references therein), and 
hence to be the lowest-redshift SMG in our sample. However, 
the source SED constructed assuming this redshift showed 
that the \textit{Herschel} flux density upper limits were inconsistent with 
the best-fit SED model, suggesting that the redshift was underestimated.

The different redshift estimates for AzTEC17a were discussed in Appendix~C of Paper~I. A spec-$z$ of $z_{\rm spec}=0.834$ for AzTEC17a is reported to be very secure in the COSMOS spec-$z$ catalogue (M.~Salvato et al., in prep.), but our initial SED fit suggested the SMG redshift to be higher. Here, we have adopted a redshift of $z_{\rm phot}=2.96^{+0.06}_{-0.06}$ reported in the COSMOS2015 catalogue (\cite{laigle2016}) for the source ID835094, which lies $1\farcs34$ away from the PdBI 1.3~mm peak position of AzTEC17a. We note that although lying outside the area covered by the slit used for the spectroscopic measurements towards AzTEC17a, there is a galaxy (ID836198) lying $1\farcs7$ NW of the PdBI 1.3~mm peak of AzTEC17a (see Fig.~E.1 in Paper~I). A likely $z_{\rm spec}$ value of $z_{\rm spec}=0.78792$ (quality flag 2) is reported for this galaxy in the COSMOS spec-$z$ catalogue (M.~Salvato et al., in prep.), while the COSMOS2015 catalogue (\cite{laigle2016}) gives a photo-$z$ value of $z_{\rm phot}=0.70^{+0.01}_{-0.01}$, in fairly good agreement with the corresponding spec-$z$. Hence, it seems possible that the spectroscopic redshift measurement towards AzTEC17a was contaminated by radiation from this foreground galaxy.

Moreover, the redshifts of AzTEC4, 5, AzTEC9, 12, 15, and 24b have been recomputed using the COSMOS2015 photometric catalogue (\cite{laigle2016}) as part of the redshift distribution analysis of the ALMA-detected ASTE/AzTEC SMGs (D.~Brisbin et al., in prep.). 

For AzTEC4, the nominal value of the updated redshift of $z_{\rm phot}=1.80^{+5.18}_{-0.61}$ is much lower than the previous estimate ($z_{\rm phot}=4.93^{+0.43}_{-1.11}$; \cite{smolcic2012}), but we note that the photometric redshift of this source suffers from very large uncertainties.

In the case of AzTEC5, the updated redshift of 
$z_{\rm phot}=3.70^{+0.73}_{-0.53}$ is consistent with the previous value within 
the uncertainties ($z_{\rm phot}=3.05^{+0.33}_{-0.28}$; \cite{smolcic2012}; 
see also Paper~I). 

For AzTEC9, we have derived a new photometric redshift of 
$z_{\rm phot}=4.60^{+0.43}_{-0.58}$. We note that a spec-$z$ of 
$z_{\rm spec}=1.357$ has been measured for AzTEC9, but it is based on 
a relatively weak spectrum (M.~Salvato et al., in prep.), and is therefore 
quite uncertain. Nevertheless, this spec-$z$ is comparable to a photo-$z$ of $z_{\rm phot}=1.07^{+0.11}_{-0.10}$ previously derived for AzTEC9 by Smol{\v c}i{\'c} et al. (2012). We also note that the COSMOS2015 catalogue (\cite{laigle2016}) gives a secondary photo-$z$ solution of $z_{\rm phot}=4.55$ for a source lying $0\farcs88$ from AzTEC9 (ID763214). Moreover, Koprowski et al. (2014) derived a redshift of $z_{\rm phot}=4.85^{+0.50}_{-0.15}$, which was based on a $K$-band counterpart lying $0\farcs67$ away from the SMA 890~$\mu$m position of AzTEC9. The latter two redshifts are comparable to our new value.

The updated photometric redshift we have obtained for AzTEC12, $z_{\rm phot}=2.90^{+0.31}_{-0.18}$, is slightly higher than the previous value of $z_{\rm phot}=2.54^{+0.13}_{-0.33}$ from Smol{\v c}i{\'c} et al. (2012).

For AzTEC15, the updated redshift of $z_{\rm phot}=2.80^{+2.45}_{-1.27}$ is somewhat lower than the previous estimate ($z_{\rm phot}=3.17^{+0.29}_{-0.37}$; 
\cite{smolcic2012}), but the large associated uncertainties make the two being consistent with each other.

Finally, for AzTEC24b we have computed a redshift of 
$z_{\rm phot}=1.90^{+0.12}_{-0.25}$ on the basis of a new counterpart 
identification enabled by our new ALMA data (cf.~Paper~I). Our 
PdBI 1.3~mm source AzTEC24b lies $2\farcs55$ NE of a 1.3 mm-detected ALMA 
SMG (M.~Aravena et al., in prep.). We can attribute this nominally large 
offset to the poorer quality of the PdBI data than the corresponding ALMA data. 

\subsection{SED fits of other sources} 

\begin{figure}[!h]
\centering
\resizebox{0.47\hsize}{!}{\includegraphics{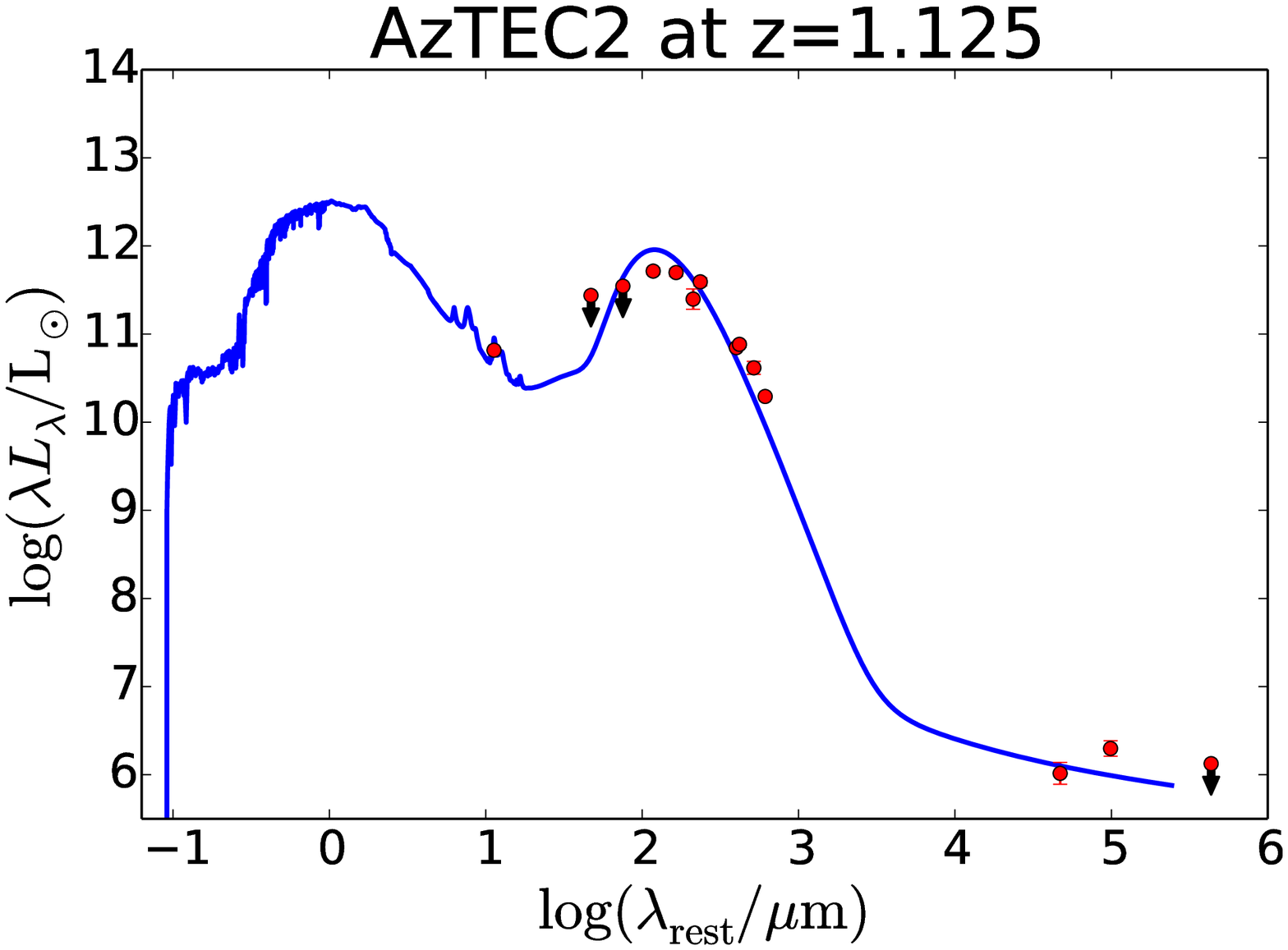}}
\resizebox{0.47\hsize}{!}{\includegraphics{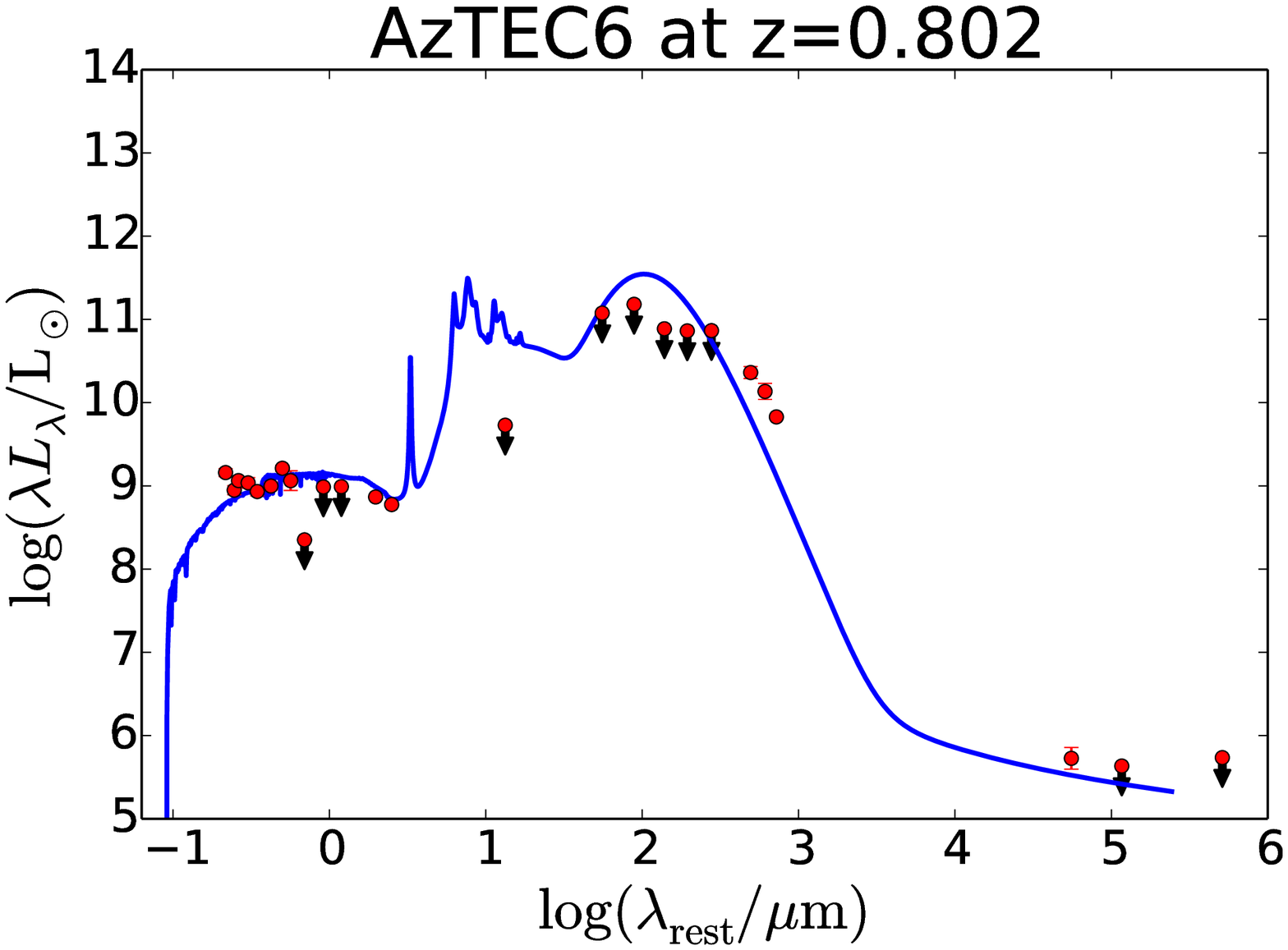}}
\resizebox{0.47\hsize}{!}{\includegraphics{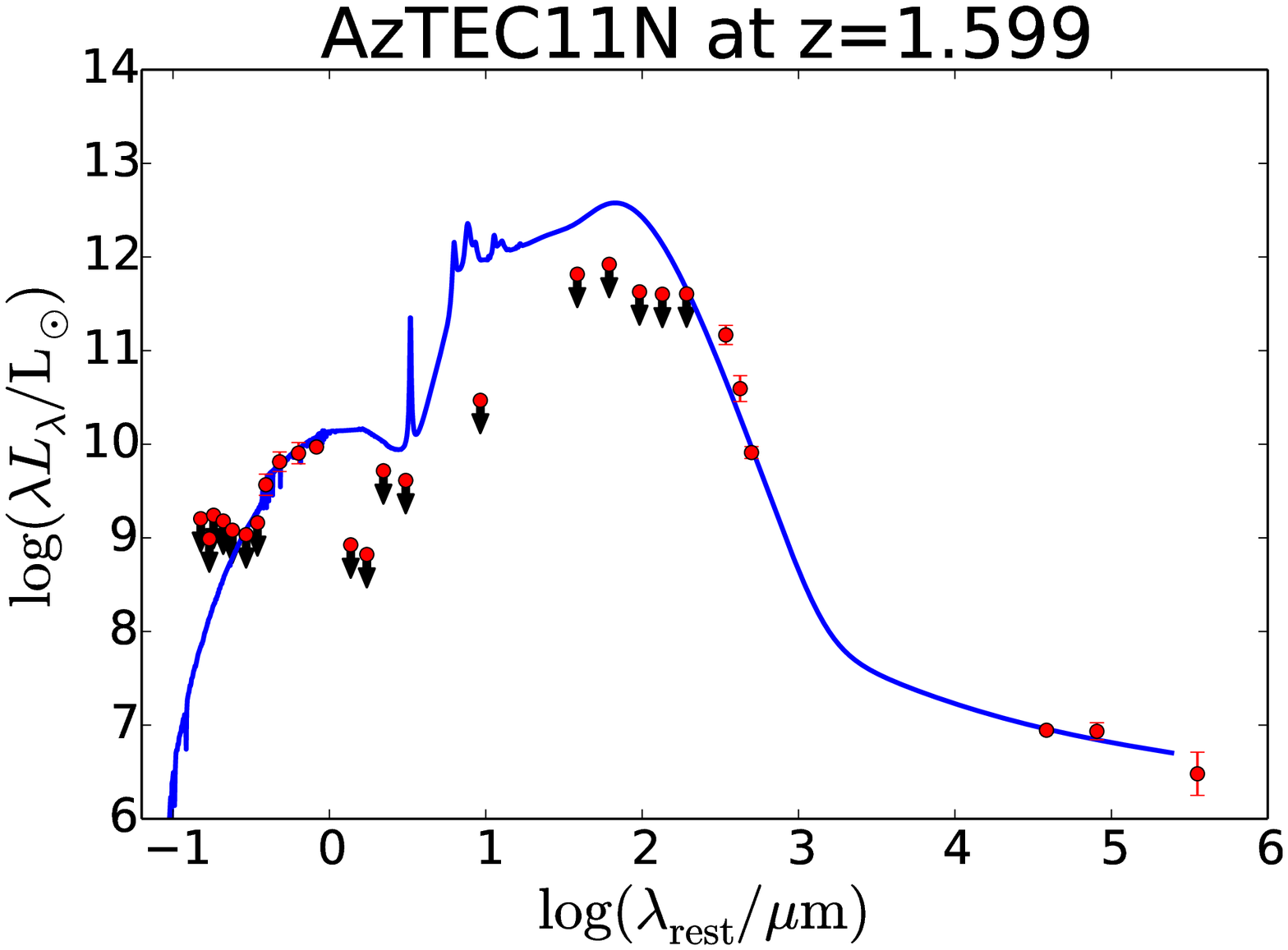}}
\resizebox{0.47\hsize}{!}{\includegraphics{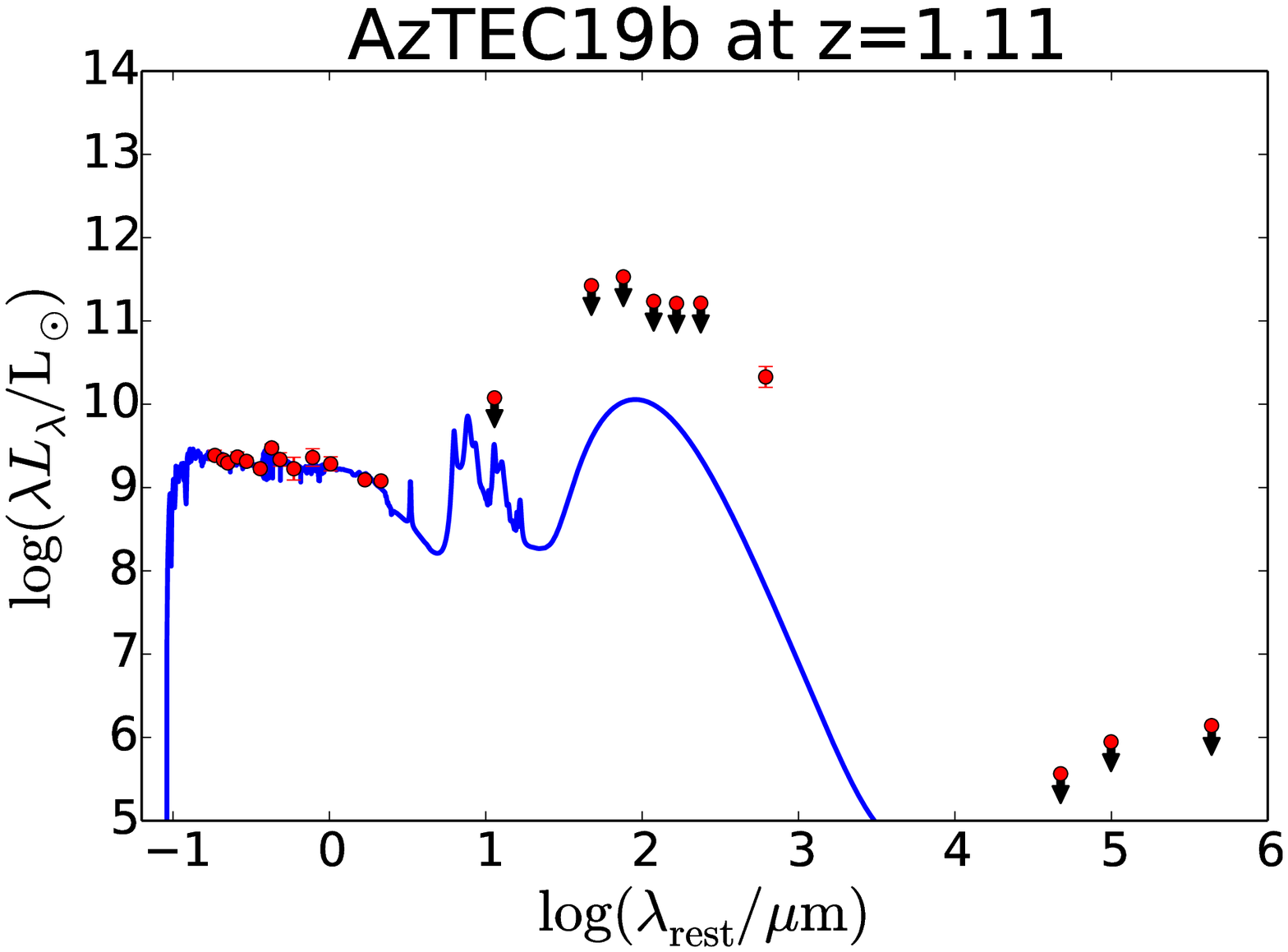}}
\resizebox{0.47\hsize}{!}{\includegraphics{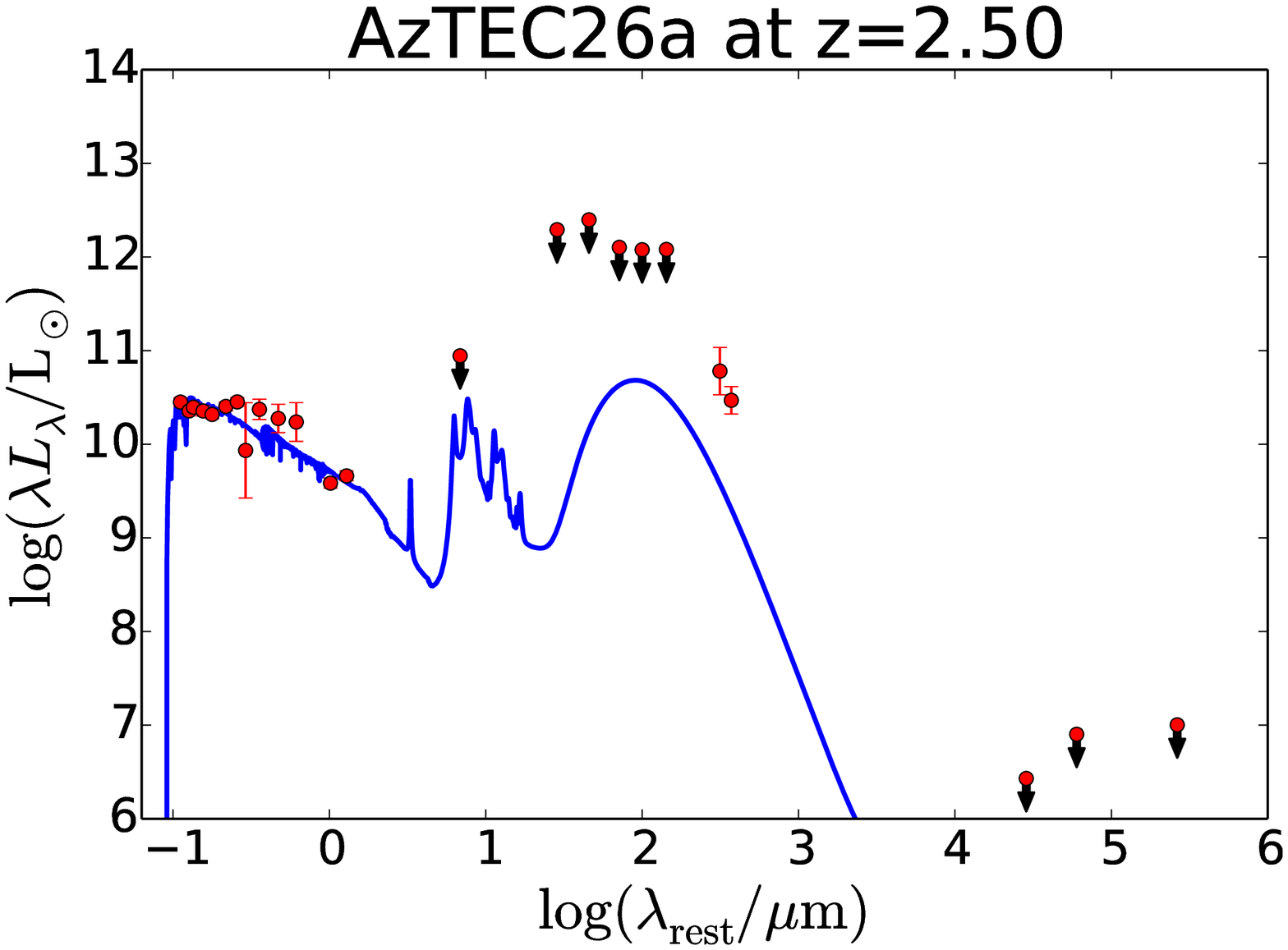}}
\caption{Best-fit panchromatic (UV--radio) rest-frame SEDs of five of our target SMGs. The source ID and redshift are shown on top of each panel. The red points with vertical error bars represent the observed photometric data, while those with downward pointing arrows mark the $3\sigma$ upper flux density limits. 
The blue line is the best-fit {\tt MAGPHYS} model SED from the high-$z$ library.}
\label{figure:otherseds}
\end{figure}

In Fig.~\ref{figure:otherseds}, we show the SED fits for AzTEC2, 6, 11-N, 19b, and 26a. In the case of AzTEC2, the lack of UV/optical to NIR data does not allow a meaningful {\tt MAGPHYS} fit, and hence the derived properties, like the stellar mass, could not be constrained. Moreover, AzTEC2 might lie at a higher redshift than $z_{\rm spec}=1.125$ because it is found to be 
the brightest 2~mm source in the IRAM/GISMO (Institut de Radioastronomie Millim\'etrique/Goddard-IRAM Superconducting 2~Millimetre Observer) 2~mm survey of the COSMOS field (A.~Karim et al., in prep.), and it is possibly affected by galaxy-galaxy lensing. As discussed in Paper~I (Appendix~B therein), the redshift of AzTEC6 is also uncertain. As shown in Fig.~\ref{figure:otherseds}, the assumption that the SMG lies at $z_{\rm spec}=0.802$ does not yield a satisfying SED fit, but it rather suggests a higher redshift (cf.~AzTEC17a). Indeed, a non-detection of AzTEC6 at 1.4~GHz suggests a redshift of $z>3.52$ (Paper~I). For AzTEC11-N, the optical to NIR photometry was manually extracted from the ALMA 1.3~mm peak position due to the nearby source AzTEC11-S. However, no reasonable SED fit could be obtained for this source. The derived SED fits for AzTEC19b and 26a are also poor. Given that AzTEC19b was identified at the border of the PdBI primary beam, and the counterpart of AzTEC26a lies $0\farcs94$ away from the PdBI emission peak (Paper~I), these sources are potentially spurious. 

\subsection{Revised SMG sample statistics} 

Given the aforementioned revised properties of our SMGs, and the fact that two more SMG candidates from our PdBI 1.3~mm survey (Paper~I), AzTEC24a and 24c, were not detected at 1.3~mm with ALMA (M.~Aravena et al., in prep.), and are hence very likely to be spurious, we are now in a position where we can revise the statistical properties of our sample. Of the 30 single-dish detected sources 
AzTEC1--30, only up to 5 appear to be resolved into multiple components at an angular resolution of about $2\arcsec$ (AzTEC29 being an uncertain case; Paper~I). This suggests a revised multiplicity fraction of $\sim17\% \pm7\%$ among our target SMGs. The revised sample mean and median redshifts are $z= 3.19\pm0.21$ and $3.00\pm0.26$, respectively, which are consistent with those reported in Paper~I ($3.19\pm0.22$ and $3.17\pm0.27$). The fraction of 3~GHz detected SMGs in our sample is revised to be $51\% \pm 12\%$ (cf.~Paper~II). Assuming the new redshift values for some of our SMGs, the median 3~GHz major axis FWHM is derived to be $4.2\pm 1.1$~kpc ($4.1\pm1.0$~kpc) for the cosmology adopted therein (in the present work), which is the same as reported in Paper~II ($4.2\pm 0.9$~kpc).


\section{Comparison with previous studies of the physical properties of the JCMT/AzTEC 1.1~mm-selected COSMOS SMGs}

The SEDs of many of our SMGs have been previously analysed to study their physical properties by Smol{\v c}i{\'c} et al. (2011; AzTEC1), 
Magnelli et al. (2012; AzTEC2, 3, 5, 11-N, and 17a), Casey et al. (2013; AzTEC2, 5, 9, 19a, 22, 28), Toft et al. (2014; 10 sources among AzTEC1--15), 
Huang et al. (2014; AzTEC1 and 3), Koprowski et al. (2014; 11 sources among AzTEC1--15), Smol{\v c}i{\'c} et al. (2015; AzTEC1 and 3), 
Yuan et al. (2015; AzTEC1), and Yun et al. (2015; AzTEC1). We note that only Smol{\v c}i{\'c} et al. (2015) used the {\tt MAGPHYS} code in their study, although with different priors as we have used (see below), and hence a direct comparison with the aforementioned studies is not feasible (Toft et al. (2014) also used {\tt MAGPHYS} to derive one of the physical parameters of their SMGs, namely the stellar mass). Moreover, Smol{\v c}i{\'c} et al. (2011, 2015), Toft et al. (2014), and Huang et al. (2014) used a redshift of $z=4.64$ for AzTEC1, 
but its revised spectroscopic redshift is $z_{\rm spec}=4.3415$ (\cite{yun2015}). In what follows, we discuss the aforementioned studies in more detail.

\underline{\textit{\cite{magnelli2012}}}. While we could not obtain a meaningful {\tt MAGPHYS} SED for AzTEC2 and AzTEC11-N, Magnelli et al. (2012) used a modified blackbody (MBB) fitting method to derive the dust properties of these sources. Out of the five common SMGs to our parent sample, Magnelli et al. (2012) used the same redshift as we for three sources. The two exceptions are AzTEC5 and AzTEC17a, for which they used $z=4.0$ and $z=0.8$, instead of the values $z_{\rm phot}=3.70^{+0.73}_{-0.53}$ and $z_{\rm phot}=2.96^{+0.06}_{-0.06}$ we have used. We also note that no interferometric (sub)mm observations of AzTEC17 existed at the time of the study by Magnelli et al. (2012). Based on their single-$T$ MBB fitting ($\beta=1.5$), Magnelli et al. (2012) derived the dust temperature values of $T_{\rm dust}=53\pm3$~K and $57\pm4$~K for AzTEC3 and 5, respectively. Within the uncertainties, the $T_{\rm dust}$ values for AzTEC3 and 5 are comparable to ours ($66.0^{+5.1}_{-8.4}$ and $48.7^{+3.3}_{-4.1}$~K, respectively) although the values were derived using different methods (and the redshift for AzTEC5 was slightly different). Our IR luminosities for these two SMGs are about 1.2 and 0.8 times the Magnelli et al. (2012) values, respectively. The dust masses we derived for AzTEC3 and 5 are 4.2 and 2.8 times higher than the corresponding Magnelli et al. (2012) values. One reason for these discrepancies is probably that Magnelli et al. (2012) did not use the SPIRE 500~$\mu$m flux density in their SED analysis. We note that for the above comparisons we have taken into account the fact that Magnelli et al. (2012) assumed a dust opacity value of $\kappa_{\rm 850 \, \mu m}=1.5$~cm$^2$~g$^{-1}$, which is 1.95 times larger than the 850~$\mu$m-normalised dust opacity of $\kappa_{\rm 850 \, \mu m}=0.77$~cm$^2$~g$^{-1}$ adopted in our {\tt MAGPHYS} analysis. 


\underline{\textit{\cite{casey2013}}}. Casey et al. (2013) used MBB SED fitting combined with a mid-IR powerlaw (see their Eq.~(3)). For AzTEC2 they used a photo-$z$ of $z_{\rm phot}=0.34^{+0.01}_{-0.01}$, which is lower than the spec-$z$ of $z_{\rm spec}=1.125$ (M.~Balokovi\'c et al., in prep.). As discussed in Paper~I, the aforementioned photo-$z$ refers to the optically visible foreground 
galaxy seen to the south of AzTEC2 (see e.g. Fig.~E.1 in Paper~I). For AzTEC9, Casey et al. (2013) adopted a redshift of $z_{\rm phot}=1.37^{+0.44}_{-0.25}$, i.e. $3.36^{+1.13}_{-1.14}$ times lower than the value of $z_{\rm phot}=4.60^{+0.43}_{-0.58}$ we have derived. For AzTEC19a, Casey et al. (2013) adopted a redshift of $z_{\rm phot}=2.86^{+0.21}_{-0.26}$, while in Paper~I, where we employed the first millimetre interferometric observations towards this source, we derived a photo-$z$ of $z_{\rm phot}=3.20^{+0.18}_{-0.45}$. The SCUBA-2 sources Casey et al. (2013) 
detected towards AzTC22 and AzTEC28, and for which they derived the SEDs, lie $7\farcs95$ and $7\farcs91$ away from our PdBI 1.3~mm detections, respectively (Paper~I). In Paper~I, 
we found that AzTEC22 and AzTEC28 have no optical, IR, or radio counterparts. Because of the aforementioned reasons, we do not compare the SED results here.

\underline{\textit{\cite{toft2014}}}. Toft et al. (2014) used $z_{\rm spec}=3.971$ as the redshift of AzTEC5. As discussed in Paper~I (Appendix~B therein), 
this spec-$z$ is based on a poor quality spectrum, but is still comparable to the revised photo-$z$ of $z_{\rm phot}=3.70^{+0.73}_{-0.53}$. For AzTEC11-S Toft et al. (2014) used a redshift of $z>2.58$, but it probably lies at $z_{\rm spec}=1.599$ (see Table~\ref{table:sample}). Although the SEDs of AzTEC13 and 14-E have not been analysed here, we note that the lower $z$ limits adopted by Toft et al. (2014) for AzTEC13 ($z>3.59$) and 14-E ($z>3.03$) are comparable to ours ($z>4.07$ and $z>2.95$, respectively). Here, we will only compare the SED results for AzTEC3, 5, 8, 10, and 15 where the redshifts used were either similar or close to ours. The present stellar masses derived for these SMGs are 0.3 to 1.6 (mean$=0.9$, median$=0.6$) times those reported by Toft et al. (2014), where the largest discrepancy is found for AzTEC15, where the discrepancy in the assumed redshift is the largest (factor of 1.13). The dust masses we have derived are lower by factors of 0.2 to 0.8 (mean$=0.5$, median$=0.4$) than those derived by Toft et al. (2014), who used the physically motivated dust model of Draine \& Li (2007) in their analysis. Our IR luminosities are 0.7--1.2 (mean$=0.9$, median$=0.9$) times the values obtained by Toft et al. (2014), and hence in fairly good agreement. This agreement also applies to the corresponding Kennicutt (1998) SFRs the authors reported.   

\underline{\textit{\cite{huang2014}}}. Based on MBB fitting, Huang et al. (2014) derived the values of $T_{\rm dust}=79\pm7$~K, $M_{\rm dust}=4.6^{+0.6}_{-0.4}\times10^9$~M$_{\sun}$, and $L_{\rm FIR}=1.5^{+1.1}_{-0.6}\times10^{13}$~L$_{\sun}$ for AzTEC3 (see their Table~4). The quoted dust mass was scaled up by a factor of 1.06 from the value reported by Huang et al. (2014) to take into account that they adopted a dust opacity of $\kappa_{\rm 250 \, \mu m}=5.1$~cm$^2$~g$^{-1}$, while our corresponding value would be 4.8~cm$^2$~g$^{-1}$ (assuming $\beta=1.5$). Our dust temperature of $66.0^{+5.1}_{-8.4}$~K for AzTEC3 is lower than the Huang et al. (2014) value by a factor of $1.2^{+0.3}_{-0.2}$, which is a reasonable agreement given that they are derived through different methods, and hence not directly comparable. The dust mass we derived is only $0.4^{+0.4}_{-0.2}$ times that obtained by Huang et al. (2014). The IR luminosity from Huang et al. (2014) refers to the FIR range (although the wavelength range was not specified), and our corresponding value of $L_{\rm FIR}= 0.34\times L_{\rm IR}=9.4^{+0.2}_{-0.2}\times10^{12}$~L$_{\sun}$ is $0.7^{+0.4}_{-0.3}$ times their value, and hence in fairly good agreement within the uncertainties. 
Huang et al. (2014) also derived the $q_{\rm FIR}$ parameter for AzTEC1 and 3 (but an old redshift value for AzTEC1 was used as mentioned above). They report the values of $q_{\rm FIR}=2.25\pm0.29$ and $q_{\rm FIR}>2.15$, respectively. Our corresponding values for AzTEC1 and 3 are $q_{\rm FIR}=2.10\pm0.17$ and $q_{\rm FIR}=1.67 \pm 0.17$, where the AzTEC1 value is comparable within the uncertainties despite the diffrent redshift used, but for AzTEC3 the value we have derived is inconsistent with the Huang et al. (2014) lower limit.

\underline{\textit{\cite{koprowski2014}}}. Koprowski et al. (2014) derived stellar masses for SMGs among AzTEC1--15 (besides the photo-$z$'s, which was the main purpose of their study; see their Table~5). They used the {\tt HYPERZ} code (\cite{bolzonella2000}) and assumed a Chabrier (2003) IMF. In all the other cases except AzTEC4, 5, 7, 10, 12 and 15 the photo-$z$'s derived by the authors are close enough to our adopted redshift values (within a factor of 1.03) that we can make a reasonable comparison between the stellar masses they obtained and those we have derived. The comparison of the stellar masses for AzTEC1, 3, 8, and 11-S showed that our values are 0.4 to 1.1 (mean$=0.8$, median$=0.8$) times the values reported by Koprowski et al. (2014), and the average agreement is fairly good. 

\underline{\textit{\cite{smolcic2015}}}. The most direct comparison we can make is with the {\tt MAGPHYS} SED of AzTEC3 from Smol{\v c}i{\'c} et al. (2015), but we note that as the prior model libraries they used those calibrated to reproduce the UV–IR SEDs of local ULIRGs (see \cite{dacunha2010}). Moreover, the authors included the UV/optical photometry in their SED fit (Fig.~4 therein), but as found in the present study, that emission can be attributed to an unrelated foreground galaxy (Appendix~D.2). The ratios of our new stellar mass, dust luminosity, and dust mass to the corresponding values from Smol{\v c}i{\'c} et al. (2015) are found to be 0.6, 2.7, and 0.3, respectively.  
They also used a MBB fit to derive a dust temperature of $43.5\pm11.5$~K for AzTEC3, which is lower than our {\tt MAGPHYS}-based value of 
$T_{\rm dust}=66.0^{+5.1}_{-8.4}$~K, and also significantly lower than the Huang et al. (2014) value. For comparison, Smol{\v c}i{\'c} et al. (2015) calculated the total IR luminosity and dust mass based on the best-fit Draine \& Li (2007) dust model, and obtained values that are $0.9\pm0.2$ and $1.5^{+2.1}_{-0.9}$ times those we have derived, respectively.
Smol{\v c}i{\'c} et al. (2015) also derived the values of $\alpha_{\rm 325\, MHz}^{\rm 1.4\, GHz}$, $\alpha_{\rm 1.4\, GHz}^{\rm 3\, GHz}$, and $q_{\rm TIR}$ for AzTEC1 and 3. For AzTEC1 the radio SED was fit using a broken power-law function with $\alpha_{\rm 325\, MHz}^{\rm 1.4\, GHz}=-1.24\pm0.28$ and $\alpha_{\rm 1.4\, GHz}^{\rm 3\, GHz}=-0.90\pm0.46$ (the plus and minus signs are reversed here to correspond to our definition of $\alpha$), which are fairly similar to the present single power-law spectral index of $-0.97\pm0.37$. For AzTEC3 the aforementioned indices were derived to be $\alpha <-1.54$ and $\alpha >-0.09$, both of which are inconsistent with a single power-law 325~MHz--3~GHz spectral index of $-1.19\pm0.32$ derived here on the basis of the deeper 3~GHz data. A $q_{\rm TIR}$ value of $2.24\pm0.18$ derived for AzTEC1 is very similar to the present value of $q_{\rm TIR}=2.19\pm0.17$ despite the differences in the analysis (e.g. a different redshift was used). Similarly to Huang et al. (2014), a lower limit of $q_{\rm TIR}>1.86$ was reported for AzTEC3. This is consistent with a value 
of $q_{\rm TIR}=2.14\pm0.17$ derived in the present work, but we note that the non-detection of AzTEC3 at 1.4~GHz means that the spectral index between 325~MHz and 1.4~GHz used in the $q$ calculation by Smol{\v c}i{\'c} et al. (2015) is only an upper limit (cf.~Fig.~\ref{figure:radio} herein). When this spectral index steepens, the rest-frame $L_{\rm 1.4\, GHz}$ increases, and hence the $q$ value should be an upper limit. 


\underline{\textit{\cite{yuan2015}}}. These authors included AzTEC1 in their study of \textit{Herschel}-detected high-$z$ galaxies. 
They derived a IR luminosity of $2.0\times10^{13}$~L$_{\sun}$ using Dale \& Helou (2002) template fitting, which is about 23\% higher than our value.

\underline{\textit{\cite{yun2015}}}. These authors revisited the SED of AzTEC1 analysed by Smol{\v c}i{\'c} et al. (2011), but utilising the correct redshift and adding the new \textit{Herschel}/SPIRE and SMA 345~GHz ($\lambda_{\rm obs}=870$~$\mu$m) photometric data. The observed SED was analysed using three different methods: \textit{i)} a MBB fitting; \textit{ii)} a starburst SED model from Efstathiou et al. (2000); and \textit{iii)} the population synthesis code GRASIL (GRAphite and SILicate; \cite{silva1998}). The total IR luminosity Yun et al. (2015) obtained from their MBB fit is in good agreement with ours; the ratio between these two values is 0.86. The best-fit SED model with a 45~Myr old starburst (\cite{efstathiou2000}) yielded a $L_{\rm IR}$ value of $1.5\times10^{13}$~L$_{\sun}$, which is even more similar to our value ($L^{\rm Yun}_{\rm IR}/L^{\rm our}_{\rm IR}=0.92$). For comparison, the mass-weighted (stellar
population) age of AzTEC1 given by our {\tt MAGPHYS} analysis, 34~Myr, is 1.3 times less than the aforementioned starburst age. Also, their best-fit GRASIL SED model provided a $L_{\rm IR}$ value, which is fully consistent with ours (the ratio between the two being $1.0\pm0.2$). The stellar mass that Yun et al. (2015) obtained from their GRASIL model -- when scaled down to a Chabrier (2003) IMF, $M_{\star}=(3.9\pm0.6)\times10^{11}$~M$_{\sun}$, is $5.1^{+1.0}_{-0.8}$ times higher than ours. 

\section{The nature of the high-$z$ SMGs AzTEC1 and AzTEC3}

In this section, we revisit and discuss some of the ISM and star formation characteristics of AzTEC1 and AzTEC3 in more detail. A summary of the physical properties discussed below 
is given in Table~\ref{table:ism}.

\subsection{AzTEC1 -- a compact starburst at $z=4.3$}

Besides spectroscopically confirming the redshift of AzTEC1 to be $z_{\rm spec}=4.3415$ (through their $\lambda_{\rm rest}=158$~$\mu$m $[\ion{C}{II}]$ observations with the SMA), 
Yun et al. (2015) derived a molecular gas mass of $M_{\rm gas}\approx M_{\rm H_2}=(1.4\pm0.2)\times10^{11}$~M$_{\sun}$ for this SMG using the CO$(4-3)$ data obtained with the Large Millimetre Telescope (LMT). Using the dust and stellar masses we have derived, we obtain a gas-to-dust mass ratio of $\delta_{\rm gdr}\equiv M_{\rm gas}/M_{\rm dust}=90^{+23}_{-19}$, 
and a gas fraction of $f_{\rm gas}=M_{\rm gas}/(M_{\rm gas}+M_{\star})\simeq65\%$. We note that Yun et al. (2015) assumed a value of 
$\alpha_{\rm CO}=0.8$~M$_{\sun}$~(K~km~s$^{-1}$~pc$^2$)$^{-1}$ for the conversion factor from CO line luminosity to molecular gas mass.
The $[\ion{C}{II}]$ fine-structure line luminosity of $L_{[\ion{C}{II}]}=(7.8\pm1.1)\times10^9$~L$_{\sun}$ reported by Yun et al. (2015) suggests 
a molecular gas mass of $M_{\rm gas}\sim5.4-10.7\times10^{10}$~M$_{\sun}$ (see \cite{swinbank2012} and references therein), the highest value being close to the aforementioned CO-based estimate. Yun et al. (2015) concluded that AzTEC1 is one of the most $[\ion{C}{II}]$ deficient objects known, and using our new IR luminosity value we derive a very low $L_{[\ion{C}{II}]}/L_{\rm IR}$ ratio of $0.0005^{+0.0001}_{-0.0001}$. 

Assuming that AzTEC1 has sustained its current SFR continuously, we can estimate its characteristic stellar mass building-up timescale to be 
$\tau_{\rm build}=M_{\rm star}/{\rm SFR}={\rm sSFR}^{-1}\simeq 47^{+2}_{-2}$~Myr (e.g. \cite{seymour2012}; \cite{tan2014}). This is comparable to the 45~Myr starburst age 
adopted by Yun et al. (2015) in their SED modelling, and a factor of $1.4^{+0}_{-0.1}$ higher than the {\tt MAGPHYS}-based mass-weighted age of 34~Myr. We can also use the gas mass and SFR to calculate the gas depletion timescale of $\tau_{\rm dep}=M_{\rm gas}/{\rm SFR}\simeq 86^{+15}_{-14}$~Myr. Hence, $\tau_{\rm build} \simeq 0.5^{+0.2}_{-0.1} \times \tau_{\rm dep}$. Yun et al. (2015) also concluded that $\tau_{\rm build}\sim \tau_{\rm dep}$ for AzTEC1, and they suggested that the stellar content of this SMG could have been fully built up during the current starburst phenomenon. The value of $\tau_{\rm dep}$ can be used to calculate the star formation efficiency as ${\rm SFE}={\rm SFR}/M_{\rm gas}=\tau_{\rm dep}^{-1}$. We obtain a value of ${\rm SFE}=12^{+2}_{-2}$~Gyr$^{-1}$. The Eddington-limited SFE is believed to be ${\rm SFE}=L_{\rm IR}/M_{\rm gas}=500$~L$_{\sun}$~M$_{\sun}^{-1}$, above which the radiation pressure from high-mass stars expels the gas out of the system (\cite{scoville2003}, 2004). A similar Eddington limit can also be set by cosmic-ray pressure feedback (\cite{socrates2008}). For AzTEC1 we derive a value of ${\rm SFE}=116^{+22}_{-17}$~L$_{\sun}$~M$_{\sun}^{-1}$, which is a factor of $4.3^{+0.8}_{-0.7}$ lower than the Eddington limit determined by the aforementioned negative feedback. However, if star formation is proceeding in compact, individual regions, those regions can be Eddington limited although the galaxy-integrated SFR appears sub-Eddington (e.g. \cite{simpson2015} and references therein).

In Paper~II, we derived the rest-frame FIR size (deconvolved FWHM at $\lambda_{\rm rest}=163$~$\mu$m) of AzTEC1 to be $0\farcs39^{+0.01}_{-0.01} \times 0\farcs31^{+0.01}_{-0.01}$, which is 
based on ALMA 870~$\mu$m observations mentioned in Sect.~2.2. This corresponds to a physical extent of $2.6^{+0.1}_{-0.0}\times 2.1^{+0.0}_{-0.1}$~kpc$^2$. Based on a dynamical mass analysis ($M_{\rm dyn}=M_{\star}+M_{\rm gas}$, if the dark matter contribution is not important), Yun et al. (2015) concluded that the gas disk of AzTEC1 has to be viewed nearly face-on with an inclination angle of $i \lesssim12\degr$. If we use the aforementioned major axis FWHM as the diameter, and our new stellar mass value for AzTEC1 to recalculate the dynamical mass from Yun et al. (2015), we obtain $M_{\rm dyn}=1.5\times10^{10}\, {\rm M}_{\sun}\times (\sin \, i)^{-2}$, which implies a similar constraint on the inclination angle, namely $i \lesssim16\degr$.
Yun et al. (2015) pointed out that the nearly face-on orientation of AzTEC1 offers a natural explanation for this SMG being so bright in the optical bands.
The fact that the ALMA-based FIR size obeys $\theta_{\rm min}\simeq0.79\times \theta_{\rm maj}$ is in fairly good agreement with the conclusion of AzTEC1 being seen nearly face-on: 
for a simple disk-like geometry, $\theta_{\rm min}=\theta_{\rm maj}\times \cos(i)$, which yields $i=37\fdg4$, i.e. about a factor of 2.3 larger inclination angle than the upper limit we obtain from the dynamical mass constraint. Besides the intrinsic uncertainties in the mass estimates, some of this discrepancy can be partly caused by the fact that Yun et al. (2015) used the $[\ion{C}{II}]$ linewidth to calculate the rotation velocity of the disk, while the region responsible for the FIR continuum emission is probably more compact and can have a somewhat different tilt angle (cf.~\cite{riechers2014}; see also Paper~II and O.~Miettinen et al., in prep.). For comparison, the 3~GHz radio size of AzTEC1 obeys $\theta_{\rm min}\simeq0.64\times \theta_{\rm maj}$ (Paper~II), which suggests $i=50\fdg1$. The final stellar mass of AzTEC1 can be estimated to be $M_{\rm \star,\, final}=M_{\star}+M_{\rm gas}\simeq M_{\rm dyn}\sim2.2\times10^{11}$~M$_{\sun}$ assuming that the galaxy does not undergo gas replenisment by accreting from the circumgalactic and/or intergalactic medium and does not expel the existing gas out. In this scenario, AzTEC1 would evolve into a massive elliptical galaxy in the present-day universe.

Using the aforementioned rest-frame FIR size, we derive a SFR surface density of 
$\Sigma_{\rm SFR}={\rm SFR}/A=378^{+28}_{-22}$~M$_{\sun}$~yr$^{-1}$~kpc$^{-2}$, where the area is defined by 
$A=\pi \times {\rm FWHM}_{\rm maj}/2\times{\rm FWHM}_{\rm min}/2$. Our $\Sigma_{\rm SFR}$ value is a factor of $2.6^{+0.2}_{-0.1}$ below the Eddington limit for a radiation pressure supported starburst of 
$\Sigma_{\rm SFR}^{\rm Edd}\sim10^3$~M$_{\sun}$~yr$^{-1}$~kpc$^{-2}$ (e.g. \cite{andrews2011}), and consistent with the aforementioned Eddington-limited SFE comparison.
Similarly, using our IR luminosity value, we can derive a $L_{\rm IR}$ surface density of 
$\Sigma_{\rm IR}=L_{\rm IR}/A=3.8^{+0.3}_{-0.2}\times10^{12}$~L$_{\sun}$~kpc$^{-2}$. 
This value lies a factor of $2.6^{+0.2}_{-0.2}$ below the $L_{\rm IR}$ surface density expected for an optically thick starburst disk ($\sim10^{13}$~L$_{\sun}$~kpc$^{-2}$; \cite{thompson2005}). 

\subsection{AzTEC3 -- a compact, protocluster-member starburst at $z=5.3$}

\begin{figure}[!h]
\centering
\resizebox{0.9\hsize}{!}{\includegraphics{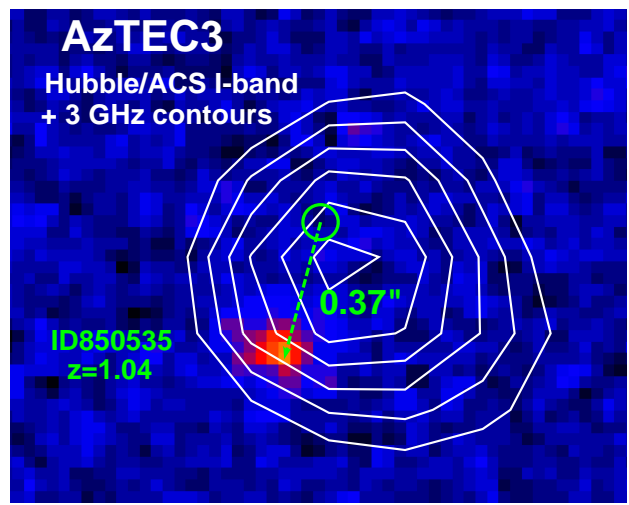}}
\caption{\textit{Hubble}/ACS F814W ($I$-band) image towards AzTEC3 shown with a logarithmic colour scaling, overlaid with contours showing the 3~GHz radio emission (starting at $3\sigma$ and increasing in steps of $1\sigma$; see Paper~II). The foregound galaxy, seen $0\farcs4$ to the SE of the SMA 890~$\mu$m position of AzTEC3 (marked with a green circle; see \cite{younger2007}), lies at $z_{\rm phot}=1.04^{+0.05}_{-0.02}$.}
\label{figure:aztec3}
\end{figure}

AzTEC3 is the best-studied, highest-redshift, and one of the most peculiar SMG in our sample. In Fig.~\ref{figure:aztec3}, we illustrate the position 
of the foreground galaxy with respect to AzTEC3 that has probably contaminated the UV/optical photometry in previous studies (\cite{smolcic2015}) as mentioned in Sect.~3.1 
(see also \cite{younger2007}; their Fig.~1). In the present work, this was revealed by the inconsistency between the optical photometry and the Lyman continuum discontinuity at 
$\lambda_{\rm rest}=911.8$~$\AA$ in our initial SED analysis (cf.~Fig.~4 in \cite{smolcic2015}). A visual inspection of the optical--NIR images towards AzTEC3 suggests that NIR emission 
at and longward of $Y$ band ($\lambda_{\rm rest}=1\,620$~\AA) can be (partly) attributed to the SMG, while the shorter wavelength emission is due to the aforementioned interloper. 
Below, we revisit some of the ISM and star formation properties of AzTEC3 in tandem with the present results. 

Based on the VLA and PdBI observations of multiple rotational lines of CO emission, Riechers et al. (2010) derived a 
gas mass of $M_{\rm gas}=5.3\times10^{10}$~M$_{\sun}$ for AzTEC3 ($\alpha_{\rm CO}=0.8$~M$_{\sun}$~(K~km~s$^{-1}$~pc$^2$)$^{-1}$ was assumed). Using the dust and stellar masses we have derived, we obtain a gas-to-dust mass ratio of $\delta_{\rm gdr}\equiv M_{\rm gas}/M_{\rm dust}=33^{+28}_{-18}$, and a gas fraction of $f_{\rm gas}=M_{\rm gas}/(M_{\rm gas}+M_{\star})\simeq37\%$. 
 We note that the $[\ion{C}{II}]$ luminosity reported by Riechers et al. (2014), $L_{[\ion{C}{II}]}=(6.69\pm0.23)\times10^9$~L$_{\sun}$, suggests a molecular gas mass of $M_{\rm gas}\sim5.2-8.3\times10^{10}$~M$_{\sun}$ (see \cite{swinbank2012}), i.e. only up to a factor of about 1.6 higher than the CO-based value from Riechers et al. (2010). The $[\ion{C}{II}]$-to-IR luminosity ratio for AzTEC3, $L_{[\ion{C}{II}]}/L_{\rm IR}=0.0002^{+0.0001}_{-0}$, is even lower than for AzTEC1, and makes this SMG an object with an extreme $[\ion{C}{II}]$ deficiency.

Assuming that AzTEC3 has sustained its current SFR continuously, its stellar mass building-up timescale is $\tau_{\rm build}=M_{\rm star}/{\rm SFR}\simeq32^{+2}_{-2}$~Myr, while the gas depletion timescale is $\tau_{\rm dep}=M_{\rm gas}/{\rm SFR}\simeq19^{+1}_{-0}$~Myr. For comparison, the {\tt MAGPHYS}-based mass-weighted age of AzTEC3 is about 22~Myr.
  A similarly short gas consumption timescale (20--30~Myr) is found for the $z=4.5$ SMG J1000+0234 (\cite{schinnerer2008}), and 
the $z=4.05$ SMG GN20, which is also a protocluster member (\cite{daddi2009}) similar to AzTEC3 (\cite{capak2011}). 
Given the $\tau_{\rm dep}/\tau_{\rm build}$ ratio of $\sim0.6^{+0.1}_{-0}$ we derive for AzTEC3, its present stellar mass could have been fully built up during the current 
starburst episode, like concluded for AzTEC1 above. 
  The value of ${\rm SFE}={\rm SFR}/M_{\rm gas}=\tau_{\rm dep}^{-1}$ for AzTEC3 is ${\rm SFE}=52^{+1}_{-1}$~Gyr$^{-1}$, and ${\rm SFE}=L_{\rm IR}/M_{\rm gas}=520^{+12}_{-12}$~L$_{\sun}$~M$_{\sun}^{-1}$. The latter is consistent with the aforementioned Eddington limit of ${\rm SFE}_{\rm Edd}=500$~L$_{\sun}$~M$_{\sun}^{-1}$. 

Based on ALMA $\lambda_{\rm rest}=158$~$\mu$m $[\ion{C}{II}]$ observations, Riechers et al. (2014) derived a dynamical mass of 
$M_{\rm dyn}=9.7\times10^{10}$~M$_{\sun}$ for AzTEC3. We can use the aforementioned gas mass and our new stellar mass to estimate the dynamical mass as 
$M_{\rm dyn}\simeq M_{\rm gas}+M_{\star}\simeq1.4\times10^{11}$~M$_{\sun}$. This is only about 1.4 times higher than the $[\ion{C}{II}]$-based value from Riechers et al. (2014), and given the uncertainty in the mass values, this discrepancy does not seem very significant. The final stellar mass of AzTEC3 can be estimated to be $M_{\rm \star,\, final}=M_{\star}+M_{\rm gas}\simeq M_{\rm dyn}\sim1.4\times10^{11}$~M$_{\sun}$, and as found for AzTEC1 above, AzTEC3 could evolve into a massive elliptical galaxy 
in the present-day universe. 

The continuum size at $\lambda_{\rm obs}=1$~mm ($\lambda_{\rm rest}=159$~$\mu$m) derived for AzTEC3 by Riechers et al. (2014) through their ALMA observations is 
$0\farcs40^{+0.04}_{-0.04} \times 0\farcs17^{+0.08}_{-0.17}$ (deconvolved FWHM). This corresponds to a physical rest-frame FIR FWHM size of $2.4^{+0.3}_{-0.2}\times 1.0^{+0.5}_{-1.0}$~kpc$^2$. 
Using this size scale, we derive a SFR surface density of $\Sigma_{\rm SFR}={\rm SFR}/A=1\,461^{+170}_{-615}$~M$_{\sun}$~yr$^{-1}$~kpc$^{-2}$, where the nominal value is 2.8 times higher than reported by Riechers et al. (2014; 530~M$_{\sun}$~yr$^{-1}$~kpc$^{-2}$). Our $\Sigma_{\rm SFR}$ value is consistent with the Eddington limit for a radiation pressure supported starburst of $\Sigma_{\rm SFR}^{\rm Edd}\sim10^3$~M$_{\sun}$~yr$^{-1}$~kpc$^{-2}$ (e.g. \cite{andrews2011}), and hence in agreement with the aforementioned Eddington-limited SFE comparison.
Similarly, using our IR luminosity value, we can derive a $L_{\rm IR}$ surface density of 
$\Sigma_{\rm IR}=L_{\rm IR}/A=1.5^{+0.1}_{-0.6}\times10^{13}$~L$_{\sun}$~kpc$^{-2}$. 
This value is similar to that expected for an optically thick starburst disk ($\sim10^{13}$~L$_{\sun}$~kpc$^{-2}$; \cite{thompson2005}). Besides the compact FIR size derived by Riechers et al. (2014), the high dust temperature we have derived for AzTEC3, $T_{\rm dust}=66.0^{+5.1}_{-8.4}$~K, suggests that star formation takes place on small spatial scales, and hence the dust grains are subject to a strong radiation field (e.g. \cite{yang2007}; \cite{dacunha2015}).\footnote{We searched for a correlation between the 3~GHz radio size and the {\tt MAGPHYS}-based dust temperature among our SMGs, but only a weak relationship was found (the Pearson correlation coefficient was found to be $r= -0.28$, while 0 implies no correlation).}

In their study of AzTEC3, Dwek et al. (2011) concluded that the dust content of this SMG 
formed over a period of $\sim200$~Myr, with a constant SFR of $\sim500$~M$_{\sun}$~yr$^{-1}$ and a top-heavy IMF. We note that the dust mass they adopted in their analysis, 
$M_{\rm dust}=(2\pm1)\times10^{9}$~M$_{\sun}$, is only $1.2^{+2.2}_{-0.9}$ times higher 
than the value we have derived, but an upper limit of $<5\times10^{10}$~M$_{\sun}$ to the stellar mass they assumed is lower than the value of $M_{\star}=8.9^{+0.2}_{-0.4} \times10^{10}$~M$_{\sun}$ we have derived. Since AzTEC3 is seen when the universe was just 1.07~Gyr old, its high dust mass we have derived suggests an efficient dust production, but inefficient dust destruction, by supernovae (cf.~the 'dust-budget crisis' in high-$z$ SMGs; \cite{rowlands2014b}). 

\begin{table}
\renewcommand{\footnoterule}{}
\caption{The physical characteristics of AzTEC1 and AzTEC3.}
{\small
\begin{minipage}{1\columnwidth}
\centering
\label{table:ism}
\begin{tabular}{c c c}
\hline\hline 
Parameter & AzTEC1 & AzTEC3 \\
\hline
$z_{\rm spec}$ & 4.3415 & 5.298 \\
$M_{\rm gas}$ [M$_{\sun}$]\tablefootmark{a} & $(1.4\pm0.2)\times10^{11}$ & $5.3\times10^{10}$ \\[1ex]
$\delta_{\rm gdr}\equiv M_{\rm gas}/M_{\rm dust}$\tablefootmark{b} & $90^{+23}_{-19}$ & $33^{+28}_{-18}$ \\[1ex]
$f_{\rm gas}=M_{\rm gas}/(M_{\rm gas}+M_{\star})$\tablefootmark{c} & $65\%$ & $37\%$ \\[1ex]
$L_{[\ion{C}{II}]}/L_{\rm IR}$\tablefootmark{d} & $0.0005^{+0.0001}_{-0.0001}$ & $0.0002^{+0.0001}_{-0}$ \\[1ex]
$\tau_{\rm build}=M_{\rm star}/{\rm SFR}$ [Myr] & $47^{+2}_{-2}$ & $32^{+2}_{-2}$ \\[1ex]
$\tau_{\rm dep}=M_{\rm gas}/{\rm SFR}$  [Myr] & $86^{+15}_{-14}$ & $19^{+1}_{-0}$ \\[1ex]
${\rm SFE}={\rm SFR}/M_{\rm gas}$ [Gyr$^{-1}$] & $12^{+2}_{-2}$ & $52^{+1}_{-1}$ \\[1ex]
${\rm SFE}=L_{\rm IR}/M_{\rm gas}$ [L$_{\sun}$~M$_{\sun}^{-1}$] & $116^{+22}_{-17}$ & $520^{+12}_{-12}$ \\[1ex]
$M_{\rm dyn}$ [M$_{\sun}$]\tablefootmark{e} & $1.5\times10^{10}\times (\sin \, i)^{-2}$ & $1.4\times10^{11}$ \\[1ex]
$\Sigma_{\rm SFR}={\rm SFR}/A$ [M$_{\sun}$~yr$^{-1}$~kpc$^{-2}$]\tablefootmark{f} & $297^{+22}_{-22}$ & $1\,148^{+133}_{-483}$ \\[1ex]
$\Sigma_{\rm IR}=L_{\rm IR}/A$ [$10^{12}$~L$_{\sun}$~kpc$^{-2}$]\tablefootmark{f} & $3.0^{+0.2}_{-0.2}$ & $11.5^{+1.3}_{-4.9}$ \\[1ex]
\hline 
\end{tabular} 
\tablefoot{\tablefoottext{a}{The gas mass of AzTEC1 was adopted from Yun et al. (2015), while that of AzTEC3 is from Riechers et al. (2010). A ULIRG-type conversion factor of $\alpha_{\rm CO}=0.8$~M$_{\sun}$~(K~km~s$^{-1}$~pc$^2$)$^{-1}$ was assumed in the calculation.}\tablefoottext{b}{The gas-to-dust mass ratio based on our new dust mass value. }\tablefoottext{c}{The gas fraction based on our new stellar mass value. }\tablefoottext{d}{The $[\ion{C}{II}]$ luminosity for AzTEC1 was taken from Yun et al. (2015), while that for AzTEC3 was taken from Riechers et al. (2014).}\tablefoottext{e}{The dynamical mass of AzTEC1 is modified from Yun et al. (2015), while that of AzTEC3 is taken from Riechers et al. (2014). }\tablefoottext{f}{The area $A$ is defined by $A=\pi \times {\rm FWHM}_{\rm maj}/2\times{\rm FWHM}_{\rm min}/2$.}      }
\end{minipage} }
\end{table}

\end{document}